\documentclass[superscriptaddress,prd,nofootinbib,preprintnumbers,longbibliography,11pt,eqsecnum]{revtex4-1}

\usepackage{amsmath} 
\usepackage{graphicx} 
\usepackage{amsthm}
\usepackage{amssymb} 
\usepackage{dsfont}
\usepackage{yfonts}
\usepackage{hyperref}
\usepackage{array,xcolor,graphicx}
\usepackage{booktabs,multirow}
\usepackage[utf8]{inputenc}
\usepackage{mathtools}
\usepackage{etoolbox}
\patchcmd{\section}
  {\centering}
  {\raggedright}
  {}
  {}
\patchcmd{\subsection}
  {\centering}
  {\raggedright}
  {}
  {}

\usepackage[normalem]{ulem}

\usepackage[all]{xy}
\usepackage{tikz}
\usetikzlibrary{arrows.meta}

\hypersetup{colorlinks=true,linkcolor=blue,citecolor=red,urlcolor=blue, linktocpage=true}

\newcommand{\be}{\begin{equation}}
\newcommand{\ee}{\end{equation}}
\newcommand{\bea}{\begin{eqnarray}}
\newcommand{\eea}{\end{eqnarray}}

\newcommand{\nn}{\nonumber\\}

\def\la{\langle}
\def\ra{\rangle}
\def\Tr{{\mathrm{Tr}}}

\def\d{\partial}
\def\grad{\nabla}
\def\vecb#1{\mathbf{#1}}

\def\tp{\tau_{\Pi}}

\def\im{\text{Im}}

\def\sig{{\sigma}}

\def\CA{\mathcal{A}}
\def\CB{\mathcal{B}}

\def\CD{\mathcal{D}}

\def\CF{\mathcal{F}}

\def\CH{\mathcal{H}}

\def\CJ{\mathcal{J}}

\def\CM{\mathcal{M}}

\def\CO{\mathcal{O}}

\def\wfr{\mathfrak{w}}

\def\fD{\mathfrak{D}}

\def\lgb{\lambda_{\scriptscriptstyle GB}}
\def\ggb{\gamma_{\scriptscriptstyle GB}}

\begin{document}

\preprint{MIT-CTP/5075}
    
\title{Holography and hydrodynamics with weakly broken symmetries}

\author{Sa\v{s}o Grozdanov}
\affiliation{Center for Theoretical Physics, Massachusetts Institute of Technology, Cambridge, MA 02139, USA}

\author{Andrew Lucas}
\affiliation{Department of Physics, Stanford University, Stanford, CA, 94305 USA}

\author{Napat Poovuttikul}
\affiliation{University of Iceland, Science Institute, Dunhaga 3, IS-107, Reykjavik, Iceland}

\begin{abstract}
Hydrodynamics is a theory of long-range excitations controlled by equations of motion that encode the conservation of a set of currents (energy, momentum, charge, etc.) associated with explicitly realized global symmetries. If a system possesses additional weakly broken symmetries, the low-energy hydrodynamic degrees of freedom also couple to a few other ``approximately conserved" quantities with parametrically long relaxation times.  It is often useful to consider such approximately conserved operators and corresponding new massive modes within the low-energy effective theory, which we refer to as {\em quasihydrodynamics}. Examples of quasihydrodynamics are numerous, with the most transparent among them hydrodynamics with weakly broken translational symmetry. Here, we show how a number of other theories, normally not thought of in this context, can also be understood within a broader framework of quasihydrodynamics: in particular, the M\"uller-Israel-Stewart theory and magnetohydrodynamics coupled to dynamical electric fields. While historical formulations of quasihydrodynamic theories were typically highly phenomenological, here, we develop a holographic formalism to systematically derive such theories from a (microscopic) dual gravitational description. Beyond laying out a general holographic algorithm, we show how the M\"uller-Israel-Stewart theory can be understood from a dual higher-derivative gravity theory and magnetohydrodynamics from a dual theory with two-form bulk fields. In the latter example, this allows us to unambiguously demonstrate the existence of dynamical photons in the holographic description of magnetohydrodynamics. 
\end{abstract}

\maketitle

\newpage
\begingroup
\tableofcontents
\endgroup

\newpage 

\section{Introduction}

The past decade has seen a resurgence of interest in developing a systematic understanding of hydrodynamics as an effective field theory, describing the relaxation of locally conserved quantities towards global equilibrium in terms of long-lived (low-energy) degrees of freedom \cite{Dubovsky:2011sj,Grozdanov:2013dba,Haehl:2015foa,Crossley:2015evo,Haehl:2015uoc,Torrieri:2016dko,Glorioso:2016gsa,Jensen:2017kzi,Glorioso:2018wxw}. While the formulation of a dissipative hydrodynamic theory from an action principle was a long-outstanding problem, the rapid development of its reformulation in terms of effective field theory, along with other formal approaches to its classification \cite{Baier:2007ix,Bhattacharyya:2008jc,Romatschke:2009kr,Grozdanov:2015kqa,Haehl:2015pja}, were largely ignited and accelerated by the advent of gauge-string duality (holography) \cite{Maldacena:1997re}, in particular, its ability to describe hydrodynamics of strongly interacting states \cite{Policastro:2002se}.

The lines of research that have sprung from these developments have led to a number of important applications, ranging from a vastly improved understanding of thermalization and hydrodynamization of the quark-gluon plasma resulting from heavy-ion collisions \cite{CasalderreySolana:2011us,Grozdanov:2016zjj,Florkowski:2017olj,Romatschke:2017ejr}, a comprehensive formulation of the theory of magnetohydrodynamics \cite{Grozdanov:2016tdf,Hernandez:2017mch}, to an understanding of the dynamics of electrons in exotic `strange' metals \cite{Bandurin1055,Crossno1058,Moll16,Lucas:2017idv}.  For recent overviews see \cite{zaanen2015holographic,hartnoll2018holographic}.   Many of the applications listed here pertain to old  problems in physics. As a result, various phenomenological hydrodynamic approaches to their resolution have been known for decades.  Unfortunately, these phenomenological approaches often lack rigor. In particular, a strategy common to many of these attempts is a rather ad hoc coupling of fluid degrees of freedom (particle density, momentum density, velocity, etc.) to {\em non-hydrodynamic} degrees of freedom (magnetic field, chemical reactant concentration, etc.). Another is an explicit breaking of various conservation laws (energy, momentum, charge, etc.).  

A classic example, which we will study in detail in this work, is the textbook formulation of magnetohydrodynamics (MHD) (see e.g. \cite{davidson_2001,freidberg_2014}). MHD combines a theory of fluid degrees of freedom that obey the continuity equation and the forced Euler (or Navier-Stokes) equation, while the dynamics of electromagnetic fields obeys Ampere's law (neglecting displacement current), Faraday's law and magnetic Gauss's law. Momentum conservation of the fluid sector is explicitly broken (forced) by the addition of an external Lorentz force and the electric field is commonly expressed in terms of the magnetic field via the boosted ideal Ohm's law in the limit of infinite conductivity (see \cite{Grozdanov:2017kyl}). Standard formulation of MHD can be seen as lacking a systematic coupling between the separated fluid and electromagnetic degrees of freedom, as well an understanding of (global) symmetries by which one can organize a theory of long-range excitations in plasmas. These questions were addressed recently in \cite{Grozdanov:2016tdf} where MHD was reformulated and extended by using the language of higher-form symmetries \cite{Gaiotto:2014kfa}. As a result of the above-mentioned issues with the historical approach to MHD, the standard formulation of MHD explicitly breaks the gradient expansion; this is particularly acute when MHD is coupled to dynamical electric fields. Nevertheless, while from a formal effective field theory point of view such a theory should be viewed with suspicion, standard MHD and its simple phenomenological extensions make a number of extremely successful predictions about the dynamics of complicated astrophysical plasmas and processes in fusion reactors. 

Another classic example of a system with an explicitly broken conservation law is fluid dynamics with weak momentum relaxation \cite{Hartnoll:2012rj,Davison:2014lua,hartnoll2018holographic}. For example, such systems are known to describe both the drag of the Earth's surface on the atmospheric fluid \cite{atmosphere}, and the hydrodynamics of electrons in clean metals \cite{Lucas:2017idv}.  

In light of these and other examples, the main goal of this work is to elucidate a formal approach to studying hydrodynamic theories with weakly broken symmetries, both from the point of view of field theory and holography.  We will show that a large number of existing phenomenological theories can be precisely understood within this framework.

Hydrodynamics is a theory which is valid on length and time scales that are long compared to the mean free time $\tau_{\mathrm{mft}}$ and mean free path $\ell_{\mathrm{mfp}}$ \cite{LLfluid}.  Since hydrodynamics is a gradient expanded theory, it is convenient to express these statements formally as
\begin{align}
  \tau_{\mathrm{mft}} \partial_t \ll 1,\qquad   \ell_{\mathrm{mfp}} \partial_i \ll 1,   \label{eq:hydroineq}
\end{align}
embodying the fact that gradients of relevant fields need to be small compared to the inverse time and length scales. The energy scale set by $1/\tau_{\mathrm{mft}}$ should be though of as the ultra-violet (UV) cut-off of the effective theory of hydrodynamics. The hydrodynamic equations of motion take the schematic form 
\begin{equation}
 \partial_t \langle\rho_A\rangle + \partial_i J_A^i \left[\langle \rho_A\rangle, \partial_j \langle\rho_A\rangle, \cdots\right] = 0, \label{eq:hydroformal}
\end{equation}
where $\rho_A$ correspond to conserved densities, and $J_A^i$ to conserved currents which can be expanded in a gradient expansion series of higher-order hydrodynamics (see e.g. \cite{Kovtun:2012rj,Grozdanov:2015kqa}). The expectation values $\la \rho_A \ra$ are computed in the equilibrium state of the system, e.g. the thermal equilibrium $\la \rho_A \ra = \Tr[\rho_A \, \mathrm{e}^{-\beta H}]$, with $\beta =1 / T$  the inverse temperature.

\begin{figure*}[t]
\centering
\includegraphics[width=1\textwidth]{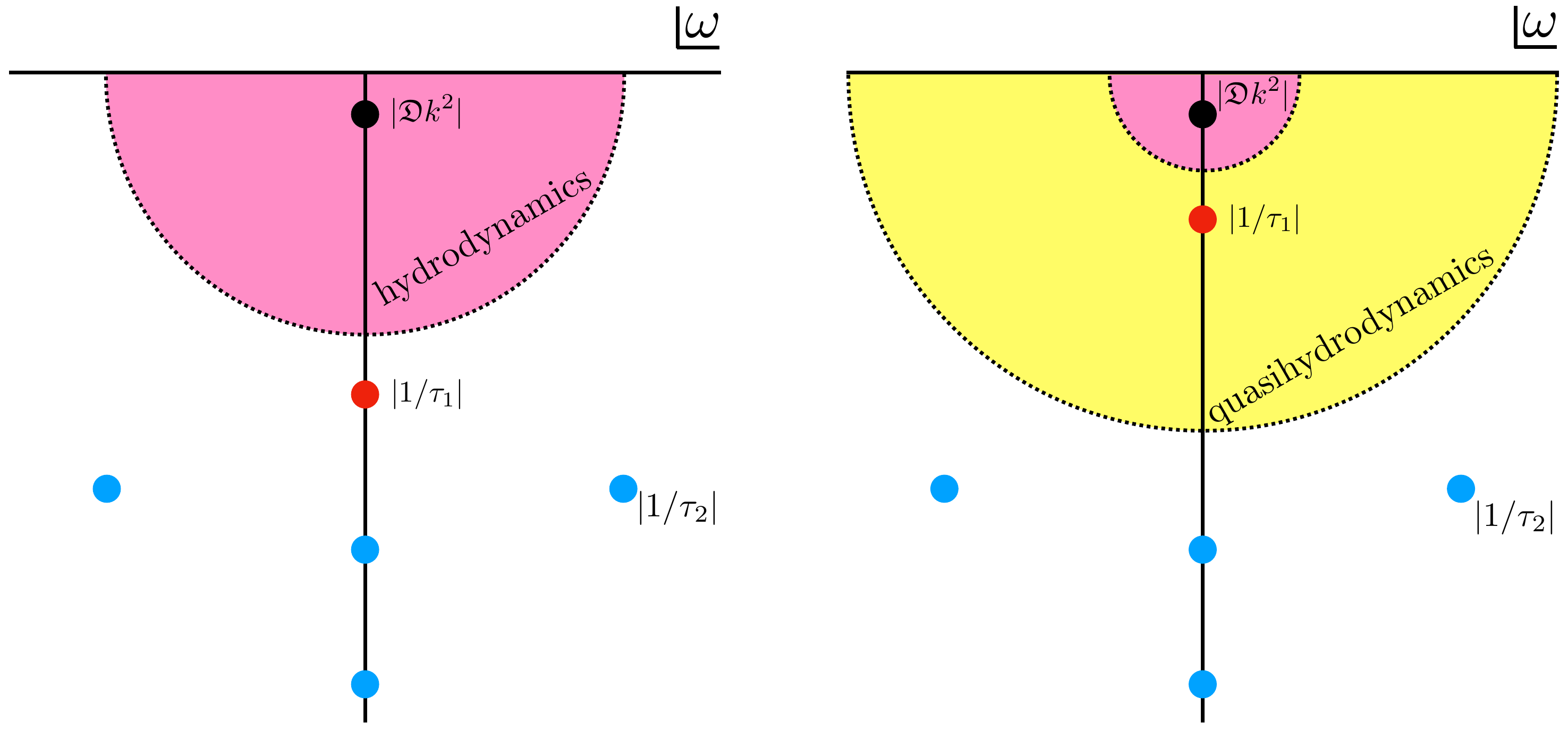}
\caption{A schematic depiction of the ranges of validity of \textit{hydrodynamics} and \textit{quasihydrodynamics} for two distinct spectra plotted on the complex frequency $\omega$ plane. In the left panel, relaxation times of all non-hydrodynamic modes are comparable and the IR part of the spectrum is dominated by hydrodynamic modes.   There is no quasihydrodynamic regime as $\tau_1 \sim \tau_2$.   In the right panel, $\tau_1$ is much larger than the other relaxation times ($\tau_1 \gg \tau_2$) and so the hydrodynamic regime (shaded pink) has a greatly reduced regime of validity.   Because $\tau_1 \gg \tau_2$, if we include the resulting long-lived mode in our effective theory, we obtain an improved quasihydrodynamic theory which is valid in a parametrically larger regime (shaded yellow). In each of the plots, the black circle depicts the hydrodynamic mode with the usual diffusive decay rate $\fD k^2$.   The red circle is the massive mode $\la P \ra$ with the longest relaxation time $\tau_1$;  blue circles depict modes with faster relaxation times $\tau_2,\,\tau_3,$ etc.}
\label{fig:decayRateOfP}
\end{figure*}

Now,  suppose that there exists a single \emph{non-conserved operator} $P$ in the theory, which has a relaxation time (inverse decay rate) $\tau_1$, so that $\langle P(t)P(0)\rangle  \sim \mathrm{e}^{-t/\tau_1}$, while all other `orthogonal' operators have a much smaller relaxation time $\{\tau_2,\tau_3, \,\ldots \} \ll \tau_1$.\footnote{The notion of operator orthogonality here can be formally understood in terms of thermodynamic susceptibilities, see e.g. \cite{hartnoll2018holographic}.} Formally speaking, the hydrodynamic degrees of freedom do not include $\langle P\rangle$.   Since $\langle P\rangle$ must relax before the theory is described by hydrodynamic modes alone, $\tau_{\mathrm{mft}} \gtrsim \tau_1$ (cf. Eq. \eqref{eq:hydroineq}). However, as shown in Figure \ref{fig:decayRateOfP}, it may be the case that the decay time $\tau_1$ of $\langle P\rangle$ is significantly larger than the decay time for all other non-hydrodynamic modes:  if $\mathcal{O}$ represents a generic local operator, orthogonal to the hydrodynamic modes and to $P$, then $\langle \mathcal{O}(t)\mathcal{O}(0)\rangle \sim \mathrm{e}^{-t/\tau_2}$, and $\tau_2 \ll \tau_1$.    In such a setting, unfortunately, hydrodynamics breaks down once $\tau_1 \partial_t \sim 1$, because on these time scales, there is still only a finite number of degrees of freedom:  the hydrodynamic modes and $\langle P\rangle$.   Rather than integrating out $\langle P\rangle$, we should keep it in our effective theory, so that our description of the dynamics remains valid all the way until $\tau_2 \partial_t \sim 1$.  At a phenomenological level, we expect that the equations of motion will appear to be hydrodynamic, except that $\langle P \rangle$ will have a small decay rate $1/\tau_1$.   Thus, we replace Eq. \eqref{eq:hydroformal} with
\begin{subequations}
  \label{eq:introlonglived2}
\begin{align}
    \partial_t \langle \rho_A\rangle + \partial_i J_A^i[\langle \rho_A\rangle, \partial \langle\rho_A\rangle, \cdots, \langle P\rangle, \partial_j \langle P\rangle, \cdots] &= 0, \\
    \partial_t \langle P \rangle + \partial_i J_P^i [\langle \rho_A\rangle, \partial \langle\rho_A\rangle, \cdots, \langle P\rangle, \partial_j \langle P\rangle, \cdots] &= -\frac{\langle P\rangle}{\tau_1}. 
\end{align}\end{subequations}
So long as $\tau_1 \gg \tau_2$, these equations of motion can describe the evolution of the system extremely accurately, even on time scales which are short compared to $\tau_1$.   Since (\ref{eq:introlonglived2}) is valid so long as $\tau_2\partial_t \ll 1$, it is  a parametric improvement over the hydrodynamic expansion without $\langle P\rangle$ as a degree of freedom, as hydrodynamics is only valid when $\tau_1 \partial_t \ll 1$.   As will be discussed in the main body of this work, there is a well-developed theory for explicitly computing $\tau_1$, either from field theory or from holography, as the answers are known to exactly match \cite{Lucas:2015vna}.  

The reason that it is useful to include $\langle P\rangle$ as a formal degree of freedom within effective theory is that in many cases, $\tau_1$ is a divergent function of a single dimensionless parameter $\lambda$:  $\tau_1(\lambda) \sim \lambda^{-\alpha}$.   Then as $\lambda \rightarrow 0$,  there is a parametric separation between $\tau_1$ and $\tau_2$, such that the resulting effective theory (\ref{eq:introlonglived2}) is local and well behaved.  For concreteness, let us imagine a conventional fluid, placed in a medium with static inhomogeneity which (weakly) breaks translational invariance.   If the dimensionless amplitude of the inhomogeneous source is $\lambda$, then momentum will not be exactly conserved;  instead, it will decay on a time scale $\tau_1\sim \lambda^{-2}$.     When $\lambda \ll 1$,   $\tau_1$ is very large and, as we will review, there is a controlled prescription for computing $\tau_1$.  Obtaining (\ref{eq:introlonglived2}) using this prescription, hydrodynamics can be improved to systematically account for weakly broken symmetries (in this example, spatial translations) and their corresponding approximate conservation laws.   It is straightforward to generalize  \eqref{eq:introlonglived2} to the case where a finite list of operators $P^\alpha$ is long-lived.  For lack of a better phrase, we will call the theory in \eqref{eq:introlonglived2}, in which an approximately conserved quantity is treated on the same footing as exactly conserved quantities, a {\em quasihydrodynamic} theory. 

Let us emphasize from the outset that the quasihydrodynamic equations (\ref{eq:introlonglived2}) are physical equations:   all quasinormal modes and poles in correlation functions which are predicted by (\ref{eq:introlonglived2}) on frequency scales $|\omega| \ll \tau_2^{-1}$ must exist in the true physical system.   Because the formalism for deriving quasihydrodynamics is only exact when $\tau_1^{-1}$ is perturbatively  small ($\lambda \ll 1$), one should not allow for the relaxation time $\tau_1$ appearing in (\ref{eq:introlonglived2}) to become small.   If $\tau_1\sim \tau_2$ (as might be expected when the symmetry breaking parameter $\lambda \gtrsim 1$), the quasihydrodynamic equations (\ref{eq:introlonglived2}) do not make sense and must be replaced with ordinary hydrodynamic equations (\ref{eq:hydroformal}).    The difference between the limits $\lambda \gtrsim 1$ (left panel) and $\lambda \ll 1$ (right panel) is illustrated in Figure \ref{fig:decayRateOfP}.

The purpose of this work is to make a three-fold contribution to an already existing theory of quasihydrodynamics, which should help in unifying a number of past results under a common language as well as establish a systematic way to study such theories in the future.  Firstly, we point out that---at least within linear response--a large number of well-known phenomenological theories are quasihydrodynamic: these include not only the momentum-relaxing fluid, but also magnetohydrodynamics and plasma physics with dynamical photons, simple models of viscoelasticity, (at least in some cases) the M\"uller-Israel-Stewart (MIS) theory of relativistic hydrodynamics, and (quantum) kinetic theory (in some respects).  In every case, we identify both the approximately conserved quantities and the perturbatively small parameters which govern their decay rates.  Some of these  identifications are novel and may lead to new insights into old phenomenolgoical theories.  Secondly, we note that many quasihydrodynamic theories exhibit a universal  ``semicircle" law in which a hydrodynamic diffusion pole collides with a quasihydrodynamic pole to create a propagating wave.  A well-understood example is the formation of sound waves in a momentum-relaxing fluid at frequencies $\gg \tau_1^{-1}$.  Examples that (to the best of our knowledge) have never yet been understood as quasihydrodynamic include transverse sound waves in elastic solids and electromagnetic waves in plasma physics.   Thirdly, we study strongly coupled quantum theories with holographic duals \cite{hartnoll2018holographic} and describe how quasihydrodynamics arises from the bulk perspective.    Within linear response, we show that the quasihydrodynamic regime exists and that quasihydrodynamics can be resummed to all orders in $\omega\tau_1$ by a controlled bulk computation of the quasihydrodynamic correlation functions and equations of motion.  This allows us to---for the first time---present an unambiguous identification of a photon in a holographic plasma, thereby justifying claims made in Refs. \cite{Grozdanov:2017kyl,Hofman:2017vwr}.  We also demonstrate how in a specific holographic model, MIS phenomenology and quasihydrodynamics with an approximately conserved stress tensor can arise. Earlier calculations along similar lines can be found in \cite{Chen:2017dsy}.

The outline of  this paper is as follows.  In Section \ref{sec:hydro}, we discuss the general quasihydrodynamic framework, and explicitly show that theories including MIS theory, magnetohydrodynamics and viscoelasticity are quasihydrodynamic in certain controllable limits.   This part of the paper will serve as a more detailed summary of our results.   Section \ref{sec:Outline-Hol} summarizes our holographic algorithm for analytically computing quasihydrodynamic equations of motion for the boundary theory, which we expect will find broad applicability to similar holographic problems.  Sections \ref{sec:MHD-Hol} and \ref{sec:MIS-Hol} deal with holographic theories of magnetohydrodynamics and higher-derivative Einstein-Gauss-Bonnet gravity, respectively.  In each case, we show analytically how the quasihydrodynamic limit arises.

\newpage
\section{Hydrodynamics with weakly broken symmetries} \label{sec:hydro}

In this section, we outline how our framework of quasihydrodynamics for systems with approximately conserved quantities can be used to understand a large number of phenomenological theories known from past literature. In particular, we begin by giving precise formulas for the phenomenological $\tau_1$ introduced in (\ref{eq:introlonglived2}).  We will then describe multiple examples of quasihydrodynamic theories and discuss the consequences of weakly broken symmetries on the quasihydrodynamic modes.

\subsection{Linear response}

Let us first summarize what is known about the equations of motion of a quasihydrodynamic theory, within linear response.  Suppose that the many-body Hamiltonian $H_0$ admits a number of local conservation laws associated with charge densities $\rho_A$ and $\rho_a$:  \begin{equation}
    \left[H_0, \int d^d\mathbf{x}\; \rho_A(\mathbf{x})\right] = \left[H_0, \int d^d\mathbf{x}\; \rho_a(\mathbf{x})\right]= 0.
\end{equation} 
One of the $\rho_A$ is always the energy density.   The operators associated with ${\rho}_a$ are also conserved, i.e. their charges commute with the Hamiltonian $H_0$.  Let us now perturb the Hamiltonian as
\begin{equation}
    H = H_0 + \epsilon H_1,
\end{equation}
so that $\rho_a$ no longer commute with the full Hamiltonian $H$: 
\begin{subequations}
    \begin{align}
      \left[H_1, \int d^d\mathbf{x}\; \rho_A(\mathbf{x})\right] &= 0, \\ 
      \left[H_1, \int d^d\mathbf{x}\; \rho_a(\mathbf{x})\right] &\ne 0.
    \end{align}
\end{subequations}
The charge densities $\rho_a$ are our approximately conserved quantities.  The quasihydrodynamic expansion is a derivative expansion in which---as we will see---$\epsilon \sim \partial^\alpha$ will scale with derivatives.  

Let $\mu_A$ and $\mu_a$ denote the thermodynamic conjugates (i.e., generalized chemical potentials) to $\rho_A$ and $\rho_a$, respectively, and let \begin{equation}
    \chi_{AB} = \frac{\partial \langle \rho_A\rangle}{\partial \mu_B}
\end{equation} 
denote the susceptibility matrix of the hydrodynamic operators.  The susceptibility matrices $\chi_{Ab}$ and $\chi_{ab}$ are defined in a similar manner.   Suppose that when $\epsilon=0$, the hydrodynamic equations read \begin{equation}
    \partial_t \langle \rho_{A,a}\rangle + \partial_i \langle J^i_{A,a}\rangle = 0,
\end{equation} 
where $\langle \rho_{A,a}\rangle$ and $\langle J^i_{A,a}\rangle$ are implicitly functions of $\mu_{A,a}$ and their derivatives.   One can show that when $\epsilon \ne 0$ \cite{Lucas:2015pxa}, 
\begin{subequations} \label{eq:sec2hydro}
    \begin{align}
        \partial_t \langle \rho_A\rangle + \partial_i \langle J^i_A\rangle &= 0, \\ 
        \partial_t \langle \rho_a\rangle + \partial_i \langle J^i_a\rangle &= -\epsilon N_{ab} \mu_b - \epsilon^2 M_{ab} \mu_b + \mathrm{O}(\epsilon^3, \epsilon \partial ), 
    \end{align}
\end{subequations}
where \begin{subequations} \label{eq:memmat}\begin{align}
    N_{ab} &= \frac{\partial \langle \dot\rho_a \rangle}{\partial \mu_b}, \\
    M_{ab} &= \lim_{\omega \rightarrow 0} \frac{1}{\omega}\mathrm{Im}\left(G^{\mathrm{R}}_{\dot\rho_a \dot\rho_b}(\mathbf{k}=\mathbf{0},\omega)\right),
\end{align}
\end{subequations}
and $\dot{X} = i[H_1,X]$ denotes the decaying part of the approximately conserved densities (up to $\epsilon$, which has been rescaled out).   Note that $N_{ab} = -N_{ba}$ \cite{Lucas:2015pxa}.   All dissipative contributions to quasihydrodynamics are contained in the matrix $M_{ab}$, proportional to the spectral weight of $\dot{\rho}_a$.   The left-hand side of (\ref{eq:sec2hydro}) can also be expanded as a power series  in $\epsilon$, but as we will see, such terms will always be subleading compared to the orders in the derivative expansion in which we are interested.

If $N_{ab}\ne 0$, then we take $\epsilon \sim \partial$ in the gradient expansion.  An example of such a system is applying a small, non-dynamical external magnetic field $B$ to a charged fluid, which breaks momentum conservation in two spatial directions.  In this case $\epsilon \propto B$, and the approximately conserved operators labeled by $a$ are two spatial momenta: e.g. $P_x$ and $P_y$.   The $M_{ab}$ corrections are subleading in the derivative expansion and contribute at the same order as viscosity.

If $N_{ab}=0$, which is the case we will focus on in this paper, then we take $\epsilon^2 \sim \partial$ in the gradient expansion.  A similar observation was made in \cite{Blake:2015epa}.   In this case, $M_{ab}$ is treated as a first-order term in the gradient expansion.

\subsection{Diffusion-to-sound crossover}\label{sec:DS}

A classic and possibly simplest quasihydrodynamic model is an example of the diffusion-to-sound crossover that interpolates from Fick's law of diffusion at low frequencies $\omega(k)$ to a propagating, sound-like linear (in $k$) waves at high frequencies $\omega(k)$.  This example is illustrative, and we will see how it arises in a diverse set of physical systems in later sections.   

One historical motivation for this model is as follows \cite{forster1995}:  consider a diffusion equation governing, e.g., the magnetization in a spin chain with a locally conserved spin $S = \langle \sigma_z\rangle$:  
\begin{equation}
    \partial_t S = \fD\partial_x^2 S + \cdots. \label{eq:fick}
\end{equation}
Here, $\fD$ is the spin diffusion constant; the hydrodynamic spin current is $J = -\fD\partial_x S + \cdots$, where $\cdots$ denotes higher-derivative corrections.   One can compute hydrodynamic Green's functions that follow from (\ref{eq:fick}) \cite{Kadanoff1963419}, which accurately describe two-point functions in the quantum theory at small $k$ and $\omega$.  These Green's functions have a pole with the dispersion relation $\omega(k)$ given to first order by  
\begin{equation}
    \omega = -i\fD k^2 + \cdots.  \label{eq:omegafick}
\end{equation} 
Such poles are often called hydrodynamic quasinormal modes. The leading-order correction to \eqref{eq:omegafick} arises from third-order hydrodynamics with a term proportional to $\CO(k^4)$ \cite{Grozdanov:2015kqa}.

There are several reasons why a dispersion relation of the form of \eqref{eq:omegafick} is problematic. Its group velocity $\partial\omega/\partial k$ is proportional to $k$, which means that at large $k$, the propagation of diffusive modes is superluminal and thus acausal. A related problem is that hydrodynamic Green's functions fail to obey microscopic sum rules, thus breaking unitarity \cite{forster1995}.   A sum rule typically fixes $\int_{-\infty}^\infty \mathrm{d}\omega \, \omega \, \mathrm{Im}[G^{\mathrm{R}}(\omega)] < \infty$, and as the integral runs over large $\omega$, we cannot trust (\ref{eq:fick}). Of course, it is clear that \eqref{eq:fick} should not be taken seriously once $\omega\tau_{\mathrm{mft}} \gtrsim 1$, so there is no true inconsistency between hydrodynamics and the sum rule.  Nevertheless, it is helpful to have a toy model which is both consistent with microscopic sum rules, and obeys (\ref{eq:fick}) on long wavelengths. This can be arranged by introducing an additional quasihydrodynamic field $\CJ$ and {\em regulating} (\ref{eq:fick}) in a manner compatible with  \eqref{eq:introlonglived2}:  
\begin{subequations} \label{eq:fick2}
\begin{align}
        \partial_t S + \partial_x \mathcal{J} &= 0, \\
    \partial_t \mathcal{J} + \frac{\fD}{\tau}\partial_x S &= -\frac{1}{\tau} \mathcal{J}.
\end{align}
\end{subequations}
The dispersion relation for a single diffusive mode (\ref{eq:omegafick}) is then replaced by a pair of modes that follow from the quadratic polynomial equation for $\omega$,
\begin{align}\label{DtoS-Poly}
\omega^2 + \frac{i}{\tau} \, \omega - \frac{\fD}{\tau} k^2 = 0,
\end{align}
and have dispersion relations 
\begin{equation}
    \omega_\pm = -\frac{i}{2\tau} \left( 1 \pm \sqrt{ 1- 4 \fD \tau k^2 }\right) . \label{eq:omegafick2} 
\end{equation}
At small momentum $k$, \eqref{eq:omegafick2} gives a diffusive mode \eqref{eq:omegafick} and an additional massive gapped mode with the gap controlled by the inverse relaxation time $1/\tau$:
\begin{align}\label{FickExpModes}
\omega_- = - i \fD k^2 + \CO(k^4), && \omega_+ = - \frac{i}{\tau} + i \fD k^2 + \CO(k^4).
\end{align}
The existence of an additional mode is a direct consequence of introducing a new dynamical field $\CJ$ with $\partial_t \CJ$ in \eqref{eq:fick2}, which makes the final (determinant) equation \eqref{DtoS-Poly} quadratic in $\omega$. At high $k$, the dispersion relation becomes linear
\begin{align}
\omega_\pm = \pm \sqrt{ \frac{\fD}{\tau} } \, k - \frac{i}{2\tau} \pm \CO(1/k),
\end{align}
which is why it is often said, somewhat imprecisely, that such modes become sound modes at large $k$. This dispersion relation is causal so long as its group velocity is
\begin{align}\label{DS-speed}
v = \lim_{k\to\infty} \left| \frac{\partial \omega }{ \partial k} \right|  = \sqrt{\frac{\fD}{\tau}} \leq 1.
\end{align}

\begin{figure*}[t]
\centering
\includegraphics[width=0.45\textwidth]{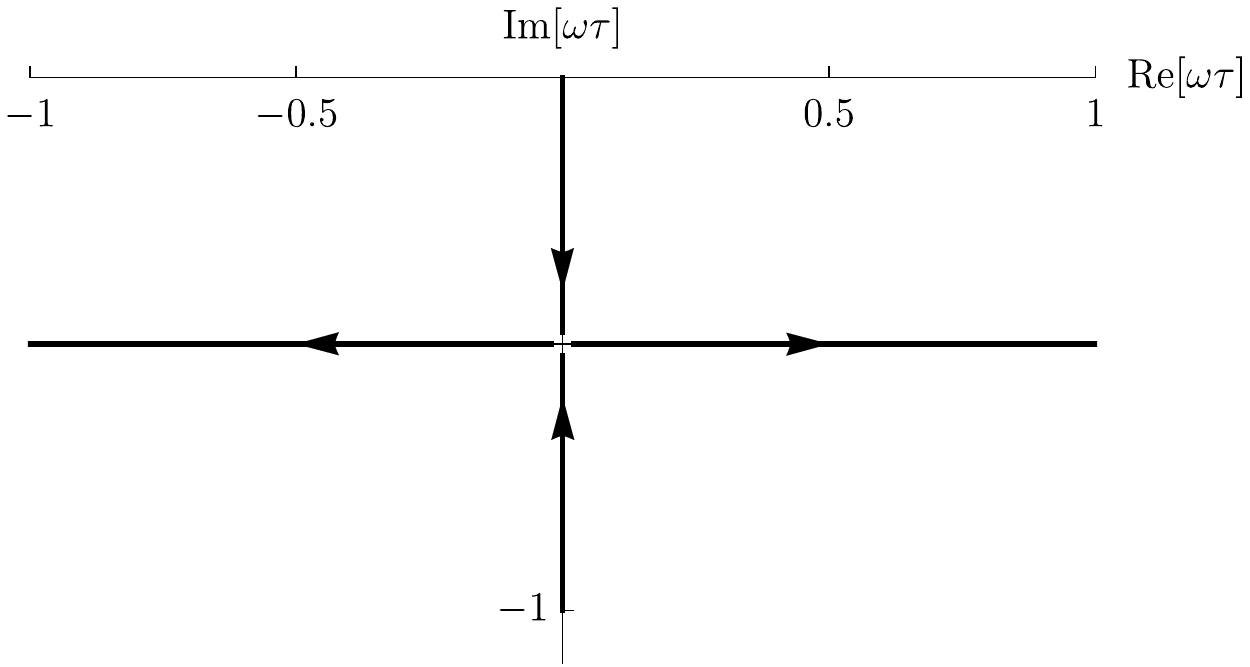}
\caption{The collision of the diffusive and gapped modes from the dispersion relation \eqref{eq:omegafick2} is plotted on the complex $\omega\tau$ plane for a choice of $\fD/\tau = 1/2$. Arrows indicate the direction of movement of the poles as $k\tau$ is increased. The poles start on the imaginary axis, collide, and move off the axis.}
\label{fig:SC-P3}
\end{figure*}

\begin{figure*}[t]
\centering
\includegraphics[width=0.45\textwidth]{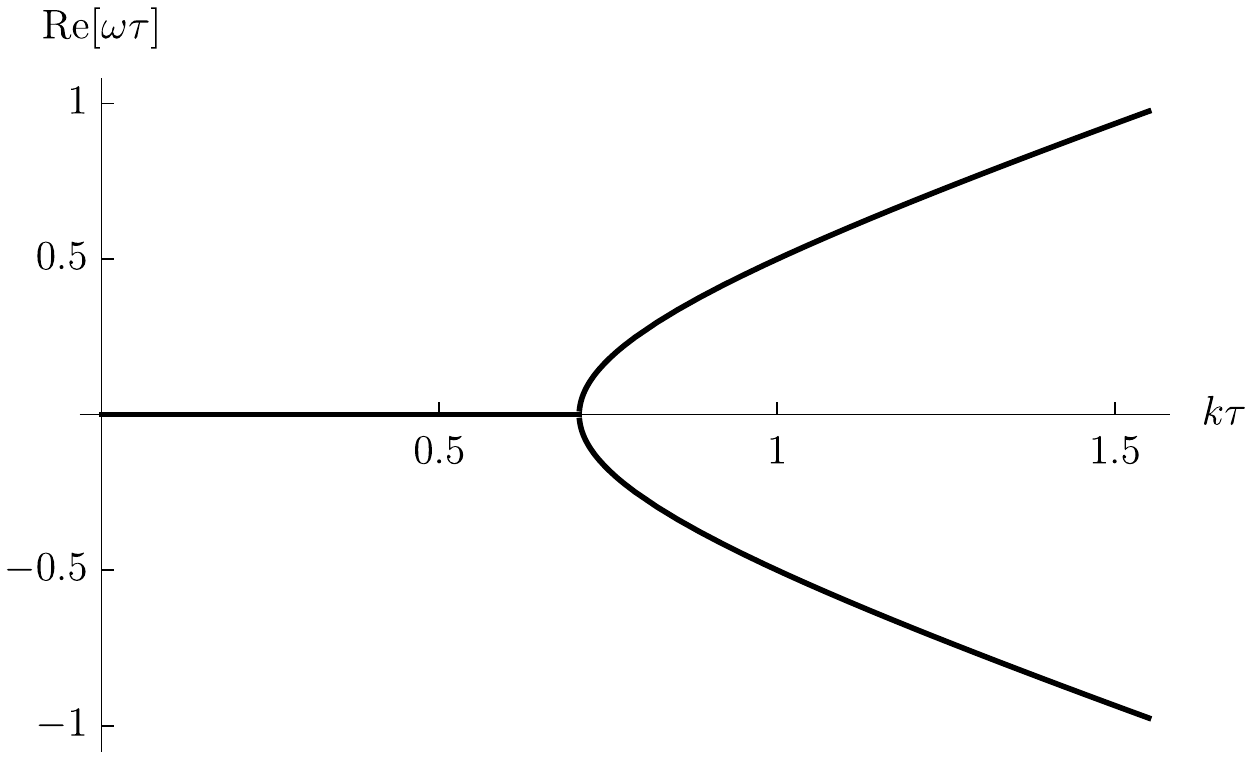}
\hspace{0.08\textwidth}
\includegraphics[width=0.45\textwidth]{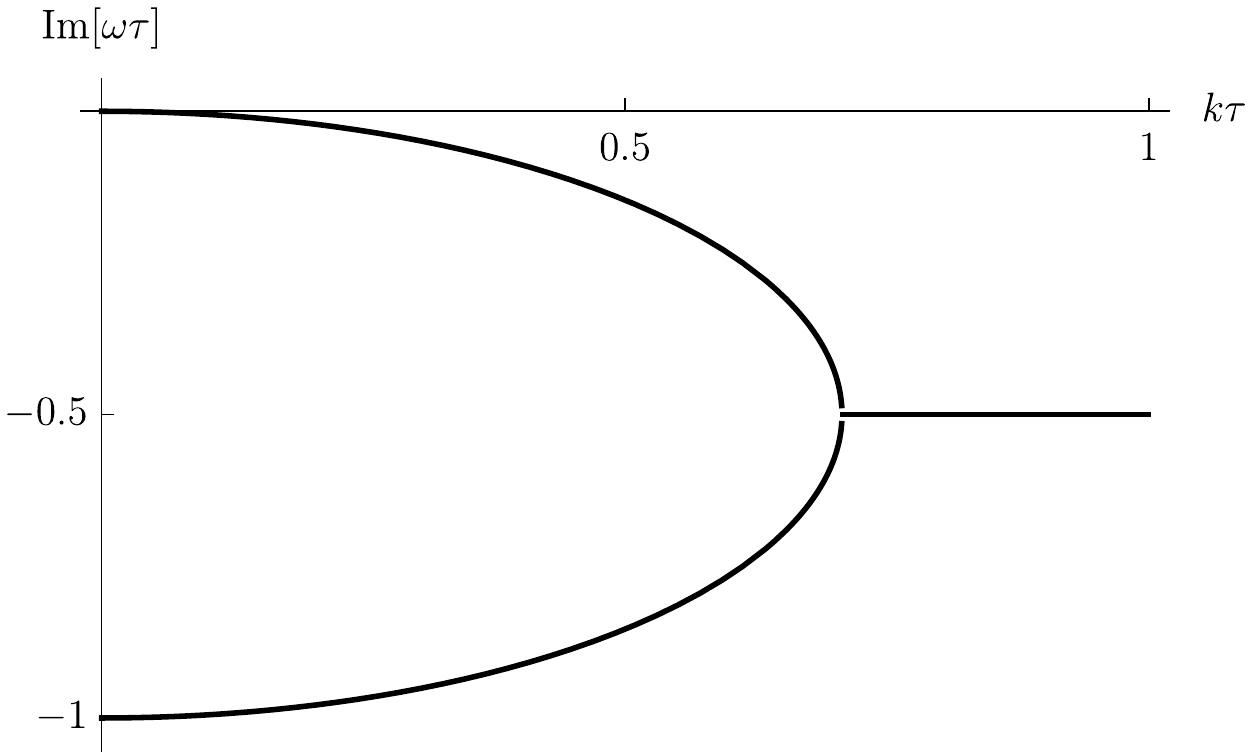}
\caption{Plots of the real and imaginary parts of the two dispersion relations in \eqref{eq:omegafick2} for a choice of $\fD/\tau = 1/2$.}
\label{fig:SC-P1-P2}
\end{figure*}

The full dispersion relations from Eq. \eqref{eq:omegafick2} exhibit the simplest signature of quasihydrodynamics: the collision of the two poles on the imaginary $\omega$ axis as a function of increasing momentum $k$.  This collision occurs at  
\begin{align}
k_c = \sqrt{\frac{1}{4 \fD \tau} },
\end{align}
which introduces a length scale into the problem above, well below which the strict hydrodynamics (\ref{eq:fick}) applies; otherwise, the quasihydrodynamics of Eq. (\ref{eq:fick2}) applies. For the theory presently under consideration, the collision is plotted on the dimensionless complex $\omega\tau$ plane in Figure \ref{fig:SC-P3}. We also plot the real and imaginary parts of dimensionless $\omega \tau$ as a function of $k\tau$ in Figure \ref{fig:SC-P1-P2}. Note that the behavior of the imaginary part of $\omega\tau$ displays the ``semicircle law" mentioned in the Introduction.

In terms of the quasihydrodynamic framework of this work, the physical motivation behind Eq. \eqref{eq:fick2} can be understood as promoting the spin current operator itself to being long-lived. There is no generic reason for this to occur, but when it does, then the present framework becomes applicable. As summarized in the Introduction, what we claim is that then, this type of a pole collision should be seen as a signature of the presence of an approximately conserved quantity.  The existence of such an approximate symmetry can then either arise from microscopic dynamics or exist accidentally at certain special points in the parameter space of couplings and other tuneable parameters. 

Let us end this section by contrasting the quasihydrodynamic description with higher-derivative hydrodynamics. Following an argument of Kovtun \cite{KovtunTalk}, one may ask whether the quasihydrodynamic pole collision is physical and universal. Adding higher derivative terms to \eqref{eq:fick}, we can write down the following expansion, valid to second order in either $\partial_x$ and $\partial_t$:
\begin{align}\label{EqKovtun1}
    \partial_t S = \fD \partial^2_x S - \tau \partial^2_t S + \cdots  .
\end{align}
When treated {\em ad verbum}, this equation gives rise to exactly the two modes of Eq. \eqref{eq:omegafick2}. However, since the above theory is defined through a gradient expansion, one may perform a field redefinition $S\to S'$ whereby the two fields are only made to differ by terms proportional to single derivatives:
\begin{align}\label{EqKovtun2}
S = S' - \alpha \partial_t S' - \beta \partial_x S'.
\end{align}
In equilibrium, $S'=S$. As a result of this redefinition, Eq. \eqref{EqKovtun1} becomes
\begin{align}
    \partial_t S' = \fD \partial^2_x S' + \beta \partial_t \partial_x S' + \left(\alpha - \tau\right) \partial_t^2 S' + \cdots .
\end{align}
Since, in hydrodynamics, $S$ has no microscopic definition, it is just as good of a field as $S'$.  The physical spectrum should thus be invariant under the field redefinition \eqref{EqKovtun2}. It is easy to see that one can choose the parameter $\alpha$ in such a way that the spectrum of the gapped imaginary mode is altered in a rather dramatic way. For example, one can choose $\alpha = \tau$, so that to order $\CO(k^2)$, the gapped mode is removed from the spectrum, leaving us with a single unambiguous diffusive mode, $\omega = - i \fD k^2 + \CO(k^3)$. The gapless diffusive mode is on the other hand invariant under the field redefinition. Similarly, one can choose $\alpha > \tau$ to make the gapped mode unstable. We note that the fact that different choices of hydrodynamic variables can generate unphysical modes is a well-documented phenomenon in hydrodynamic literature (see e.g. \cite{Hiscock1985}). In light of this discussion, it is therefore natural to ask whether and when the gapped imaginary mode $\omega_+ = -i \tau^{-1}$ describes a physical excitation.

The answer to this question depends on ``perspective".   If $S$ is the only degree of freedom that has been retained within the IR theory, then indeed the hydrodynamic diffusive pole is the only physical mode in the theory.  However, the quasihydrodynamic theory of Eq. \eqref{eq:fick2} is different: both $S$ and $\CJ$ are independent fields and both have been kept in the IR description.  The equations contain only first-order derivatives and are not a formal gradient expansion.  In fact, to the order in which we have performed the quasihydrodynamic expansion, field redefinitions are ``forbidden":   re-defining $S$ leads to second-derivative (subleading) corrections to its equation of motion, whereas $\mathcal{J}$ cannot be re-defined at all as it is not a conserved quantity.  The explicit presence of $\tau$ in the quasihydrodynamic equations partially fixes the fluid frame. Hence, in the quasihydrodynamic theory, the gapped mode $\omega_+$ is physical---the theory contains additional degrees of freedom.

\subsection{M\"uller-Israel-Stewart theory}\label{sec:MIS}

As for the second example, we turn our attention to the M\"uller-Israel-Stewart (MIS) theory of relativistic hydrodynamics \cite{Muller:1967zza, Israel:1976tn, Israel:1979wp}, in particular, to its complete form, which is a direct extension of second-order hydrodynamics \cite{Baier:2007ix} (the BRSSS theory). In the language of the Schwinger-Keldysh field theory, the study of MIS theory was initiated in Ref. \cite{Torrieri:2016dko}. In order to place the MIS theory within the context of quasihydrodynamics and theories with approximately conserved currents, let us consider the four-dimensional, neutral and conformal stress-energy tensor expanded to second order in derivative expansion \cite{Baier:2007ix}:
\begin{align}\label{tab2ndOrd}
T^{\mu\nu} =&\,\, \varepsilon u^\mu u^\nu + p \Delta^{\mu\nu} -\eta \sigma^{\mu\nu} + \eta \tp \left[ {}^{\langle}D\sigma^{\mu\nu\rangle} + \frac{1}{3} \sigma^{\mu\nu} \left(\nabla \cdot u \right) \right]  \nn
&+ \kappa \left[ R^{\langle \mu\nu \rangle} - 2 u_\rho R^{\rho \langle \mu\nu \rangle \sigma} u_\sigma  \right]  +\lambda_1 \sigma^{\langle \mu}_{~~\rho} \sigma^{\nu\rangle \rho}  +\lambda_2 \sigma^{\langle \mu}_{~~\rho} \Omega^{\nu\rangle \rho}  +\lambda_3 \Omega^{\langle \mu}_{~~\rho} \Omega^{\nu\rangle \rho} ,
\end{align}
where $D = u^\lambda \nabla_\lambda$ is the longitudinal derivative, $\Delta_{\mu\nu} = u_\mu u_\nu + g_{\mu\nu}$, $\sigma_{\mu\nu} = 2 \nabla_{\la \mu} u_{\nu\ra}$ is the shear stress tensor, $\Omega_{\mu\nu} = \nabla_\mu u_\nu - \nabla_\nu u_\mu$ the vorticity, $R^{\rho\mu\nu\sigma}$ the Riemann tensor of the manifold the fluid occupies, and the bracket $A^{\la \mu\nu \ra}$ denotes the transverse, symmetric and traceless part of $A^{\mu\nu}$:
\begin{align}
A^{\la \mu\nu \ra} \equiv \frac{1}{2} \Delta^{\mu\rho} \Delta^{\nu\sigma} \left(A_{\rho\sigma} + A_{\sigma\rho} \right) - \frac{1}{3} \Delta^{\mu\nu} \Delta^{\rho\sigma} A_{\rho\sigma}.
\end{align}
The coefficients $\varepsilon$ and $p$ are the thermodynamic energy density and pressure, with $\varepsilon = 3p$, $\eta$ is the shear viscosity and $\eta\tau_\Pi$, $\kappa$, $\lambda_1$, $\lambda_2$ and $\lambda_3$ are the second-order transport coefficients.

The conservation of the stress-energy tensor $T^{\mu\nu}$, 
\begin{align}\label{MIS-TCon}
\nabla_\mu T^{\mu\nu} = 0,
\end{align} 
generates a coupled set of four hydrodynamic partial differential equations for the three independent components of $u^\mu$ and any scalar function, e.g. $\varepsilon$, or alternatively, the near-equilibrium temperature field $T(x)$, which in general have acausal solutions, just as in the simple diffusive case in Section \ref{sec:DS}.\footnote{For some recent progress on formal discussions of causality in fluid dynamics, see \cite{Bemfica:2017wps}.} These equations correspond to the purely hydrodynamic set of equations of the \eqref{eq:hydroformal} type, with only strictly conserved energy and momentum. Hereon, we will only consider linearized solutions of \eqref{MIS-TCon}---the quasinormal modes---in flat space. 

As in Section \ref{sec:DS}, the full set of diffusive and propagating sound modes suffers from acausal behavior. Similarly, this problem can be cured by treating $\Pi^{\mu\nu} = - \eta \sigma^{\mu\nu}$ as an independent set of five degrees of freedom ($\Pi^{\mu\nu}$ is transverse, symmetric and traceless) and rewriting $T^{\mu\nu}$ in terms of $\Pi^{\mu\nu}$ as
\begin{align}
T^{\mu\nu} = \varepsilon u^\mu u^\nu + p \Delta^{\mu\nu} + \Pi^{\mu\nu} ,
\end{align}  
and introducing an independent equation of motion for $\Pi^{\mu\nu}$.  In this work, we will only consider the linearized stress-energy tensor.  What we find is the following set of nine quasihydrodynamic differential equations:  
\begin{subequations}\label{ISlin}
\begin{align}
\partial_\mu \left[( \varepsilon + p) u^\mu u^\nu \right] + \partial^\nu p + \partial_\mu \Pi^{\mu\nu} &= 0 ,  \label{ISlin1} \\
 {}^{\langle}D \Pi^{\mu\nu \rangle} + \frac{2 \eta}{\tau} \partial^{\langle\mu} u^{\nu\rangle} &= - \frac{1}{\tau} \Pi^{\mu\nu}, \label{ISlin2}
\end{align}  
\end{subequations}
Eq. \eqref{ISlin2} is the approximate conservation law, which we will, in analogy with our example from Section \ref{sec:DS}, interpret as the equation of type \eqref{eq:introlonglived2}.
 This interpretation is only justified in certain theories, such as for example in the holographic model we describe in Section \ref{sec:MIS-Hol}. 

From the quasihydrodynamic perspective, we interpret the linearized Eq. (\ref{ISlin2}) as a weakly broken conservation law for an approximately conserved quantity associated with $\Pi^{ij}$. To make this statement more explicit, we would like to recast Eq. \eqref{ISlin2} into an expression, which, as we take $\tau\to\infty$, becomes a conservation equation---an equation expressing a vanishing divergence of some current. To this end, and to first order in the fluctuations of all quasihydrodynamic fields, we can write
\begin{equation}\label{eq:MIS-3indexJ}
\partial_t \Pi^{ij} +  \partial_k J_\Pi^{kij} = -\frac{1}{\tau}\Pi^{ij},
\end{equation}
where the purely spatial part of the associated current $J_\Pi^{\lambda\mu\nu}$ is given within linear response by 
\begin{equation}\label{Def-3indexJ}
\delta J_\Pi^{kij} = v^2(\epsilon+P)\left[   \delta^{kj}\delta u^{i}+  \delta^{ki}\delta u^{j}- \frac{2}{3} \delta^{ij}\delta u^{k}  \right]
\end{equation} 
and 
\begin{align}\label{MIS-speed-Shear}
v = \sqrt{ \frac{\eta}{\left(\varepsilon+p\right)\tau} }.   
\end{align}
Note that $J_\Pi^{kij}$ is not symmetric in $(ki)$ or $(kj)$ indices. Moreover, it is natural to define $J_\Pi^{tij} \equiv \Pi^{ij}$ so that the Eq. \eqref{Def-3indexJ} becomes 
\begin{equation}\label{eq:MIS-3indexJ-v2}
\partial_t J_\Pi^{tij} +  \partial_k J_\Pi^{kij} = -\frac{1}{\tau} J_\Pi^{tij}.
\end{equation}
Within linear response, this explicitly quasihydrodynamic equation reproduces Eq. (\ref{ISlin2}).  Beyond linear response, it is less obvious how to define $J_\Pi^{kij}$, as the derivatives in the equation of motion can now act on $v^2(\epsilon+P)$, along with the projectors.   We expect that the existence of quasihydrodynamics at the nonlinear level demands additional constraints on these ``thermodynamic" prefactors which we will not explore in this paper.   Nonlinear effects will not be relevant for the holographic calculations that follow.

The coefficient $\tau$ plays the role of the relaxation time for the additional quasihydrodynamic modes $\Pi^{\mu\nu}$. In general, the relaxation time $\tau$ and the second-order transport coefficient $\tau_\Pi$ in \eqref{tab2ndOrd} need not be the same. However, they are equal if the hydrodynamic series is truncated at second order, as in Eq. \eqref{tab2ndOrd}. We will return to this point in Section \ref{sec:MIS-Hol}.  We emphasize that $v$ is held fixed as $\tau$ is taken to be parametrically large, such that these equations admit a quasihydrodynamic interpretation.   As $\eta$ is the shear viscosity of the fluid in the true hydrodynamic limit, this implies that $\eta \propto \tau$, as is indeed the case in ordinary kinetic theory (see e.g. Ref. \cite{Grozdanov:2016vgg}). Therefore, as $\tau\to\infty$, Eq. \eqref{ISlin2} with the help of an introduction of $J_\Pi^{\lambda ij}$ becomes, within linear response, a conservation equation
\begin{equation}
\partial_\mu J_\Pi^{\mu ij } = 0.
\end{equation}

We emphasize that if the quasihydrodynamic equations (\ref{ISlin}) are taken seriously, they encode arbitrarily high-order derivative corrections to hydrodynamics, not just second-order hydrodynamics in Eq. \eqref{tab2ndOrd}.  They make physical predictions:  the transverse diffusion of momentum morphs into a ballistically propagating mode on a parametrically long time scale $\tau$.  In a generic fluid, we cannot identify any (approximate) symmetry that enforces this.   Hence, these quasihydrodynamic MIS equations are not a correct description of the second-order hydrodynamics of generic fluids.   Nevertheless, we will see that there are special holographic fluids for which the quasihydrodynamic MIS equations are a quantitatively correct description of the dynamics on time scales much shorter than $\tau$.  It would be interesting to obtain a better understanding of why such fluids exhibit an approximately conserved quantity $\Pi^{\mu\nu}$.

We now look for the dispersion relations $\omega(k)$ of the quasinormal modes by solving the system of equations \eqref{ISlin1}--\eqref{ISlin2} in the transverse (shear) and longitudinal (sound) channels. The quadratic and cubic polynomial equations for shear and sound channels, respectively, are 
\begin{subequations}\label{ShearSoundISPoly}
\begin{align}
&\text{shear:} &&  \omega^2 + \frac{i}{\tau} \omega - \frac{\eta}{\left(\varepsilon + p\right)\tau} k^2 = 0 , \label{ShearISPoly} \\
&\text{sound:} && \omega^3 + \frac{i}{\tau} \omega^2 - \left( c_s^2   + \frac{4}{3} \frac{\eta}{\left(\varepsilon+p\right)\tau}\right) \omega k^2 - \frac{ i c_s^2}{\tau} k^2 = 0 , \label{SoundISPoly} 
\end{align}
\end{subequations}
where $c_s = 1/\sqrt{3}$ is the speed of sound of a conformal fluid in four dimensions. Notice that the equation \eqref{ShearISPoly} in the shear channel is identical to Eq. \eqref{DtoS-Poly} from Section \ref{sec:DS}, up to the identification of %$\tau = \tp$ and 
the diffusion constant with the usual ratio
\begin{align}\label{MISDiffCoeff}
\fD = \frac{\eta}{\varepsilon + p } = \frac{\eta}{s T},
\end{align}
where $s$ is the entropy density. The solutions of \eqref{ShearISPoly} are thus again 
\begin{equation}\label{NewHydroShearIS0}
\omega_\pm = - \frac{i}{2\tau} \left( 1 \pm \sqrt{1 - \frac{4\eta\tau}{\varepsilon + P} k^2} \right) ,
\end{equation}
with the same small-$k$ and large-$k$ characteristics as the pair of modes in Section \ref{sec:DS}---the collision of the diffusive and gapped poles occurs at the critical momentum 
\begin{align}
k_c = \sqrt{\frac{\varepsilon+p}{4 \eta \tau} } .
\end{align}
After the collision, at large $k$, the speed of propagation is given by $v$, defined in (\ref{MIS-speed-Shear}).  We refer the reader to Figures \ref{fig:SC-P3} and \ref{fig:SC-P1-P2}.\footnote{For a recent discussion of this behavior in experiments and numerical simulations done in various liquid phases, see e.g. the review in Ref. \cite{Trachenko2016review}.}

\begin{figure*}
\centering
\includegraphics[width=0.45\textwidth]{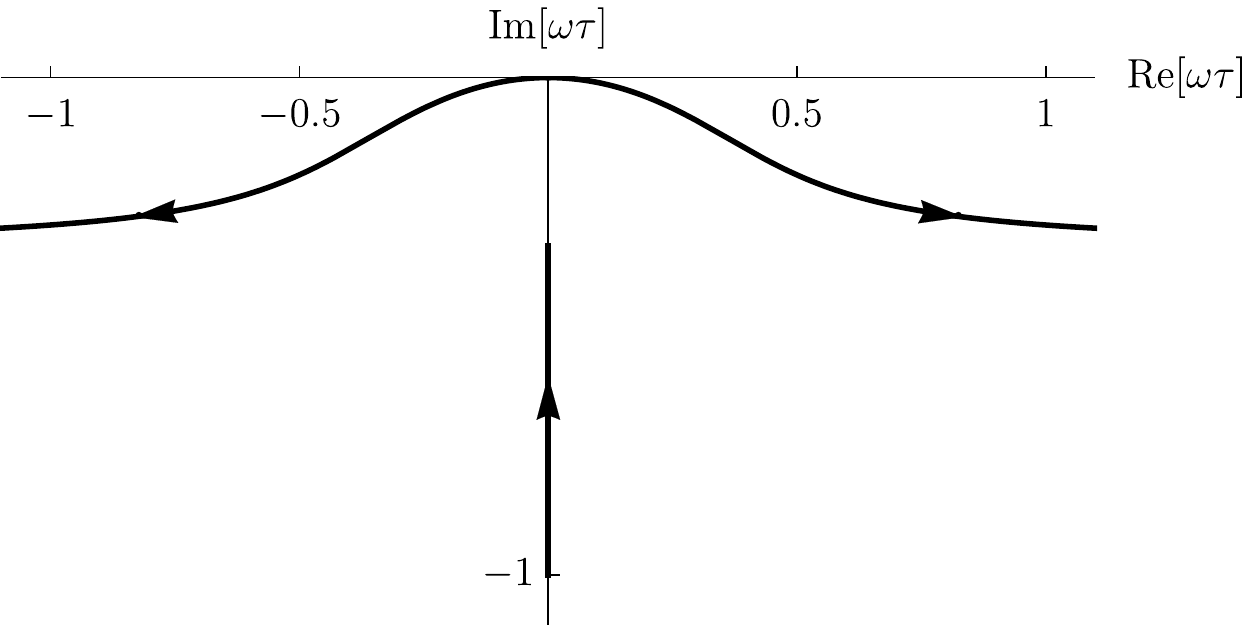}
\caption{The collision of the MIS sound channel dispersion relations plotted on the complex $\omega\tau$ plane for a choice of $\mathfrak{D}/\tau = \eta/((\varepsilon + p) \tau)= 1/2$. Arrows indicate the direction of movement of the poles as $k\tau$ is increased.}
\label{fig:MIS-P3}
\end{figure*}

\begin{figure*}[h]
\centering
\includegraphics[width=0.45\textwidth]{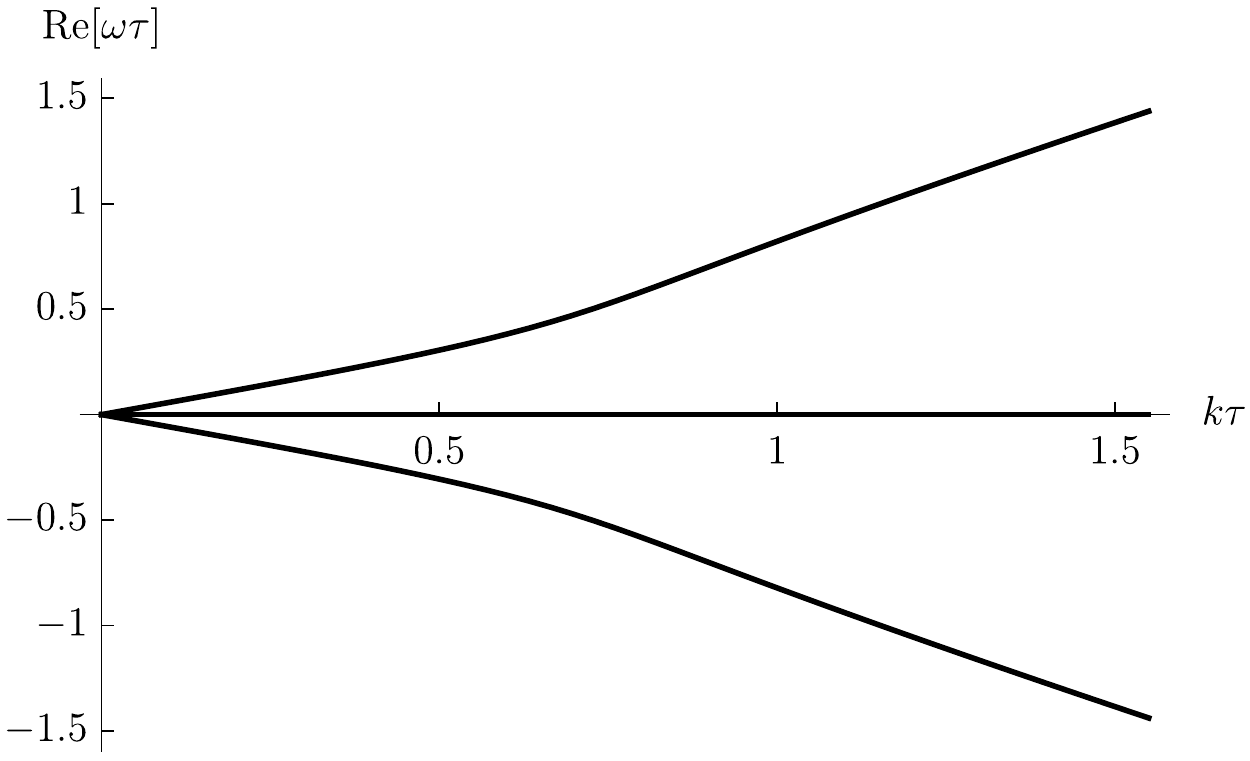}
\hspace{0.08\textwidth}
\includegraphics[width=0.45\textwidth]{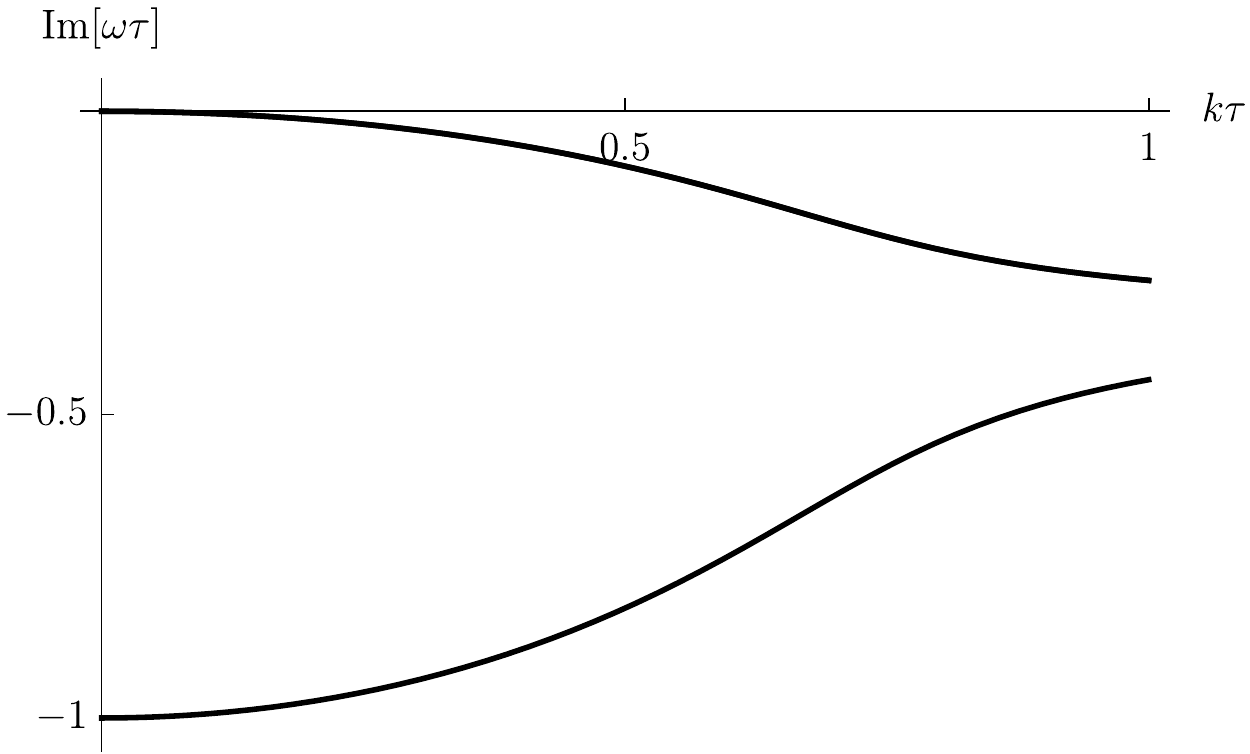}
\caption{Plots of the real and imaginary parts of the sound dispersion relations for a choice of $\fD/\tau = \eta/((\varepsilon + p) \tau)= 1/2$. Note that $\im[\omega_1] = \im[\omega_2]$.}
\label{fig:MIS-P1-P2}
\end{figure*}

In the sound channel, the cubic (in $\omega$) polynomial  equation \eqref{SoundISPoly} gives three modes with dispersion relations that can be found explicitly, but we do not state them here in full. The modes consist of a pair of hydrodynamic modes with the sound dispersion relation at small $k$ and a gapped mode. In the small-$k$ expansion, they are
\begin{subequations}\label{NewHydroSoundIS}
  \begin{align}
\omega_{1,2} &=  \pm \frac{k}{\sqrt{3}} -  \frac{2}{3} i \frac{\eta}{\varepsilon +P}\, k^2 \pm \CO(k^3) , \label{NewHydroSoundIS1}\\ 
\omega_3 &= - \frac{i}{\tau}  + \frac{4}{3} i \frac{\eta}{\varepsilon +P}\, k^2 + \CO(k^4).\label{NewHydroSoundIS2}
\end{align}
\end{subequations}
We note that because of a single relaxation time $\tau$, the gap is the same in both channels. As a result, at $k=0$, when the $SO(3)$ rotational invariance is restored, both shear and sound channels exhibit only a single gapped mode each with $\omega = - i/\tau$. We also note that if the hydrodynamic modes are matched to those derived from second-order hydrodynamics, then $\tau = \tp$. At large $k$,
\begin{subequations}
\begin{align}\label{NewHydroSoundIS-largeK}
\omega_{1,2} &=  \pm \left( 1+\frac{4 \eta}{\left(\varepsilon+p\right)\tau} \right)^{1/2} \frac{k}{\sqrt{3}} -  \frac{2 i \eta }{\left(\varepsilon+p\right) \tau^2}   \left( 1+\frac{4 \eta}{\left(\varepsilon+p\right)\tau} \right)^{-1} \pm \CO(1/k), \\
\omega_3 &= - \frac{i}{\tau}  \left( 1+\frac{4 \eta}{\left(\varepsilon+p\right)\tau} \right)^{-1} + \CO(1/k) .
\end{align}
\end{subequations}
We plot the full dispersion relations in Figures \ref{fig:MIS-P3} and \ref{fig:MIS-P1-P2}.  The precise motion of the quasihydrodynamic modes in the complex plane depends on $\mathfrak{D}$ and $\tau$ \cite{Grozdanov:2016vgg}. 

As in the previous subsection, it is tempting to associate this theory with quasihydrodynamics, yet, a priori, there is no reason why $\Pi^{\mu\nu}$ would correspond to an approximately conserved quantity.  Moreover, since the ``UV completion" of the MIS theory is highly phenomenological, it is difficult to understand the precise microscopic origin of $\Pi^{\mu\nu}$ or how the theory should be correctly extended. For this reason, it would be highly desirable to have a well-defined microscopic way of deriving the MIS theory. As we will show in Section \ref{sec:MIS-Hol}, the structure of the MIS theory and its equations can in fact be systematically derived from holography, using dual higher-derivative theories, such as the Einstein-Gauss-Bonnet theory \cite{Brigante:2007nu,Grozdanov:2016vgg,Grozdanov:2016fkt}. This will be done to first order in fluctuations of the equilibrium fields. Indeed, the fact that the structure of MIS should be reproduced by higher-derivative theories could be anticipated from the studies of coupling-dependent properties of thermal spectra in \cite{Grozdanov:2016vgg,Grozdanov:2016fkt}. It was shown there that the coupling between hydrodynamic and new, purely relaxing modes at intermediate coupling gives rise to the polynomial equations \eqref{ShearSoundISPoly} and thus dispersion relations \eqref{NewHydroShearIS0} and \eqref{NewHydroSoundIS}. The relaxation time $\tau$ is controlled by the coupling constant, and in the regime of large higher-derivative terms---weak field theory coupling---the new modes become long-lived with $\tau T \gg 1$. In Section \ref{sec:MIS-Hol}, we will explicitly demonstrate how $\Pi^{\mu\nu}$ arises from dual gravitational perturbations, and derive the (linearized) structures through which it couples to $T^{\mu\nu}$ in equations of motion---i.e., the MIS equations \eqref{ISlin}.

\subsection{Magnetohydrodynamics}\label{sec:MHD}

We now turn our attention to the second main example of a quasihydrodynamic theory studied in this work: magnetohydrodynamics of a plasma with dynamical photons. Plasma is an ionized gas, which is charge-neutral at long distances---the long-range electric force is Debye screened and in the plasma phase, photons are massive. Nevertheless, the constituents of the plasma interact electromagnetically. While its long-distance charge neutrality implies that the equilibrium electric field $\vecb E = 0$, the equilibrium magnetic field $\vecb B$ can be arbitrarily strong. In its standard formulation \cite{davidson_2001,freidberg_2014} (see also Ref. \cite{Grozdanov:2017kyl}), magnetohydrodynamics (MHD) is a theory of fluid  motion (continuity equation and Euler or Navier-Stokes equations) coupled to Maxwell's equations of electromagnetism. The equations of ideal MHD are
\begin{subequations}
\begin{align}
\partial_t \rho + \vecb \nabla \cdot \left(\rho\, \vecb v\right) &= 0  , \label{MHD1} \\
\rho \left( \partial_t + \vecb v \cdot \nabla \right) \vecb v &= - \vecb \nabla p + \frac{1}{\mu_0} \left( \vecb \nabla \times \vecb B \right)\times \vecb B  , \label{MHD2} \\ 
\partial_t \vecb B &= \vecb\nabla \times \left(\vecb v \times \vecb B \right)  , \label{MHD3} \\
\vecb \nabla \cdot \vecb B &= 0 , \label{MHD5} \\
\left( \partial_t + \vecb v \cdot \nabla \right) \left( \frac{p}{\rho^\gamma} \right) &= 0  , \label{MHD4}
\end{align}
\end{subequations}
where $\vecb v$ is the velocity, $\rho$ is the mass density, $p$ the pressure and $\mu_0$ the vacuum permeability. Eq. \eqref{MHD1} is the continuity equation, Eq. \eqref{MHD2} the (Lorentz) forced Euler equation, Eq. \eqref{MHD3} is Faraday's law,  \eqref{MHD5} is magnetic Gauss's law constraint equation and Eq. \eqref{MHD4} is the adiabatic equation of state with $\gamma = 5/3$. The electric field is not treated a dynamical variable. Thus, neglecting $\partial_t \vecb E$ in Ampere's law fixes the current in the Lorentz force contribution to \eqref{MHD2}. In Faraday's law, $\vecb E$ is completely fixed by the assumption of the idea Ohm's law $\vecb E = - \vecb v \times \vecb B$ and the leading-order dissipative correction to this relation are suppressed by $1/\sigma$ where $\sigma$ is the conductivity.  Gauss's law plays no role in ideal MHD. For details, see  Ref. \cite{Grozdanov:2017kyl}.

The theory of MHD was recently reformulated by using the language of higher-form (generalized) global symmetries in \cite{Grozdanov:2016tdf}.\footnote{The relation between MHD formulated in the language of higher-form symmetries and MHD written directly in terms of electromagnetic fields was discussed in \cite{Hernandez:2017mch}. A recent work \cite{Armas:2018atq} claimed that writing down a consistent hydrodynamic partition function for MHD (on a thermal circle) requires the addition of an extra scalar degree of freedom. For a hydrodynamic theory with such a mode, the relation between the two different formalisms requires additional considerations, which were worked out in \cite{Armas:2018atq}.}  All known MHD equations and their systematic dissipative extensions follow from the realization that plasma possesses a global $U(1)$ one-form symmetry associated with the conservation of the number of magnetic flux lines. As a result, the theory has a two-form conserved current $J^{\mu\nu}$ and the full set of equations of motion is \cite{Grozdanov:2016tdf}\footnote{For the construction of a theory of fluids with $q$-form symmetries, see Ref. \cite{Armas:2018ibg}.}
\begin{subequations}
  \begin{align}
\nabla_\mu T^{\mu\nu} &= 0,  \label{EOM1}\\
\nabla_\mu J^{\mu\nu} & = 0 .  \label{EOM2} 
\end{align} 
\end{subequations}
In terms of the microscopic photon field $A_\mu$, the two-form current is $J^{\mu\nu} = \frac{1}{2} \epsilon^{\mu\nu\rho\sigma} F_{\rho\sigma} $, with $F_{\mu\nu} = \partial_\mu A_\nu - \partial_\nu A_\mu$.   Eq. \eqref{EOM2} is the Bianchi identity. The holographic dual of MHD, which we will study in Section \ref{sec:MHD-Hol}, was constructed in Ref. \cite{Grozdanov:2017kyl}.\footnote{For a study of different aspects of the same holographic theory dual to a state with a one-form symmetry, see \cite{Hofman:2017vwr}.}

Unlike standard MHD, a symmetry-based classification of the stress-energy tensor $T^{\mu\nu}$ and the two-form $J^{\mu\nu}$ permits  a systematic gradient expansion with all possible transport coefficients. Moreover, the equation of state of the plasma can be an arbitrary function of temperature and the strength of the magnetic field, parametrized by the chemical potential $\mu$ conjugate to the number density of the magnetic flux lines $\rho$. In particular, the pressure can be an arbitrary function $p(T,\mu)$. Beyond $\mu$, the other hydrodynamic fields are the usual $T$, $u^\mu$ and the spacelike vector field $h^\mu$. The vectors are normalized in the following way: $u_\mu u^\mu = -1$, $h_\mu h^\mu = 1$, $u_\mu h^\mu = 0$. The relevant thermodynamic relations are $\varepsilon + p = s T + \mu \rho$ and $d p = s \, dT+ \rho \, d\mu$. Explicitly, to first order in the gradient expansion (see Ref. \cite{Grozdanov:2016tdf}), 
\begin{subequations} \label{eq:MHDtot}
\begin{align}
T^{\mu\nu} & = (\varepsilon + p)\, u^{\mu}u^{\nu} + p \, g^{\mu\nu} - \mu\rho\, h^{\mu}h^{\nu} + \delta f \, \Delta^{\mu\nu} + \delta \tau \, h^{\mu}h^{\nu}  + 2 \, \ell^{(\mu}h^{\nu)} + t^{\mu\nu} ,\label{MHDstress-energy} \\
J^{\mu\nu} & = 2\rho \, u^{[\mu}h^{\nu]} + 2m^{[\mu}h^{\nu]} +  s^{\mu\nu}  \label{MHDcurrent}, 
\end{align}
where
\begin{align}
\delta f & = -\zeta_{\perp} \Delta^{\mu\nu}\nabla_{\mu}u_{\nu} -\zeta_{\times} h^{\mu}h^\nu \grad_\mu u_\nu ,\label{visc1} \\
\delta \tau & = -\zeta_\times  \Delta^{\mu\nu}\grad_\mu u_\nu - \zeta_{\parallel} h^{\mu}h^{\nu} \nabla_{\mu} u_{\nu}  ,\\ 
\ell^{\mu} & = -2\eta_{\parallel}\Delta^{\mu\sig}h^{\nu} \nabla_{(\sig}u_{\nu)} ,\label{visc4} \\
t^{\mu\nu} & = -2\eta_{\perp}\left(\Delta^{\mu\rho}\Delta^{\nu\sig}- \frac{1}{2} \Delta^{\mu\nu}\Delta^{\rho\sig}\right)\nabla_{(\rho}u_{\sig)}  ,\\
m^{\mu} & = -2 r_{\perp}  \Delta^{\mu\beta}h^{\nu} T  \nabla_{[\beta}\left(\frac{h_{\nu]} \mu}{T}\right) ,\label{mdef}\\
s^{\mu\nu} & = -2 r_{\parallel}  \Delta^{\mu\rho} \Delta^{\nu\sigma} \mu\nabla_{[\rho} h_{\sigma]}   , \label{sdef}
\end{align}
\end{subequations}
with $\eta_\perp$, $\eta_\parallel$, $\zeta_\perp$, $\zeta_\parallel$, $\zeta_\times$, $r_\perp$ and $r_\parallel$ the seven transport coefficient in a charge conjugation symmetric state (see also Ref. \cite{Hernandez:2017mch}). The projector transverse to $u^\mu$ and $h^\mu$ is $\Delta^{\mu\nu} = g^{\mu\nu} + u^\mu u^\nu - h^\mu h^\nu$.

As in standard MHD, the electric field is not a dynamical variable. In the formalism of \cite{Grozdanov:2016tdf}, using the microscopic relation between $J^{\mu\nu}$ and $F^{\mu\nu}$, we can write the (relativistic) magnetic and electric fields in the fluid's {\it comoving frame} as
\begin{subequations}
  \begin{align}
\mathtt{B}^\mu &\equiv J^{\mu\nu} u_\nu  = \rho \, h^\mu ,  \label{MagneticFieldMHD} \\
\mathtt{E}^\mu &\equiv - \frac{1}{2} \varepsilon^{\mu\nu\rho\sigma} u_\nu J_{\rho\sigma} = - \frac{1}{2} \varepsilon^{\mu\nu\rho\sigma} u_\nu \left( 2 m_{[\rho} h_{\sigma]} + s_{\rho\sigma} \right), \label{ElectricFieldMHD} 
\end{align}
\label{eq:EMfieldMHD}
\end{subequations}
where the electric field $\mathtt{E}^\mu$ is proportional to one-derivative tensors $m^\mu$ and $s^{\mu\nu}$ and thus, the two resistivities $r_\perp$ and $r_\parallel$---a generalized (inverse) Ohm's law. Note that in the above relativistic definition, the electric field $\mathtt{E}^\mu$ is defined in the frame where the fluid is at rest. For this reason, in equilibrium, $\mathtt{E}^\mu = 0$. This should be contrasted with electric and magnetic fields measured in equilibrium of the plasma, i.e. when $u^\mu = (1,\textbf{0})$. Adopting the definitions in \eqref{eq:EMfieldMHD}, we write 
\begin{equation}\label{eq:EMfieldMHD-nonRela}
     B^i  = J^{ti},\qquad  E^i  =-\frac{1}{2} \varepsilon^{ijk}J_{jk}.
\end{equation}
When perturbed away from the equilibrium, i.e. for $u^\mu = (1,\textbf{v})$ with $|\textbf{v}|^2 \ll 1$, then the constitutive relations for $J^{\mu\nu}$ together with the definitions \eqref{eq:EMfieldMHD-nonRela} give rise to the ideal Ohm's law used above, $\vecb{E} = -\vecb v \times \vecb B$. The conservation law $\d_\mu J^{\mu\nu} =0$ automatically gives Faraday's and magnetic Gauss's laws. Also, note that the identifications \eqref{MagneticFieldMHD} and \eqref{ElectricFieldMHD} require the state to be perturbatively connected to the vacuum so that the ``concept" of a microscopic photon field $A_\mu$ exists.

It is clear from the present discussion that the effective theory of MHD is only valid so long as the electric field in a plasma is small---of the order of momentum. Therefore, for example, an initial state with a separation of positive and negative charges, and thus a large initial electric field that may lead to plasma oscillations cannot be described by MHD. Moreover, such scenarios would necessarily break the gradient expansion. Additional degrees of freedom are required in the effective theory. These degrees of freedom are gapped massive photons. Consider for example a thermal plasma with $\sqrt{B} / T \ll 1$ in quantum electrodynamics (QED). Then, to leading order, the Debye mass is $m_D^2 = \frac{1}{3} e^2  T^2$. In the (extreme) low-energy limit, $ k / m_D \sim k / T \ll 1$, so the spectrum of the effective theory contains no gapped photons. This is MHD. However, when either the electric field or $k$ become comparable to $T$, the inclusion of massive photons is needed. 

One may attempt to use the MIS-type procedure from Section \ref{sec:MIS} to perform a simple phenomenological extension of the MHD equations \eqref{EOM1}--\eqref{EOM2} to include additional massive degrees of freedom. Since the dynamical electric field arises from the charge sector, we can extend the two-form as
\begin{align}
J^{\mu\nu} = 2 \rho u^{[\mu} h^{\nu]} + \CJ^{\mu\nu} ,
\end{align}
where $\CJ^{\mu\nu}$ is an anti-symmetric two-form, which we treat as encoding massive degrees of freedom that are independent of $u^\mu$, $h^\mu$, $T$ and $\mu$. In order to keep the definition of electric and magnetic fields consistent with Eqs. \eqref{ElectricFieldMHD} and \eqref{MagneticFieldMHD}, we introduce the following constraint
\begin{align}
\CJ^{\mu\nu} u_\nu = 0 ,
\end{align}
which enforces Eq. \eqref{MagneticFieldMHD}. Hence, $\CJ^{\mu\nu}$ contains three independent degrees of freedom and governs the dynamical electric field (cf. Eq. \eqref{ElectricFieldMHD}):
\begin{align}
\mathtt{E}^{\mu} = - \frac{1}{2} \varepsilon^{\mu\nu\rho\sigma} u_\nu \CJ_{\rho\sigma} .
\end{align}

In the vacuum state, there are no thermally charged particles. Hence, by electric/magnetic duality, if the magnetic flux density is a conserved quantity, so is the electric flux density.  At finite temperature, the electric flux density is no longer conserved, yet at sufficiently low temperature it may be the case that only a few thermally excited charges are present.  In this limit, it is natural to expect that the breaking of the electric conservation law is weak and that the dynamical photon is a quasihydrodynamic excitation.  It is more convenient to write such quasihydrodynamic equations in terms of $\CJ^{\mu \nu}$ instead of $\mathtt{E}^\mu$:
\begin{subequations}
\begin{align}
\nabla_\mu T^{\mu\nu} &= 0  ,  \label{MHD-MIS-EOM1} \\
\nabla_\mu \left( 2\rho u^{[\mu} h^{\nu ] } \right) + \nabla_\mu \CJ^{\mu\nu}  & = 0  , \label{MHD-MIS-EOM2} \\
u^\lambda \nabla_\lambda  \CJ^{\mu\nu} - \frac{2}{ \tau } m^{[\mu} h^{\nu ] } - \frac{1}{ \tau } s^{\mu\nu} &= - \frac{1}{ \tau } \CJ^{\mu\nu} , \label{MHD-MIS-EOM3}
\end{align}  
\end{subequations}
While the form of the MIS-like equation in \eqref{MHD-MIS-EOM3} may appear unfamiliar, its extension of the MHD equations \eqref{EOM1}--\eqref{EOM2} is precisely the desired dynamical version of Ampere's law in a plasma with Ohm's law fixing the current. In fact, the main consequence of Eq. \eqref{MHD-MIS-EOM3} is a reinstated time derivative of the electric field in Ampere's law. To see this, we need to perform a small perturbation of the (extended) hydrodynamics variables $u^\mu$, $T$, $\mu$, $h^\mu$ and $\CJ^{\mu\nu}$ around equilibrium. Here, for reasons of Section \ref{sec:MHD-Hol}, we will only focus on transverse fluctuations. If we assume that the equilibrium magnetic field is pointing along the $z$-direction, then the resulting changes of various components of conserved operators in the transverse channel become 
\begin{subequations}
  \begin{align}
      \delta T^{ti} &= (\varepsilon+p) \delta u^i,\\
      \delta T^{zi} &= -\mu\rho \delta h^i + \eta_\parallel \d_z \delta u^i,\\
      \delta J^{ti} &= \rho h^i,\\
      \delta J^{zi} &= -\rho v^i + \CJ^{zi},
  \end{align}
  \label{MHD-perturbed-Currents}
\end{subequations}
with $J^{tz} = \rho = B^z$ setting the strength of the magnetic field in equilibrium. Using Eq. \eqref{eq:EMfieldMHD-nonRela}, it is then straightforward to show that at the leading order in derivatives, Eq. \eqref{MHD-MIS-EOM3} becomes the advertized Ampere's law in a plasma:
\begin{align}
\d_t \vecb{E} - \left( \frac{\mu r_\perp}{\rho \tau} + \frac{\mu\rho}{\varepsilon+p} \right) \left(\nabla\times \vecb{B}  \right)  = -\frac{1}{\tau} \left(\vecb{E} + \vecb{v}\times \vecb{B}  \right) + \CO(\d^2),
\end{align}
for perturbations that only depends on $t$ and $z$. The right-hand side is nothing but the boosted Ohm's law in the presence of a magnetic field.  The relaxation time $\tau$ is inversely proportional to the conductivity of the plasma. It turns out that the holographic model of \cite{Grozdanov:2017kyl,Hofman:2017vwr} precisely encodes the above equation for any strength of the magnetic field, including $\vecb B = 0$. This will be shown in Section \ref{sec:MHD-Hol}. 

As in Section \ref{sec:MIS}, it is helpful to explicitly re-write (\ref{MHD-MIS-EOM3}) in a quasihydrodynamic form (the other two equations of MHD are manifestly conservation laws, as previously discussed). In analogy with our discussion in Section \ref{sec:MIS}, one can write the left-hand side of Eq. \eqref{MHD-MIS-EOM3} as the divergence of a three-index current $J^{\mu ij}_{\CJ}$, where $J^{tij}_{\CJ} =\CJ^{ij}$.  We again emphasize that this procedure is completely well behaved within the linear response regime, and that the procedure is not yet developed beyond linear response.   The spatial components of the current $J^{kij}_{\mathcal{J}}$ that corresponds to the approximately conserved number of electric flux lines can be explicitly written as
\begin{equation}
  \delta J^{kij}_{\CJ} = \chi_\perp \left( \Delta^{ k [i}h^{j]}\left(\frac{\delta \mu}{\mu} - \frac{\delta T}{T}\right) + h^k h^{[i}\Delta^{j]m} \delta h_m \right) + \chi_\parallel \Delta^{k[i}\delta h^{j]},
\end{equation}
where $\delta h^\mu$, $\delta \mu$ and $\delta T$ denote the linearized fluctuations, and $\chi_{\perp, \parallel}$ become
\begin{subequations}\begin{align}
\chi_\perp &= \frac{2r_\perp \mu }{\tau}, \\
\chi_\parallel &= \frac{2r_\parallel \mu}{\tau},
\end{align}\end{subequations}
but more generally, they may be more complicated functions, which have the property that within linear response, their total derivatives of $\mu$ and $T$ reproduce the coefficients of $h^\nu \partial_\rho h^\nu$, $\partial_\rho \mu$ and $\partial_\rho T$ respectively in (\ref{eq:MHDtot}) and (\ref{MHD-MIS-EOM3}).   At the fully nonlinear level, the existence of such a construction places very strong constraints on the functions $\tau$, $r_\perp$ and $r_\parallel$, which have yet to be fully explored. At the fully nonlinear level, it seems likely that the requirement that $J^{\mu ij}_{\mathcal{J}} $ is conserved in the limit of $\tau\to\infty$ fixes much of the functional form of $\chi_\perp$ and $\chi_\parallel$.

To investigate the simplest prediction of this theory, we study the corrections from new modes to the transverse Alfv\'{e}n waves (see \cite{Grozdanov:2016tdf}). To this end, we perturb to first order the quasihydrodynamic fields with the Fourier decomposition $e^{-i \omega t + i k x \sin\theta + i k z \cos\theta }$ and compute the spectrum in this channel. Note that $\CJ^{\mu\nu}$ has a zero equilibrium value ($\vecb E = 0$ in equilibrium). Again, the magnetic field is chosen to point in the $z$-direction.   $\theta$ denotes the angle between the background magnetic field and momentum; without loss of generality we set $k_y=0$. In the (transverse) Alfv\'{e}n channel, the only non-zero fluctuations are $\delta u^y$, $\delta h^y$, $\delta \CJ^{xy}$ and $\delta \CJ^{yz}$. Instead of a pair of Alfv\'{e}n modes, we now obtain a cubic polynomial for the quasinormal modes, which has the form
\begin{align}
&\omega^3  + i \left(\frac{1}{\tau} +   \frac{\mathfrak{V}(\theta)}{\varepsilon + p}  k^2     \right)\omega^2 - \left[ \frac{1}{\tau} \left( \frac{\mathfrak{V}(\theta)}{\varepsilon+p} + \frac{\mu \mathfrak{R}(\theta)}{\rho} \right) + \frac{\mu\rho\cos^2\theta}{\varepsilon+p}   \right] \omega k^2  \nn
&- \frac{i}{\tau} \left( \frac{ \mu \rho^2\cos^2\theta + \mu \, \mathfrak{V}(\theta) \mathfrak{R}(\theta) k^2  }{(\varepsilon+p)\rho} \right) k^2 = 0 ,
\end{align}
where the anisotropic combinations of viscous and resistive transport coefficients are defined as
\begin{subequations}
\begin{align}
\mathfrak{V}(\theta) &\equiv \eta_\perp \sin^2\theta + \eta_\parallel \cos^2\theta , \\
\mathfrak{R}(\theta) &\equiv r_\parallel \sin^2\theta + r_\perp \cos^2 \theta.
\end{align}
\end{subequations}
In the small $k$ expansion, the three modes take the form
\begin{subequations}
\begin{align}
\omega_{1,2} &= \pm \sqrt{\frac{\mu\rho}{\varepsilon+p}} \, k \cos\theta - \frac{i}{2} \left( \frac{\mathfrak{V}(\theta)}{\varepsilon+p} + \frac{\mu \mathfrak{R}(\theta)}{\rho}  \right) k^2  \pm \CO(k^3) , \\
\omega_3 &= - \frac{i}{\tau} + \frac{i\mu \mathfrak{R}(\theta)}{\rho} k^2 + \CO(k^4) . 
\end{align}
\end{subequations}
The pair of modes $\omega_{1,2}$ is the pair of (non-perturbative) Alfv\'{e}n waves from \cite{Grozdanov:2016tdf} and $\omega_3$ is the new gapped mode. The form of the dispersion relations is completely analogous to the sound modes in MIS theory, cf. Eqs. \eqref{NewHydroSoundIS1}--\eqref{NewHydroSoundIS2}, with the gapped mode $\omega_3$ having the $k^2$ term be twice the dissipative $\CO(k^2)$ contribution to $\omega_{1,2}$. Note that because we only extended $J^{\mu\nu}$ and not $T^{\mu\nu}$, this contribution is purely resistive. Because of the lack of UV completion of $T^{\mu\nu}$, which would introduce a second relaxation time, the large $k$ limit is still acausal. 

The conclusion that can be drawn is precisely the same as in the MIS theory. While the hydrodynamic part of the theory---MHD---can be constructed unambiguously, the construction of its MIS-type quasihydrodynamic extension, while consistent with expectations, is highly phenomenological. In an analogous manner, \cite{Hernandez:2017mch} extended MHD by adding the simplest $E^2$ terms to the partition function and treating $E^\mu$ as a dynamical field. This allowed them to find Langmuir waves in a plasma with non-zero charge density. However, as we will see in Section \ref{sec:MHD-Hol}, the theory of MHD and its systematic extensions can be derived directly from holography, in this case by using the holographic dual of MHD constructed in \cite{Grozdanov:2017kyl}. 

Now that we have discussed the need to extend MHD to a quasihydrodynamic theory, the question that remains is what is the approximately conserved current when $\tau \sim \tau_{mft}$? The answer is simple. In such systems, there exists an approximate one-form symmetry associated with an approximate conservation of a number of electric flux lines crossing a co-dimension two-surface. In the limit of $\tau \to \infty$, in which the mass of the new mode becomes zero---i.e., the photon becomes massless---we arrive at an electromagnetic vacuum state. In terms of Eqs. \eqref{MHD-MIS-EOM1}--\eqref{MHD-MIS-EOM3}, the two terms in \eqref{MHD-MIS-EOM2} combine into the full dual electromagnetic field strength $\tilde F^{\mu\nu} = \frac{1}{2} \epsilon^{\mu\nu\rho\sigma} F_{\mu\nu}$ ($u^{[\mu} h^{\nu]} = u^{\mu\nu}$ is the magnetic part of $\tilde F^{\mu\nu}$ \cite{Grozdanov:2016tdf}) with both electric and magnetic components, we can write $F=d A$, so that both $d F = 0 $ and $ d \star F = 0$. 

\subsection{Theory of elasticity, higher-form symmetries and topological aspects of quasihydrodynamic currents}\label{sec:Elasticity}

It is instructive to look at another well-known example of a quasihydrodynamic theory, which is in fact similar to magnetohydrodynamics of Section \ref{sec:MHD}. This is the theory of elasticity, a historically extensively explored subject. While not normally phased in this language, the recent reformulation of elasticity from the point of view of higher-form symmetries \cite{Grozdanov:2018ewh} makes the connection to MHD (through the symmetry structure) and quasihydrodynamics rather transparent. Moreover, it will enable us to further elaborate on the  meaning of some approximately conserved currents, which can be interpreted from a purely symmetry-based perspective. 

The theory of elasticity (see e.g. Ref. \cite{landau1986theory}) describes the dynamics of an elastic medium in a $d$-dimensional Euclidean space, embedded into the $(d+1)$-dimensional spacetime $x^\mu =(t,x^i)$. In its most common incarnation, the theory uses a dynamical variable, which is the displacement field $\phi^I$, where $I=\{1,\ldots,d\}$. One can think of $\phi^I$ as describing the coordinates of a given piece of the solid.  The dynamics of $\phi^I$ is governed by the conservation of momentum
\begin{equation}
    \d_\mu P^{\mu}_I=0,\qquad P^\mu_I \equiv C^{\mu\nu}_{IJ}\d_\nu \phi^J ,
\end{equation}
where $P^0_I = \rho \d_t \phi^I$ and $P^i_I = C^{ij}_{IJ}\d_i \phi^J$. The tensor $C_{IJ}^{ij}$ is known as the elastic tensor (see e.g. Refs. \cite{chaikin2000principles,kleinert1989gauge,Beekman:2016szb} for reviews). One may also think of this theory as a theory of massless Goldstone boson that arise from spontaneously broken translational symmetry due to the presence of a lattice.   Elasticity theory is then the low-energy limit in which the inhomogeneity of the state can be neglected \cite{Leutwyler:1996er}. For example, when all translational symmetries are broken by the lattice with some corresponding spatial group $\mathbb{L}$, then the Goldstones $\phi^I$ parametrize the quotient space $\mathbb{R}^d/\mathbb{L}\simeq U(1)^d$. 

In the solid free of crystalline defects, such as dislocations and disclinations, there exists an additional conserved current when $\phi^I$ are single-valued. In particular, 
\begin{equation}\label{eq:Elastic-conservedDomainWall}
    \d_{\mu_1} J_I^{\mu_1...\mu_d} = 0,\qquad J_I^{\mu_1...\mu_d }= \epsilon^{\mu_1...\mu_{d}\nu}\d_\nu \phi^I .
\end{equation}
This property plays the role of the topological Bianchi identity. As in the case of electromagnetism and MHD discussed in Section \ref{sec:MHD}, where the Bianchi identity should be thought of as encoding the conservation of the number of magnetic flux lines \cite{Gaiotto:2014kfa,Grozdanov:2016tdf}, Ref. \cite{Grozdanov:2018ewh} argued that the topological current $J^I$ should also be treated as a genuine conserved global current, which can facilitate the dynamics of an effective field theory of elasticity. 

In fact, the periodicity of the Goldstones, i.e. $\phi^I(x) \sim \phi^I(x+\ell)$, where $\ell$ is the lattice spacing, gives a straightforward interpretation to $J_I^{\mu_1...\mu_d}$ as the current corresponding to a conserved number of domain walls of $\phi^I$ in a direction perpendicular to the $(\mu_1,\ldots,\mu_d)$ plane. Their conserved number density (the charge) is $\int_{S^1} \star J_I$, which in conventional language counts the number density of the lattice sites of the elastic medium. The conservation of the higher-form current implies that there can be no crystalline defects in the system, which is consistent with the interpretation of \eqref{eq:Elastic-conservedDomainWall} noted above and stems from the analytic properties of $\phi^I$. By focusing on the theory in $2+1$ dimensions, one may work with a dualized description of the momentum operator in which $P^\mu_I = \epsilon^{\mu\nu\lambda} F^I_{\nu\lambda}$. In this picture, the conservation of momentum is equivalent to the conservation of magnetic flux lines in electromagnetism (now point-like objects in $2+1$ dimension) and the conservation of $J^I$ play a role of the conservation of electric flux lines \cite{kleinert1989gauge,Beekman:2016szb}. 

For a theory with a non-vanishing equilibrium domain wall density $\la J^I\ra \ne 0$, it is clear that the current $J^I$ and the momentum $P_I$ are overlapped (in the terminology of the memory matrix formalism).\footnote{In this example, where $J^I\sim \star P_I$, this statement is rather obvious. However, one can also show that even when $P_I \ne \star J^I$, the susceptibility of $P_I$ and $J^I$ is also non-zero by using the consistency of the equilibrium partition function \cite{Grozdanov:2018ewh}.} Among other things, this gives rise to a propagating mode in the transverse channel captured by linearized conservation laws. To be more explicit, consider the fluctuation of the transverse $P_\perp(t,x) = T_{\mu y}(t,x)\, dx^\mu $ in $2+1$ dimensions. Then,
\begin{equation}
J_\perp(t,x) = \epsilon_{\mu\nu\lambda}\left( C^{-1} \cdot T \right)^{\lambda y} dx^\mu \wedge dx^\nu   ,\quad \text{where}~ \left(C^{-1}\cdot T \right)_\mu^J = (C^{-1})^{IJ}_{\mu\nu} P^\nu_I ,
\end{equation}
and where $C^{-1}$ is the inverse of the elastic tensor. The conservation equation for $J_\perp$ then couples to the conservation of momentum in the following way:
\begin{subequations}
\begin{align}
\d_\mu P^\mu_\perp = \d_t T^{ty} + \d_x T^{xy} &=0 ,\\
\d_\mu (J_\perp)^{\mu y} =  \d_t \left( C^{-1}\cdot T \right)^y_x - \d_x \left( C^{-1}\cdot T\right)^y_t &=0  ,
\end{align}
\end{subequations}
which can be combined into a wave equation for $T^{ty}$ upon decomposition of $C^{-1}$ (see e.g. \cite{Beekman:2016szb}). This is in contrast with the spectrum of transverse fluctuations of fluids, which are controlled only by the conservation of momentum and, as a result, exhibit only (non-propagating) diffusive modes (see also \cite{Grozdanov:2018ewh} and \cite{Baggioli:2018vfc,Baggioli:2018nnp}).

The higher-form current $J^I$ can become approximately conserved in the presence of small non-dynamical dislocations. This scenario can occur in the melting of two-dimensional crystals \cite{Halperin1978,Zippelius1980,Delacretaz2017,Delacretaz2017b}. In this situation, one can write down the broken Ward identity for $J^I$ as 
\begin{equation}
    \d_\mu J^{I\mu\nu} = -\frac{1}{\tau} J^{I t \nu}.
\end{equation}
Depending on the time scale $\tau$, the transverse excitation of the system with an approximately conserved $J^I$ can exhibit features of either a solid (propagating transverse sound) or a fluid (pure diffusion). In the regime of $\omega \tau \ll 1$, the system exhibits a diffusive mode and a decaying non-hydrodynamic mode is located at $\omega = -i \tau^{-1}$ at zero $k$. In the large $\omega \tau$ regime, where one can neglect the relaxation time of the higher-form current, the two purely imaginary poles collide and turn into transverse sound poles, just as in the examples discussed in previous sections. This system therefore also fits into the wide class of  quasihydrodynamic theories. Moreover, in the $\tau \to \infty$ limit, the approximately conserved current has a clear interpretation in terms of the known higher-form conserved current. 

Interestingly, the approximately conserved currents discussed in previous subsections can often also be written as a Hodge dual of an exactly conserved (hydrodynamic) current in the system (perhaps after rescaling the time coordinate). We end this section by listing some instructive examples:
 
\begin{itemize}
    \item \textit{A spin system:} In $1+1$ dimensions, consider a system with a conserved zero-form $U(1)$ current $\d_\mu j^\mu = 0$, where $j^\mu = (S,\CJ)$, in which there exists an additional approximately conserved current conservation equation stated in Eq. \eqref{eq:fick2} of Section \ref{sec:DS}. One can define the current $\tilde j^\mu = (S/v,\CJ)$ where the characteristic velocity is defined via $v^2 = \fD/\tau$. We further define $Q^\mu = \epsilon^{\mu\nu} \tilde j_\nu$. With these definitions in hand, we can write the system of quasihydrodynamic equations  \eqref{eq:fick2} as
    \begin{equation}
        \d_\mu \tilde j^{\mu} =0,\qquad \d_\mu Q^\mu =-\frac{1}{\tau}Q^t,
    \end{equation}
where the operator $\int Q = \int \star\, j $ counts the number density of the domain walls in the system. The number is conserved when $\tau \to \infty$. 

     \item \textit{Higher-dimensional systems and domain walls:} One can straightforwardly generalize the $(1+1)$-dimensional discussion to a $(d+1)$-dimensional system. Imagine a theory with a conserved one-form $j$. Then, we define $Q^{\mu_1...\mu_d }\equiv \epsilon^{\mu_1... \mu_d \nu} \tilde j_\nu$, which can be conserved or approximately conserved: 
    \begin{equation}\label{eq:elastic-Spin-higherform}
      \d_{\mu_1} Q^{\mu_1...\mu_d} = -\frac{1}{\tau}Q^{t\mu_2...\mu_d}.
    \end{equation}
Eq. \eqref{eq:elastic-Spin-higherform} expresses the fact that the number density of domain walls is only approximately conserved.  This discussion can be trivially extended to higher-form currents. Note also that the form of the higher-form currents \eqref{eq:elastic-Spin-higherform} implies that the equation governing $\d_t j^i$ in the directions transverse to the wave vector cannot depend on the momentum. For example, in a $(2+1)$-dimensional theory, the equation of motion for $\tilde j^y(t,x)$ following from \eqref{eq:elastic-Spin-higherform} is
\begin{equation}\label{eq:elastic-Spin-puredecay}
  \d_t \tilde j^y = -\frac{1}{\tau} \tilde j^y.
\end{equation}
We will discuss a class of holographic models which exhibit this property in Section \ref{sec:Outline-Hol}.
    
     \item \textit{MIS theory:} In the shear channel of the MIS theory, the linearized equation of motion is identical to the theory with an approximately conserved $d$-form that appeared above in the context of the theory of elasticity. The higher-form current is $J_\perp = \star P_\perp$ (there is no appearance of an analogue of the elastic tensor $C^{\mu\nu}_{IJ}$). Similarly to the spin system discussed above, the antisymmetric structure of $J_\perp$ also implies that the spectrum in the scalar channel does not depend on the wave vector. In particular, in $3+1$ dimensions, we have 
     \begin{equation}
         \d_t \delta T^{xy}(t,z) = -\frac{1}{\tau} \delta T^{xy}(t,z).
     \end{equation}
In the sound channel, the identification of the approximately conserved operator does not follow from such simple topological considerations.  
     
     \item \textit{Magnetohydrodynamics:} The meaning of the approximately conserved current was already briefly discussed in Section \ref{sec:MHD}. For completeness, we only restate here the main conclusion that in the vacuum, there exist two one-form symmetries: the conservations of both numbers of magnetic and electric flux lines. They are encoded in conserved $J$ and $\star J$, where $J$ is the magnetic two-form current. In a plasma with long-lived photons, the electric flux density becomes approximately conserved and there is a quasihydrodynamic description.
    
\end{itemize}

\subsection{Kinetic theory}
A more subtle appearance of the quasihydrodynamic formalism arises in (quantum) kinetic theory, within linear response  \cite{chapman-book,kvasnikov-book,saint-raymond-book,ferziger-kaper-book,ford-book,DeGroot-book,silin-book,grad-1963,gross-1959}.  Let $f_{\mathbf{p}}(x)$ be the one-particle distribution function of a kinetic theory for particles of momentum $\mathbf{p}$ at position $x$, and let $\delta f_{\mathbf{p}}(x)$ denote perturbations away from thermal equilibrium.   The linearized kinetic equations read \begin{equation}
    \partial_t \delta f_{\mathbf{p}} + v_{\mathbf{p}}^i \partial_i \delta f_{\mathbf{p}} = -\sum_{\mathbf{q}} W_{\mathbf{pq}} \delta f_{\mathbf{q}},   \label{eq:boltzmann}
\end{equation}
where $W_{\mathbf{pq}}$ is the linearized collision integral, whose precise form we will not write here.  As noted in \cite{Lucas:2017vlc} for a (quantum) kinetic theory of fermions, $W_{\mathbf{pq}}$ is related to the spectral weight of the quasiparticle density operators.   Thus,  (\ref{eq:boltzmann}) exactly reproduces (\ref{eq:memmat}), suggesting that kinetic theory also fits into our broader framework of hydrodynamics with weakly non-conserved operators.   Here, the set of conserved and approximately conserved operators includes all quasiparticle density operators:  e.g. $c^\dagger_{\mathbf{p}}c_{\mathbf{p}}$, where $c^\dagger_{\mathbf{p}}$ and $c_{\mathbf{p}}$ are quasiparticle creation/annihilation operators at momentum $\mathbf{p}$.  

There is an important physical distinction  between (general) kinetic theory described here and other examples of quasihydrodynamic theories studied above.   As emphasized in the Introduction, quasihydrodynamic theories typically have a small number of approximately conserved quantities, along with a ``soup" of many degrees of freedom with much shorter lifetimes.  In the present example, however, \emph{all} of the interesting degrees of freedom obtain lifetimes that are comparable to the eigenvalues of $W_{\mathbf{pq}}$.  

To the extent that kinetic theory is quasihydrodynamic, we anticipate that the quasihydrodynamic analogy will also persist to other integrable models in $1+1$ dimensions, following the recent advances in ``generalized hydrodynamics" \cite{Castro-Alvaredo:2016cdj,joelmoore,DeNardis:2018omc}.  We further expect that the quasihydrodynamic formalism may allow for a proper description of dynamics in such theories when they are weakly perturbed by integrability-breaking deformations.

One common approximation in kinetic theory is that the relaxation to equilibrium is dominated by only a few non-conserved modes \cite{Romatschke2009,Baier:2007ix}.  In other words, $W_{\mathbf{pq}}$ will have a hierarchy of non-trivial eigenvalues and we focus on the smallest ones.  In the context of the linearized MIS theory, the stress tensor itself is an approximately conserved quantity.  This implies that \begin{equation}
    \delta T^{\mu\nu} = \int \left(\frac{d^4p}{(2\pi)^4}\right)_{\mathrm{on-shell}} \; p^\mu p^\nu \delta f
\end{equation}
should ``almost" be a null vector of the linearized collision integral---namely, the corresponding eigenvalue of $W_{\mathbf{pq}}$ should be very small.   In general, there is  no reason for this to occur.   Nevertheless, one can explicitly reproduce the MIS equations by evaluating the Boltzmann equation integrated over $p^\mu p^\nu$ and assuming that \begin{equation}
X^{\mu\nu\rho} =     \int\left(\frac{d^4p}{(2\pi)^4}\right)_{\mathrm{on-shell}} p^\mu p^\nu p^\rho \delta f  \approx X^{\mu\nu\rho}(T,u,\Pi)
\end{equation} can be approximately written in terms of only $u^\mu$, $T$ and  $\Pi^{\mu\nu}$.  In this expression, one conventionally approximates the collision integral in the relaxation time approximation \cite{Baier:2007ix, Romatschke2009}.  Let us emphasize that in a generic kinetic theory, there is no reason to expect that $X^{\mu\nu\rho}(T,u,\Pi)$ is not just as long lived as $\Pi$.   What is special about the holographic models we describe below is that for certain parameter regimes and due to special microscopic reasons, $\Pi$ is (alone) a long-lived quantity.

An alternative approximation in kinetic theory is that all non-zero eigenvalues of the collision integral are identical (see e.g. Ref. \cite{Lucas:2017vlc}).   Such models will not generically recover the MIS phenomenology.   To the extent that such models are quasihydrodynamic, every single mode $\delta f_{\mathbf{p}}$ is quasihydrodynamic.

\subsection{Other examples}

There are a few further examples of quasihydrodynamics that we will not address in this paper, but which we note here for reference:  the ``zero-sound" physics of low temperature holographic probe brane models \cite{Karch:2008fa, Davison:2011ek, Chen:2017dsy, Gushterov:2018spg}, and quantum-fluctuating superconductivity \cite{Davison:2016hno}. Other systems of interest that may be re-examined in the future from the point of view of quasihydrodynamics include theories such as the one studied in \cite{Davison:2018ofp}.  

As mentioned in the Introduction, recent experiments \cite{Bandurin1055,Crossno1058,Moll16,Lucas:2017idv} have observed evidence for hydrodynamic flow of electrons.   Unfortunately, this hydrodynamic flow is not exact due to the presence of impurities, phonons and Umklapp scattering, all of which can relax momentum away from the electronic fluid.  As derived in \cite{Hartnoll:2012rj,hartnoll2018holographic}, in the presence of momentum relaxation, the linearized hydrodynamic equations of motion for momentum density $g_x$ and energy density $\epsilon$ in a charge neutral fluid become exactly analogous to (\ref{eq:fick2}), with $S$ replaced by $\epsilon$ and $\mathcal{J}$ replaced by $g_x$.   The conventional sound wave of a fluid is split into a diffusive mode and a ``gapped" mode, as depicted previously in Figure \ref{fig:SC-P1-P2}.

\section{Outline and summary of the holographic method}\label{sec:Outline-Hol}
The purpose of this section is to introduce a simple holographic calculation of quasihydrodynamics.   The physical system that we study is a transition from diffusion to ballistic physics, analogous to Section \ref{sec:DS}.   Our focus here is not on the physics, but on the technical strategy that we use to analyze the holographic model.  The methods that we develop below are a streamlined version of the algorithm of \cite{Lucas:2015vna,Chen:2017dsy}.   For the remainder of the paper, we assume that the reader is familiar with holographic theories; we will not try to explain the basics of the correspondence.

Our goal is to study linear response in a large-$N$ (matrix) field theory in $d+1$ spacetime dimensions.  We postulate that this theory is holographically dual to classical gravity in $d+2$ spacetime dimensions.   Single-trace operators which are rank-$s$ tensors in the field theory are dual to rank-$s$ fields in the bulk.   In this section, we will be studying a gauge field $A_a$ in the bulk with field strength $F_{ab} = \partial_a A_b - \partial_b A_a$.   The dual operator of this gauge field is a conserved $U(1)$ current associated to a conventional (zero-form) symmetry.   Suppose that the electromagnetic part of the action is
\begin{align}\label{eq:higherF-toybulk}
     S_{\mathrm{EM}} = - \frac{1}{4}\int d^4x \sqrt{-g}\, {X_{ab}}^{cd} F^{ab} F_{cd},
\end{align}
where the tensor $X_{abcd}$ is anti-symmetric under $a \leftrightarrow b$ and $c\leftrightarrow d$ and symmetric under $ab\leftrightarrow cd$. In the four-dimensional bulk, there are only six components which are
\begin{equation}
    \{ X_1,\ldots,X_6\} = \{X_I^{\;\; I} \},\qquad \text{where}\qquad I =\{tx,ty,tr,xy,xr,yr \} .
\end{equation}
These tensors $X$ are functions of the background bulk fields.   We will assume that the background geometry is isotropic and translationally invariant in the boundary spacetime directions: 
\begin{align}\label{eq:HigherF-coord}
    ds^2 = \frac{L^2}{r^2}\left( -a(r)b(r)dt^2 + \frac{b(r)}{a(r)}dr^2 + dx^2 + dy^2 \right).
\end{align} 
We also assume that on the background geometry, $A_a=0$.   These assumptions suggest that  $X$ should only be functions of the background metric, i.e. of $a(r)$ and $b(r)$.   Isotropy implies that $X_1 =X_2$ and $X_5=X_6$ everywhere in the bulk. Furthermore, the regularity of the background solution implies a relation between $t$ and $r$ components of $X_{abcd}$ at the horizon:  $X_1(r_h) =X_5(r_h)$ and $X_2(r_h) = X_6(r_h)$.  Holographic electromagnetic bulk actions that can be cast in this form include the probe brane theory \cite{Chen:2017dsy} as well as theories with higher-derivative couplings for $F_{ab}$ \cite{Myers:2010pk,Witczak-Krempa2014a,Bai:2013tfa,Grozdanov:2016fkt}.   Our convention will be that the boundary theory, which lives in the UV, is at $r = 0$;  the geometry ends in the IR at a planar black hole horizon at $r=r_{\mathrm{h}}$.   From general principles, \begin{equation}
    a(r) = 4\pi T(r_h - r) + \CO\left((r_h-r)^2\right)  \label{eq:anearhorizon}
\end{equation}   
near the horizon, while $b(r)$ is finite.   If the UV theory is conformal, then $a(r)\sim b(r)\sim r^{-2}$ as $r\rightarrow 0$.   For simplicity, we can assume these UV scalings in the discussion that follows, though the general algorithm we describe is not sensitive to this assumption.

Since $A_a$ couples quadratically in $S_{\mathrm{EM}}$, we conclude that the linearized equations of motion for bulk fields depend only on fluctuations $\delta A_a$.   Using boundary spacetime translational invariance, we may write \begin{equation}
    \delta A_a(r,x^\mu) = \int \frac{d\omega d^dk}{(2\pi)^{d+1}}\; \delta A_a (r)e^{-i\omega t + i \vec{k} \cdot \vec{x}}\,
\end{equation}
and need only solve ordinary differential equations for $\delta A_a(r)$.   For convenience, we will assume that $\vecb{k}$ points in the $x$-direction.  Let us first consider the equations of motion for $\delta A_t$ and $\delta A_x$, which couple and determine the one-point functions $\la \delta j^t \ra$ and $\la \delta j^x \ra$: 
\begin{subequations}
\label{eq:HigherF-eom}
 \begin{align}
      \partial_r \left( \frac{X_3}{b} \delta A_t' \right) + \frac{k X_1}{a} \left( \omega \delta A_x + k \delta A_t \right) &= 0,\label{eq:HigherF-eomAt}\\
       \partial_r \left( a X_5 \delta A_x'  \right) - \frac{\omega X_1}{a} \left( \omega \delta A_x + k \delta A_t \right) &=0, \label{eq:HigherF-eomAx} \\
       \frac{\omega X_3}{b} \delta A_t' + ka X_5 \delta A_x' &= 0.  \label{eq:HigherF-eomr}
 \end{align} 
\end{subequations}
In general, even for the simplest geometries, these equations cannot be solved exactly.

However, we do not need the exact solution.   We are only interested in the quasihydrodynamic regime, which (at least in holography) necessitates \begin{equation}
    \omega \ll T.  \label{eq:omegalessT}
\end{equation}
Roughly speaking, in holographic models, the radial coordinate $r$ corresponds to the ``energy scale" at which physics takes place.   At energy scales large compared to temperature, a field obeying (\ref{eq:omegalessT}) appears approximately static.   Indeed, if we look at \eqref{eq:HigherF-eom}, as $r\rightarrow 0$, all of the $\omega$ and $k$ dependence is negligible.   So we could perturbatively construct a solution in $\omega$ and $k$ near the boundary.   In contrast, very close to the horizon, (\ref{eq:anearhorizon}) dominates the equations of motion.    The near boundary expansion fails and instead the solutions must obey suitable infalling boundary conditions. Indeed, (\ref{eq:HigherF-eom}) again becomes a simpler differential equation to solve in the limit $r\rightarrow r_h$.

\begin{figure}[t]
\includegraphics[width=4in]{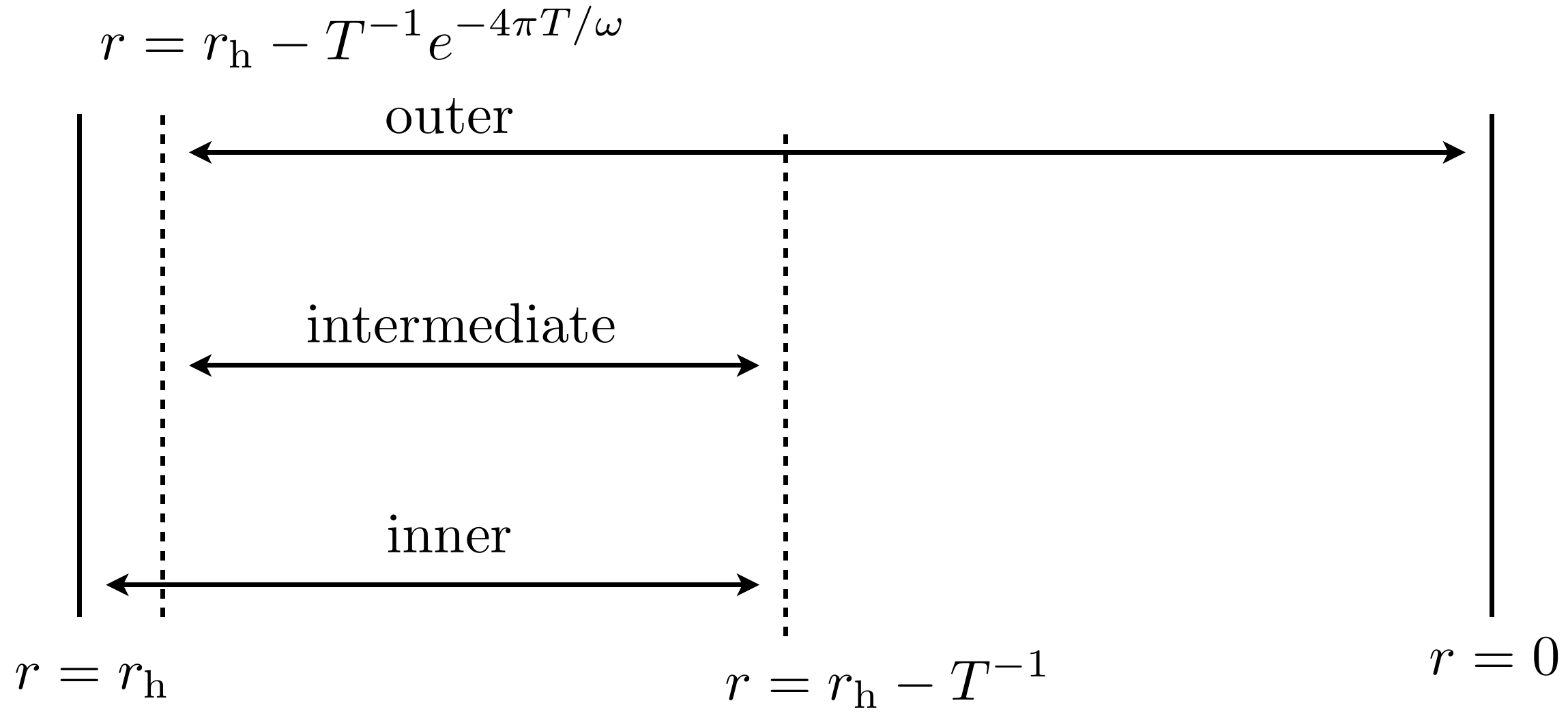}
\caption{The matched asymptotic expansion relies on simplifying the bulk equations in two separate limits (near-horizon and near-boundary), and on the overlap of the regions where these two different expansions are valid.   The overlap appears very close (but not too close) to the horizon.}
\label{figure:matchedexpansions}
\end{figure}

Since we have two separate regimes in which the solution to (\ref{eq:HigherF-eom}) appears simple, we may attempt to perform a matched asymptotic expansion, as sketched in Figure \ref{figure:matchedexpansions}.   We will first determine the complete set of solutions to the bulk equations of motion---with $\omega=k=0$---in the \emph{outer region}, which begins at $r=0$ and extends in the direction towards the horizon. Then, we will construct a solution to (\ref{eq:HigherF-eom}) in the \emph{inner region}, which extends from the horizon $r=r_h$ outwards, usually by a distance $\sim T^{-1}$.   This solution will be non-perturbative in $\omega$, but also assume $k=0$.   Our claim is that the outer and inner regions interlap in an \emph{intermediate region}:  $r_h - T^{-1} e^{-4\pi T/\omega} > r > r_h - T^{-1}$.   We do not have a proof that this is universally an intermediate region in which to match solutions, but this inequality will hold in every example studied in this paper.  This intermediate region is close to the horizon, and so the matching coefficients between inner and outer solutions will generally depend on the near-horizon geometry.    

We emphasize that ``matching procedures" have been widely used in the literature \cite{Kovtun:2003wp,Iqbal:2008by,Edalati:2010hk,Edalati:2010pn,Davison:2013bxa,Donos:2017ihe,Blake:2018leo}.  Nevertheless, there are a few subtle differences between our implementation and in those that exist in the literature, which we will comment on as we proceed.   The matching algorithm that we describe generalizes straightforwardly to much more sophisticated problems than have previously been tractable in the literature; indeed, our discussion of holographic magnetohydrodynamics in Section \ref{sec:MHD-Hol} would be a highly non-trivial calculation using other previous holographic matching methods.

%------------
\paragraph{Outer region:}
In the region away from the horizon, we expand the solutions for $\delta A_t$ and $\delta A_x$ as $\omega,k\rightarrow 0$:
\begin{subequations}
  \begin{align}
      \delta A_t(r,x^\mu) &= \hat a_t(x^\mu) + \hat j_t(x^\mu) \Phi_1(r) + \CO(\omega,k),\\
      \delta A_x(r,x^\mu) &= \hat a_x(x^\mu) + \hat j_x(x^\mu) \Phi_2(r) + \CO(\omega,k).
  \end{align}
\end{subequations}
The radial functions $\Phi_{1,2}(r)$ can be obtained by setting $\omega = k =0$ in (\ref{eq:HigherF-eom}), as mentioned previously:
\begin{align}
    \Phi_1(r) = \int_{s=0}^{r}ds \,\frac{b(s)}{X_3(s)},\qquad \Phi_2(r) = \int_{s=0}^r ds\, \frac{1}{a(s)X_5(s)}\,.
\end{align}  
Both of these functions vanish at the boundary, while at the horizon, $\Phi_1(r_h)$ is finite and $\Phi_2(r_h)$ diverges logarithmically. For future purposes,  it is convenient to denote the finite part of the functions $\Phi_{1,2}$ by $\phi_{1,2}$, and subtract out the logarithmic divergence as follows: 
\begin{align}
    \Phi_2(r) =\phi_2(r) + \frac{1}{a'(r_h) X_5(r_h)} \ln a(r) .
\end{align}
Hence,
  \begin{align}
  \phi_1(r) = \Phi_1(r),\qquad \phi_2(r) =  \int^r_{s=0} ds\, \frac{1}{a(s)X_5(s)}\left(1-\frac{a'(s)X_5(s)}{a'(r_h)X_5(r_h}  \right) .
  \end{align}

To determine the value of $r$ at which this outer region ends, we expand to first order in $\omega$ and $k$. This can be done either in the scheme where $\omega \sim k\sim \epsilon$ or $\omega\sim k^2\sim \epsilon$, where the small parameter $\epsilon \ll 1$. In either scheme, we can write 
\begin{align}
    \delta A_a = \delta A_a^{[0]} + \epsilon\, \delta A_a^{[1]} + \CO(\epsilon^2), \label{eq:higherF-smallFreqExpansion}
\end{align}
substitute this expansion into the equation of motion and solve for $\delta A_a^{[n]}$, order by order in $\epsilon$, with $\delta A_a^{[n]}(r\to 0) =0$ for $n>0$. 
Focusing on the equation of motion for $\delta A_x$ in \eqref{eq:HigherF-eomAx}, we find that the equation for $\delta A_x^{[1]}$ is identical to those at zeroth order in $\epsilon$, namely, 
\begin{align}
    \d_r\left( a X_5 \left(\delta A_x^{[1]}\right)' \right) = 0  .
\end{align}
This implies that $\delta A_x^{[1]} \propto \Phi_2(r)$. The fact that $\Phi_2(r)$ diverges deeper in the bulk implies that the expansion breaks down when 
\begin{equation}
    \frac{\epsilon}{a'(r_h)X_5(r_h)} \ln a(r) \approx 1  .
\end{equation}
In other words, in the low frequency limit $\omega/T$, the regime of the validity of the outer region solution can be expanded up to at least $r\approx r_h - T^{-1} e^{ -4\pi T/\omega }$, as depicted in Fig. \ref{figure:matchedexpansions}. 

Using (\ref{eq:HigherF-eomr}) along with the results of the previous paragraph, we immediately  see that (after performing an inverse Fourier transformation from $\omega$ and $k$): \begin{equation}
    \partial_t \hat j^t + \partial_x \hat j^x = 0.  \label{eq:exactconssec3}
\end{equation}
This is the exact Ward identity associated with charge conservation.  It is one of the two equations of motion in the quasihydrodynamic regime.   The other, approximate conservation law cannot be fixed without matching the solution into the near-horizon regime.

A few points are in order.  Firstly, we have not used gauge-invariant variables.  While this approach is easy to use here, it becomes increasingly difficult in more complicated systems, where gauge-invariant variables may become difficult to relate to physical quantities of the boundary theory.  Moreover, the equations of motion of the fluctuations in the radial gauge can be easily written down, even in a more complicated problem (as we discuss in Section \ref{sec:MHD-Hol}). Secondly, our aim is to find the explicitly broken Ward identity and not simply the dispersion relation of the quasinormal modes.   Thus, it is easier to work directly with variables whose boundary conditions relate to the boundary observables of interest.   This also helps us avoid ambiguities which arise at quasihydrodynamic pole collisions.   These features will be explicitly visible in the discussion below.

%------------
\paragraph{Inner region:} In the region close to the horizon, the solution to (\ref{eq:HigherF-eom}) can be written as 
\begin{equation}\label{eq:HigerF-nearHorizon}
    \delta A_a(r,x^\mu) = \CA_a^{(1)}(r,x^\mu) + \CA_a^{(2)}(r,x^\mu)\, a(r)^{-i\omega/(4\pi T)}\, ,
\end{equation}
where the functions $\CA_a^{(1)}(r)$ and $\CA_a^{(2)}(r)$ are either vanishing or finite at the horizon, $r=r_h$.
The second term $\CA_a^{(2)}(r)$ contains ingoing solutions which are familiar in holography.  The first term $\CA^{(1)}_a(r)$ is a pure gauge solution: it is absent in computations with gauge-invariant variables, but it is nevertheless useful to keep track of it.  For example, we can determine both the source and the response when interpolating to the boundary in the radial gauge computation \cite{Policastro:2002tn}. One may argue that since $\CA_a^{(1)}(r)$ is pure gauge, it has to satisfy a first-derivative constraint and have no effect on physical quantities. However, on the operational level, one may also just substitute the ansatz \eqref{eq:HigerF-nearHorizon} into the equation of motion \eqref{eq:HigherF-eom} near the horizon and solve for $\CA_a^{(1)}(r)$ and $\CA_a^{(2)}(r)$. In order to do this, one can start by demanding that, upon the above substitution, the coefficients of divergent terms, such as $\{ a(r)^{-1},a(r)^{-1-i\omega/4\pi T},\ldots\}$, have to vanish. This turns out to give all constraints on the near-horizon solutions at the order in $\omega$ and $k$ to which we are working. In this case, the relevant constraints are  
  \begin{align}
  \d_t \CA_x^{(1)}(r_h,x^\mu) - \d_x \CA_t^{(1)}(r_h,x^\mu) = 0,\qquad \CA_t^{(2)}(r_h,x^\mu) = 0 . 
  \end{align}
Observe that the gauge-invariant part of $\CA^{(1)}_a$ vanishes when evaluated at the horizon, as regularity demands.  Note also that we will be using the shorthand notation $\CA_a^{(n)} = \CA_a^{(n)}(r_h)$ in the rest of the paper.

\paragraph{Intermediate region:} This is the overlapping region in which both the outer and the inner region solutions are valid. Here, we may approximate
\begin{align}
\exp\left\{-i\frac{\omega}{4\pi T}\ln a(r)  \right\} \approx 1 + \frac{\ln a(r)}{4\pi T} \d_t +\CO(\d^2) .
\end{align}
 We first match the logarithmic pieces in the solutions in both regions.  In the inner region, this is the $\mathcal{O}(\omega)$ contribution arising from the infalling contribution to (\ref{eq:HigerF-nearHorizon}). We obtain
\begin{align}
    \d_t\CA_x^{(2)} = \frac{4\pi T}{a'(r_h) X_5(r_h)} \left(\hat j^x + \CO(\omega,k)\right)= -\frac{1}{X_5(r_h)}\left( \hat j^x+\CO(\omega,k) \right), 
\end{align}
where we used the fact that $a(r) = 4\pi T(r_h-r) +\CO(r_h-r)^2$ in our coordinates. Similarly, the matching condition for the finite part implies that 
\begin{subequations}
  \begin{align}
    \CA_{t}^{(2)} &= \hat a_t + \phi_1(r_h) \hat j_t + \CO(\omega,k),\\
    \CA_{x}^{(1)}+ \CA_x^{(2)} &= \hat a_x + \phi_2(r_h) \hat j_x + \CO(\omega,k).
\end{align}
\end{subequations}
These matching conditions together with the regularity condition imply that, up to first order in derivatives, we have 
\begin{equation}\label{eq:higherF-derivedAlmostConserved-jx}
    \phi_2(r_h) \d_t \hat j_x - \phi_1(r_h) \d_x \hat j_t = -(d\hat a)_{tx} - \left(\frac{1}{X_5(r_h)}  \right)\hat j_x .
\end{equation}
One can recast this relation in the form of a weakly broken conservation law as in Eq. \eqref{eq:fick2}, with
\begin{align}\label{eq:HigherF-v2andTau}
\tau = X_5(r_h) \phi_2(r_h),\qquad v^2 \equiv \frac{\fD}{\tau} = \frac{\phi_1(r_h)}{\phi_2(r_h)}.
\end{align}
This computation is almost parallel to the charge diffusion in Einstein-Maxwell theory presented in \cite{Kovtun:2003wp} where one finds that $\tau T \sim 1$ and thus $1/\tau \gg \omega$ in the hydrodynamic limit of $\omega/T \ll 1$. In this regime, it simply means that the relaxation time of $\hat j_x$ is very small at large temperature and $j_t$ is not a long-lived operator anymore. Thus
it is sensible to drop the time derivative term in \eqref{eq:higherF-derivedAlmostConserved-jx} and obtain the constitutive relation $\hat j_x = \fD \d_x \hat j_t$.  
However, in non-Maxwell theories, e.g. the DBI action or higher-derivative theories, one can show that the relaxation time can be tuned such that $(\tau T)^{-1} \ll 1$ \cite{Chen:2017dsy,Myers:2010pk,Witczak-Krempa2014a,Grozdanov:2016fkt}. In this case, one has to keep all the terms in \eqref{eq:higherF-derivedAlmostConserved-jx} to properly capture the quasihydrodynamic behavior of the system.  

\paragraph{Results:} The quasihydrodynamic equations of motion for this theory consist of (\ref{eq:exactconssec3}) and (\ref{eq:higherF-derivedAlmostConserved-jx}): \begin{subequations}
  \begin{align}
  \partial_t \hat j^t + \partial_x \hat j^x &= 0, \\
  \partial_t \hat j^x + v^2 \partial_x \hat j^t &= \chi_j (d\hat a)^t_x - \frac{1}{\tau} \hat j_x,
  \end{align}
\end{subequations}
where $\chi_j = 1/\phi_2(r_h)$ is a thermodynamic susceptibility which determines the coupling to the background gauge field $\hat a$.

\paragraph{Scalar channel:} A similar analysis can also be carried out for the scalar channel involving $\delta A_y$, which results in a finite lifetime of the operator $\hat j_y$. The relevant equation of motion is 
\begin{align}
    \partial_r \left( a X_5 \delta A_y' \right) -\frac{1}{a}\left(\omega^2X_1 - k^2ab   \right)\delta A_y = 0\, ,
\end{align}
which yields the following solution in the outer region
\begin{align}
    \delta A_y = \hat a_y + \hat j_y \Phi_2(r) .
\end{align}
Here, $\Phi_2(r)$ is the same function as the one found in the previous (longitudinal) sector (\ref{eq:HigherF-eom}).  Since here, $\mathcal{A}^{(1)}_y  =0$, we find that the matching conditions imply
\begin{equation}\label{eq:higherF-derivedAlmostConserved-jy}
    \d_t \hat j_y = -\frac{1}{\phi_2(r_h)}(da)_{ty} - \frac{1}{\tau} \hat j_y.
\end{equation}
Note the absence of spatial derivatives.  The parameter $\tau$ is the same as those in \eqref{eq:HigherF-v2andTau}. Moreover, we can also combine the two relations in both channels, \eqref{eq:higherF-derivedAlmostConserved-jx} and \eqref{eq:higherF-derivedAlmostConserved-jy}, into a single approximately conserved two-form current $Q^{\mu\nu} = \left(\star\, \tilde j\right)^{\mu \nu}$, where $\tilde j^\mu = (\hat j^t /v, \hat j^x,\hat j^y)$. Together with the conservation of $j^\mu$, we finally arrive at the full set of quasihydrodynamic equations: 
\begin{equation}
    \d_t \tilde j^t + v\d_i \tilde j^i =0,\qquad \d_t Q^{t\mu} + v\d_i Q^{i\mu} = - \frac{1}{\tau} Q^{t\mu} .
\end{equation}

Let us end this summary by quickly discussing how quasihydrodynamics arises in the presence of double-trace deformation.    Such a deformation implies that the definition of the physical source is a linear combination of the radially dependent and independent pieces (analogues of $\hat a$ and $\hat j$ in \eqref{eq:higherF-derivedAlmostConserved-jx} and \eqref{eq:higherF-derivedAlmostConserved-jy}). Depending on the double-trace coupling, which determine the linear combination, it is possible for $\tau$ to be large ($\tau T\gg 1$); in this case, the current becomes approximately conserved.  We will discuss an approximately conserved current of this type in Section \ref{sec:MHD-Hol}.

\section{Quasihydrodynamics from holography I: Magnetohydrodynamics with dynamical photons}\label{sec:MHD-Hol}

In this section, we apply the holographic algorithm of Section \ref{sec:Outline-Hol} to derive the quasihydrodynamic, low-energy excitations in a field theory with a conserved two-form current. The holographic dual of a strongly interacting theory with a one-form symmetry was proposed in \cite{Grozdanov:2017kyl,Hofman:2017vwr}. In particular, \cite{Grozdanov:2017kyl} argued that this holographic setup furnishes a description of a strongly interacting field theory with a matter sector gauged under dynamical $U(1)$ electromagnetism, thereby providing a holographic dual description of a plasma described by MHD in the limit of low energy. As a result of electromagnetism, the theory was argued to contain dynamical photons \cite{Grozdanov:2017kyl,Hofman:2017vwr}. Hence, one should be able to use this holographic setup to not only describe MHD but also its (quasihydrodynamic) extension to a theory of plasma with dynamical, and potentially strong, electric fields. At the level of linear response, we   will explicitly demonstrate that this is achieved in holography.

Before delving into the details of the calculation, we will review a few essential features of this model. The bulk theory is composed of the Einstein-Hilbert gravity action, a negative cosmological constant and the two-form gauge field $B_{ab}$, written as 
\begin{equation}
\label{eq:MHDintro-bulkaction}
\begin{aligned}
S = \int d^5x\sqrt{-g} \left[R + \frac{12}{L^2} -\frac{1}{3}H_{abc}H^{abc}   \right]  + S_{bnd}-\frac{1}{\kappa(\Lambda)} \int_{r=1/\Lambda} d^4x\sqrt{-\gamma} (n^aH_{a \mu \nu})(n_b H^{b \mu \nu})\, ,
\end{aligned}  
\end{equation}
where $H=dB$, $\Lambda$ is the UV cut-off and $n^a$ is the unit vector normal to the boundary. $S_{bnd}$ combines the boundary Gibbons-Hawking term and counter-terms for the gravitational part of the theory. A detailed exposition of the model, including holographic renormalization, can be found in \cite{Grozdanov:2017kyl}. The diffeomorphism invariance of the metric and the gauge symmetry of the two-form bulk fields imply that the boundary duals possesses a conserved stress-energy tensor $T^{\mu\nu}$ and a conserved two-form current $J^{\mu\nu}$. In particular, in the presence of an external two-form source $b_{\mu \nu}$,
\begin{equation}
  \nabla_\mu T^{\mu \nu} = H^{\nu}_{\;\;\rho \sigma}J^{\rho \sigma},\qquad \nabla_\mu J^{\mu \nu} = 0,
\end{equation}
where $H$, here, is the three-form field strength $H = db$ of an external two-form gauge field. In terms of the generating functional,
\begin{equation}
  Z[b_{\mu \nu}] = \left\la \text{exp}\left[ i\int d^4x J^{\mu \nu}b_{\mu \nu} \right] \right\ra_{\mathrm{QFT}} \, .
\end{equation}
Holographic renormalization and the bulk equations of motion imply that the boundary two-form current $J_{\mu \nu}$ corresponds to the projected bulk three-form field strength $n^a H_{a \mu \nu}$. In fact, the counter-term for the two-form gauge field, i.e. the last term in Eq. \eqref{eq:MHDintro-bulkaction}, can be thought of as a double-trace deformation. This implies that the regularized boundary source is
\begin{equation}\label{eq:MHD-Hol-defSource}
  b_{\mu \nu} = B_{\mu \nu}(r= 1/\Lambda) - \frac{1}{\kappa(\Lambda)}n^a H_{a \mu \nu}\Big\vert_{r=1/\Lambda} \, .
\end{equation}
For the source to be physical, one has to formally impose that it be cut-off independent, i.e. that $\partial b_{\mu \nu}/\partial \Lambda = 0$, which results in the logarithmic running of the double-trace coupling $\kappa$ with a Landau pole. In other words, one finds that 
\begin{equation}
  \frac{1}{\kappa(\Lambda)} = \text{finite} -  \ln (\Lambda/L),
\end{equation}
where the finite part that sets the {\em renormalized} electromagnetic coupling needs to be imposed as the renormalization condition at a new scale $L$, or equivalently, by the choice of the Landau pole scale. In practice, the value can be chosen by hand. For details, see \cite{Grozdanov:2017kyl,Grozdanov:2018ewh}. As we will see below, it is the (finite) value of the electromagnetic coupling that governs the relaxation time of the operator $\la J^{zi}\ra$, or the electric flux density, and sets the regime of validity of the quasihydrodynamic approximation with an approximately conserved electric flux density.

We take the background metric ansatz and two-form gauge field to have the following equilibrium forms (cf. Eq. \eqref{eq:HigherF-coord}):
\begin{subequations}
\begin{align}
  ds^2 &= \frac{L^2}{r^2}\left[ -a(r)b(r)dt^2 + \frac{b(r)}{a(r)}dr^2 + dx^2 + dy^2+c(r)dz^2\right],\\
  B &= h(r)\, dt \wedge dz,
\end{align}
\end{subequations}
where $h(r)$ parametrizes the density of the two-form conserved current, which equals the Hodge dual of the magnetic flux density. The functions $a(r)$, $b(r)$ and $c(r)$ in the metric all asymptote to one at the boundary, $r\to 0$, so that the geometry is asymptotically AdS${}_5$. At the horizon, $b(r)$ and $c(r)$ are regular and approach non-vanishing constants $b(r_h)$ and $c(r_h)$. The regularity conditions impose a stronger constraint on the ``emblackening factor" $a(r)$ and the gauge field $h(r)$, which are forced to behave as 
\begin{equation}
  a(r) = 4\pi T(r_h-r) + \CO(r_h-r)^2,\qquad h(r) = b(r_h)c(r_h)^{1/2}\frac{h_0}{r_h} (r_h-r) + \CO(r_h-r)^2 ,
\end{equation}
close to the horizon at $r=r_h$. 

Finally, we note that in this work, we will study the transverse fluctuations 
\begin{subequations}\begin{align}
    \delta(ds^2) &= \frac{2}{r^2}      e^{i(kz-\omega t)}   \left[h_{ti}(r)dtdx^i + h_{zi}(r) dzdx^i  \right], \\
    \delta B &=  e^{i(kz-\omega t)} \left(\delta B_{ti} dt\wedge dx^i + \delta B_{zi} dz\wedge dx^i\right), 
    \end{align}\end{subequations}
where the index $i$ denotes the $(x,y)$ plane coordinates, with the plane being perpendicular to the $z$-direction of the background magnetic field lines. 

The two subsection, which contain the bulk of the derivation of aspects of quasihydrodynamics in a holographic plasma can be summarized as follows:
\begin{itemize}
  \item First, in Subsection \ref{sec:photon}, we consider the case with a zero background magnetic flux density. At small momentum, we find that the system exhibits a diffusive mode, which will be referred to as magnetic diffusion, along with a massive, non-hydrodynamic mode with an imaginary gap set by the renormalized electromagnetic coupling. The system will then be shown to exhibit the collision of the diffusive and gapped poles described in Section \ref{sec:hydro}.  We will  find explicit analytic expressions (in terms of background bulk quantities and the $U(1)$ coupling) for the relaxation time of the approximately conserved operator $\la J^{zi}\ra$ and the asymptotic speed of the pair modes (after the collision) at large $k$. In the large-$k$ limit and for small electromagnetic coupling, this speed will tend to the speed of light. We thus explicitly show the presence of photons in the plasma.

  \item Then, in Subsection \ref{sec:AlfvenWave}, we consider the case with a non-zero background magnetic field in the MHD channel with Alfv\'{e}n waves. We demonstrate in detail how a pair of diffusive modes (shear momentum and magnetic diffusion) combines into a pair of propagating modes after the collision. We also show that operators which set the validity of the (magneto)hydrodynamic regime are linear combinations of the stress-energy tensor and the two-form current. The relaxation time can again be expressed in terms of an integral over the equilibrium bulk quantities.  

\end{itemize}

%----------------------------------

\subsection{The photon in a charge neutral plasma}\label{sec:photon}

We begin by considering a setup with $h(r)=0$.  In this limit, the function $c(r) = 1$.  Moreover, the metric and the two-form gauge field fluctuations are decoupled so we can focus only on the equations of motion for the two-form gauge field. In radial gauge, they are 
\begin{subequations}
  \begin{align}
     \d_r \left(\frac{r}{b(r)} \delta B_{ti}'  \right) - \frac{r k}{a(r)} \left( \omega \delta B_{zi} + k \delta B_{ti} \right)&=0,\label{eq:zeroBshear1}\\
    \d_r \left( r a(r) \delta B_{zi}'   \right) +\frac{r \omega }{a(r)} \left( \omega \delta B_{zi}+ k\delta B_{ti} \right)&=0\label{eq:zeroBshear2} , \\
    \omega \frac{r}{b(r)} \delta B_{ti}' + kr a(r) \delta B_{zi}' &= 0. \label{eq:sec4acons}
\end{align}
\end{subequations}
While in this case, the background solution is just the AdS${}_5$-Schwarzschild black brane, our derivation does not relying on the explicit form of background solution, so we keep $a(r)$ and $b(r)$ general. For this reason, the computation presented here could be immediately extended to a larger class of theories with more complicated $a(r)$ and $b(r)$, supported by other bulk matter which does not couple to $B_{ab}$.

\paragraph{Outer region:}
In the low frequency limit, we can show that the solution is 
\begin{subequations}
\begin{align}
    \delta B_{ti}(x^\mu,r) &=\delta \hat B_{ti}(x^\mu) +
    \delta\hat J_{ty}(x^\mu) \Psi_1 (r),\\
    \delta B_{zi}(x^\mu,r) &= \delta \hat B_{zi}(x^\mu) + \delta \hat J_{zi}(x^\mu) \Psi_2 (r),
\end{align}
\end{subequations}
where the functions $\Psi_{1,2}(r)$ satisfy the following first-order differential equations: 
\begin{equation}
    \frac{r}{b(r)}\Psi_1'(r)  = 1,\qquad ra(r)\Psi_2'(r)   = 1,
\end{equation}
with the UV boundary conditions at $r=1/\Lambda$, with $\Lambda L \gg1 $, giving rise to logarithmically divergent behavior (the Landau pole), 
\begin{equation}
    \Psi_{1,2}(r = 1/\Lambda) = -\ln(\Lambda L) + \cdots \, .
\end{equation}
This means that the solutions can be written as 
\begin{equation}
    \Psi_1(r) = -\ln(\Lambda L) + \int^{r}_{s=1/\Lambda} ds \frac{b(s)}{s},\qquad \Psi_2(r) = -\ln(\Lambda L) + \int^{r}_{s=1/\Lambda} ds \frac{1}{sa(s)}.
\end{equation}
The two functions $\Psi_{1,2}(r)$ can be expressed as 
\begin{equation}
\begin{aligned}
    \Psi_1(r) &= -\ln(\Lambda L) + \int^{r}_{s=1/\Lambda} \frac{ds}{s} + \int^{r}_{s=1/\Lambda} ds \left( \frac{b(s)-1}{s}\right) ,\\
    &= \ln(r/L) + \psi_1(r) ,\qquad 
\end{aligned}
\end{equation}
which is finite, and 
\begin{align}
    \Psi_2(r) &= -\ln(\Lambda L) + \int^{r}_{s=1/\Lambda} \frac{ds}{s} +\frac{1}{r_h a'(r_h)}\int^{r}_{s=1/\Lambda} ds \frac{a'(s)}{a(s)} +\int^{r}_{s=1/\Lambda} ds \left[  \frac{1-a -\frac{s a'(s)}{r_h a'(r_h)}  }{s a(s)}\right],\nn
    & = \ln(r/L) - \frac{1}{4\pi T r_h} \ln \left(a(r)  \right) + \psi_2(r),
\end{align}
where 
\begin{subequations}
    \begin{align}
    \psi_1(r) &= \int^r_{s=1/\Lambda} ds \left( \frac{b(s)-1}{s}\right), \\
    \psi_2(r) &= \int^{r}_{s=1/\Lambda} ds \left(  \frac{1-a -\frac{s a'(s)}{r_h a'(r_h)}  }{s a(s)}\right).
    \end{align}
\end{subequations}
Both $\psi_1$ and $\psi_2$ are constructed so that they are finite everywhere in the bulk. 

As before, (\ref{eq:sec4acons}) guarantees that \begin{equation}
    \partial_t \delta \hat J^{ti} + \partial_z \delta \hat J^{zi} = 0.  \label{eq:Jticons}
\end{equation}
This is the exact conservation law for magnetic flux lines.   The quasihydrodynamic mode for electric flux lines will arise by matching to the near-horizon region.

\paragraph{Inner region:}

Near the horizon, the general solution for the two-form field as a sum of a regular pure-gauge and infalling parts: 
\begin{subequations}
\begin{align}
  \delta B_{ti} &=\CB_{ti}^{(1)}(r,x^\mu) + \CB_{ti}^{(2)}(r,x^\mu)  a^{-i\omega/4\pi T},\\
  \delta B_{zi} &=\CB_{zi}^{(1)}(r,x^\mu) + \CB_{zi}^{(2)}(r,x^\mu) a^{-i\omega/4\pi T},
\end{align}
\end{subequations}
where $\delta \CB_{\mu i}$ are regular function at $r=r_h$. We proceed by substituting the above ansatz into the equations of motion near the horizon and demanding that the coefficients of terms containing $a(r)^{-1}$, and $a(r)^{-n-i\omega/4\pi T}$ for $n>0$ vanish. This procedure implies several constraints on $\CB_{ti}$ and  $\CB_{zi}$. Firstly, demanding that the coefficients of $a(r)^{-1}$ in both \eqref{eq:zeroBshear1} and \eqref{eq:zeroBshear2} vanish, gives
\begin{equation}\label{eq:zeroBnearhorConstrain}
  \d_t \CB^{(1)}_{zi} - \d_z \CB^{(1)}_{ti} = 0 ,
\end{equation}
where we denote $\CB_{\mu\nu}(r_h,x^\mu)$ as simply $\CB_{\mu\nu}$. Similarly, by demanding that the coefficient of $a^{-2-i\omega/4\pi T}$ in \eqref{eq:zeroBshear1} vanishes, we find that 
\begin{equation}
  \CB^{(2)}_{ti} = 0.
\end{equation}
On the other hand, the function $\CB^{(2)}_{zi}(r_h,x^\mu)$ is non-vanishing and satisfies a more complicated constraint. However, there is no need for us to solve explicitly for $\CB^{(2)}_{zi}$. We only require it to be finite at the horizon. 

\paragraph{Intermediate region:}

In the intermediate region, we match the solutions from inner and outer regions. Staring from the inner region, we isolate the logarithmically divergent piece in $\delta B_{xz}$ by writing 
\begin{equation}
\begin{aligned} 
  \delta B_{zi} &=\CB_{zi}^{(1)} + \CB_{zi}^{(2)}(x^\mu)  \left( 1-\frac{i\omega}{4\pi T} \ln a(r) + \CO(\omega^2/T^2)\right),\\
&\approx \left(\CB_{zi}^{(1)} + \CB_{zi}^{(2)}  \right) + \frac{1}{4\pi T} \d_t \CB_{zi}^{(2)} \ln a(r) ,
\end{aligned}
\end{equation}
where we take $\Phi_2(r)\approx \Phi_2(r_h)$. The matching between the solutions in the two regions implies that 
\begin{subequations}
\begin{align}
 \d_t \CB_{zi}^{(2)} & = -\frac{1}{r_h} \delta \hat J_{zi}, \label{eq:matchingzeroB-1} \\
\CB_{zi}^{(1)} + \CB^{(2)}_{zi} &= \delta \hat B_{zi} + \delta \hat J_{zi} \left(\psi_2(r_h) + \ln(r_h/L)  \right), \label{eq:matchingzeroB-2} \\
\CB_{ti}^{(1)} &= \delta \hat B_{ti} + \delta \hat J_{ti} \left( \psi_1(r_h) + \ln(r_h/L) \right). \label{eq:matchingzeroB-3} 
\end{align} 
\end{subequations}
Substituting the expression for $\CB^{(1)}$ from Eqs. \eqref{eq:matchingzeroB-1}--\eqref{eq:matchingzeroB-3} into the near-horizon constraint \eqref{eq:zeroBnearhorConstrain}, we find that 
\begin{equation}
\begin{aligned}
- \frac{1}{r_h} \delta \hat J_{zi} &= (d\delta \hat B)_{tiz} -  \left( \ln(r_h/L) + \psi_1(r_h)  \right) \d_z \delta \hat J_{ti} \\
 &+  \left( \ln(r_h/L)  + \psi_2(r_h)\right)  \d_t \delta \hat J_{xz}.
\end{aligned}
\end{equation}
To interpret this relation in the dual field theory language, we recall that the physical source is defined by the following linear combination of the coefficients of the $r$-independent term and the logarithmic term, as outlined in Eq. \eqref{eq:MHD-Hol-defSource} (see also \cite{Grozdanov:2017kyl,Hofman:2017vwr}):
\begin{equation}\label{eq:SubleadingConstrainGenMetric}
\begin{aligned}
    \delta b &= \delta \hat B(r\to 1/\Lambda) - \frac{1}{\kappa(\Lambda)} \delta \hat J =  \delta\hat B - \ln\left( L\Lambda e^{\kappa(\Lambda)} \right) \delta \hat J,
\end{aligned}
\end{equation}
where $\CM \equiv \Lambda \exp(\kappa(\Lambda)) $ is an RG-invariant scale. The constraint \eqref{eq:SubleadingConstrainGenMetric} then becomes 
\begin{equation}
  -  r_h^{-1} \delta  \hat J_{zi} = (d\delta b)_{tiz}  - \left(\ln(\CM r_h) + \psi_1(r_h)  \right) \d_z \delta \hat J_{ti} + \left( \ln(\CM r_h) + \psi_2(r_h)  \right) \d_t \delta \hat J_{zi}.
\end{equation}
By turning off the source, we finally find the Ward identity for the approximately conserved current: 
\begin{equation}
    \d_t  \la \delta J_{zi}\ra  - v^2 \d_z  \la \delta J_{ti}\ra = - \frac{1}{\tau}\la \delta  J_{zi}\ra , \label{eq:dtjyzsec4a}
\end{equation}
where
\begin{equation}
    v^2 = \frac{\ln(\CM r_h) + \psi_1(r_h)}{\ln (\CM r_h) +  \psi_2(r_h)},\qquad \frac{1}{\tau} = \frac{r_h^{-1}}{\ln (\CM r_h) + \psi_2(r_h)}.  \label{eq:v2tausec4a}
\end{equation}
The meaning of this equation is clear. This is a quasihydrodynamic relaxation equation for $\delta J_{yz}$, which corresponds to the electric field in the thermal plasma. The electric field is therefore not only induced through fluctuating magnetic fields, as in MHD, but is a genuine dynamical degree of freedom, with a relaxation time $\tau$.  A direct consequence is that the system indeed contains dynamical photons, as claimed in \cite{Grozdanov:2017kyl,Hofman:2017vwr}. At large $\tau$, the approximately conserved quantity is the conservation of the number of electric flux lines crossing a co-dimension-two surface. We note that in this expression, temperature (or $r_h$) sets the characteristic energy scale of the system.    $\CM r_h$ is the parameter that controls the renormalized electromagnetic coupling: \begin{equation}
    e_r^{-2} = \ln (\CM r_h).  \label{eq:renormalized-e}
\end{equation}

For the case at hand, the AdS${}_5$-Schwarzschild black hole, we find that $\psi_1(r_h) = \psi_2(r_h) = 0$. Hence, the speed of the propagation of photons is actually the speed of light in vacuum, which is independent of the electric charge: $v=1$.    As $e_r^{-2} \gg 1$, the time scale $\tau$ obeys $\tau T \gg 1$, and so the electric field is indeed a quasihydrodynamic mode.   For this geometry, \begin{equation}
    \frac{1}{\tau} = \frac{\pi T}{\ln \left(\frac{\mathcal{M}}{\pi T}\right)}.  \label{eq:sec4atau}
\end{equation}
This may be compared with the numerical estimate of this lifetime found in  \cite{Hofman:2017vwr}, which is about  3.5\% larger.

\paragraph{Results:}  We now collect our results.   Using (\ref{eq:Jticons}), (\ref{eq:dtjyzsec4a}) and (\ref{eq:v2tausec4a}), and defining $\delta E_x = \delta J_{yz}$, $\delta B_y = \delta J^t_y$, we conclude that \begin{subequations}
  \begin{align}
  \partial_t \delta B_y + \partial_z \delta E_x &= 0, \\
  \partial_t \delta E_x + v^2 \partial_z \delta B_y &= - \frac{\delta E_x}{\tau}.
  \end{align}
\end{subequations}
These are precisely the dynamical Maxwell's equations in the presence of an Ohmic current $\mathbf{J} \propto \mathbf{E}$, as discussed in Section \ref{sec:MHD}.   The conductivity of the plasma is proportional to $1/\tau \propto e_r^2 $ and is small in the quasihydrodynamic limit.   Indeed, as the electromagnetic coupling vanishes, the photons decouple from matter and there is no Debye screening.   Hence, as $T\rightarrow 0$ and $e_r \rightarrow 0$, the quasihydrodynamic mode---the photon---becomes arbitrarily long lived.

%%%%%%%%%%%%%%%%%%%%%%%%%%%%%
\subsection{Alfv\'{e}n waves and photons}\label{sec:AlfvenWave}
We now consider the case with a non-zero background magnetic field, i.e. with $h(r) \ne 0$. The background solution is the magnetic black brane solution \cite{DHoker2009}, which is the same background one considers in the case of external, non-dynamical magnetic field (see Ref. \cite{Grozdanov:2017kyl} for a discussion on the relation between holography with and without dynamical magnetic fields). The thermodynamic quantities, transport coefficients and the low-energy spectrum as a function of temperature and background magnetic field were computed numerically in  \cite{DHoker2010b,Grozdanov:2017kyl} (see also \cite{Ammon2017,Janiszewski2016}). Here, we show that the form of the quasihydrodynamic low-energy spectrum can in fact be obtained analytically.  

The equations of motion for the background metric, i.e. for $\{a,b,c\}$, are not particularly illuminating and we will not write them here. However, it is important to note that the background equation for the background two-form gauge field is 
\begin{equation}
  \partial_r \left( \sqrt{-g} H^{rtz} \right) =\frac{d}{dr}\left( \frac{r h'}{b\sqrt{c}} \right) = 0.
\end{equation}
In other words, we can write the radial derivative of $h(r)$ as 
\begin{equation}\label{eq:Alfven-conservedBG-1}
  h'(r) = h_0 \frac{b\sqrt{c}}{r},
\end{equation}
where $h_0$ is a constant. We choose a gauge with $h(r_h)=0$ and, as a result, it is convenient to write down the solution to \eqref{eq:Alfven-conservedBG-1} as 
\begin{equation}\label{eq:Alfven-hProfileFixedhrh}
  h(r) = -h_0 \ln(r_h \Lambda) + \phi(r_h) + h_0 \int_{s=1/\Lambda}^r ds \frac{b(s)\sqrt{c(s)}}{s},
\end{equation}
where $\phi(r_h)$ is a finite integral
\begin{equation}
  \phi(r_h) = h_0\int^{r_h}_{s=0} ds \left( \frac{b(s)\sqrt{c(s)}}{s}-\frac{1}{s} \right).
\end{equation}
This form is particularly convenient for analyzing the solution near the boundary, $r\approx 1/\Lambda \ll 1$. The result is 
\begin{equation}\label{eq:Alfven-hNearBnd}
  h(r) =  h_0 \ln\left( \frac{r}{r_h} \right) + \text{finite}.
\end{equation}
Similarly, there are two additional radially conserved quantities, $Q_1$ and $Q_2$, such that $dQ_1/dr = dQ_2/dr = 0$, which arise from linear combinations of the $xx$- and $tt$-components, and $yy$- and $xx$-components, respectively. Namely,
\begin{equation}\label{eq:Alfven-conservedBG-2}
Q_1= \frac{\sqrt{c}}{br^3}(ab)' + \frac{4rhh'}{b\sqrt{c}} ,\qquad Q_2= \frac{a}{r^3\sqrt{c}}c' +\frac{4r h h'}{b\sqrt{c}} .
\end{equation}
Using $h(r_h)=0$, we find that \begin{subequations}
  \begin{align}
  Q_1 &= -\frac{4\pi T\sqrt{c(r_h)}}{r_h^3} = -Ts, \label{eq:Q1s} \\
  Q_2 &= 0,
  \end{align}
\end{subequations}
where $s$ is the entropy density of the system, as measured by the horizon area.

These relations are crucial to simplify the equations of motion for the metric $h_{\mu\nu}$ and the gauge field fluctuation $\delta B_{\mu\nu}$, which consist of four second-order ODEs:
\begin{subequations}
\begin{align}
  \left( \frac{\sqrt{c}}{r^3b} h'_{ti} \right)' - 4h_0 \delta B'_{zi} - \frac{k}{r^3a\sqrt{c}} \left( \omega h_{zi} + k h_{ti} \right) & = 0,\\
  \left(  \frac{a}{r^3\sqrt{c}}h'_{zi} \right)' - 4h_0 \delta B'_{ti} + \frac{\omega}{r^3a\sqrt{c}} \left( \omega h_{zi} + k h_{ti} \right) &=0,\\
  \left(  \frac{r\sqrt{c}}{b} \delta B_{ti}\right)' - h_0 h'_{zi} - \frac{kr}{a\sqrt{c}} \left( \omega \delta B_{zi} + k \delta B_{ti} \right) &=0,\\
  \left( \frac{ra}{\sqrt{c}} \delta B'_{zi} \right)' - h_0 h'_{ti} + \frac{\omega r}{a\sqrt{c}} \left( \omega \delta B_{zi} + k \delta B_{ti} \right) &=0,
\end{align}
\label{eq:Alfven-dyeom1}
\end{subequations}
and two first-order (constraint) equations: 
\begin{subequations} \label{eq:Alfven-constrain}
    \begin{align}
                \omega \left(\frac{\sqrt{c}}{br^3}h_{ti}^\prime - 4h_0\delta B_{zi}\right)+k \left(\frac{a}{\sqrt{c}r^3}h_{zi}^\prime - 4h_0\delta B_{ti}\right) &= 0, \label{eq:Alfven-constrain-1}\\
                \omega \left(\frac{r\sqrt{c}}{b}\delta B_{ti}^\prime - h_0h_{zi}\right) + k\left(\frac{ra}{\sqrt{c}}\delta B_{zi}^\prime - h_0 h_{ti} \right)  &=  0.\label{eq:Aflven-constrain-2}
  \end{align}\end{subequations}
%

%-----------------------------------
\paragraph{Outer region:}
The equations in the outer region can be organized into two decoupled sets:
\begin{subequations}
\begin{align}
   \frac{\sqrt{c}}{r^3b} h'_{ti}(r,x^\mu)  - 4h_0 \delta B_{zi}(r,x^\mu) &= \hat \pi_{ti}(x^\mu),\label{eq:Alfven-pair1-1}\\
   \frac{ra}{\sqrt{c}} \delta B_{zi}'(r,x^\mu) - h_0 h_{ti}(r,x^\mu) &= \delta \hat J_{zi}(x^\mu), \label{eq:Alfven-pair1-2}
\end{align}
\end{subequations}
and 
\begin{subequations} 
\begin{align}
   \frac{a}{r^3\sqrt{c}} h'_{zi}(r,x^\mu)  - 4h_0 \delta B_{ti}(r,x^\mu) &= \hat \pi_{zi} (x^\mu), \label{eq:Alfven-pair2-1}\\
   \frac{r \sqrt{c}}{b} \delta B_{ti}'(r,x^\mu) - h_0 h_{zi}(r,x^\mu) &= \delta \hat J_{ti}(x^\mu),\label{eq:Alfven-pair2-2} 
\end{align}
\end{subequations}
where $\{ \hat \pi_{ti},\hat \pi_{zi},\hat J_{ti},\hat J_{zi} \}$ are integration constants. The first-order equations \eqref{eq:Alfven-constrain} imply that 
\begin{equation}
  \d_t \hat \pi^{ti} + \d_z \hat \pi^{zi} = 0,\qquad \d_t \hat \delta J^{ti} + \d_z \delta \hat J^{zi} = 0, \label{eq:sec4bcons}
\end{equation}
which are nothing but the exact conservation laws for transverse momentum and the two-form current (magnetic flux). 

Let us first look at the first pair of equations, i.e. Eqs. \eqref{eq:Alfven-pair1-1}--\eqref{eq:Alfven-pair1-2}. Solving for $h_{ti}$, we find
\begin{equation}\label{eq:Alfven-pair1-3}
  \frac{ra}{\sqrt{c}}\left( \frac{\sqrt{c}}{r^3b} h'_{ti} \right) -4h_0 \left( h_0 h_{ti} + \delta \hat J_{zi} \right) = 0.
\end{equation}
This implies that 
\begin{equation}
  h_{ti}(r,x^\mu) =- \frac{\delta \hat J_{zi}(x^\mu)}{h_0} + \hat C_1^{(1)}(x^\mu) \Psi_1^{(1)}(r) + \hat C^{(2)}(x^\mu) \Psi_1^{(2)}(r),
\end{equation}
where $\hat C_1^{(n)}$ are two integration constants of the equation \eqref{eq:Alfven-pair1-3}. Similarly, using \eqref{eq:Alfven-pair1-1}--\eqref{eq:Alfven-pair2-2}, we find that 
\begin{equation}
  \delta B_{zi}(r,x^\mu) = -\frac{\hat \pi_{ti}(x^\mu)}{4h_0}  + \hat C_1^{(1)}(x^\mu) \Phi_1^{(1)}(r) + \hat C_1^{(2)}(x^\mu) \Phi_1^{(2)}(r),
\end{equation}
where $\Psi_1$ and $\Phi_1$ satisfy
\begin{equation}\label{eq:Alfven-pair1-radialeom}
  \frac{\sqrt{c}}{r^3b} \Psi_1' -4h_0 \Phi_1 =0,\qquad \frac{ra}{\sqrt{c}} \Phi_1' -h_0 \Psi_1 = 0, 
\end{equation}
which are solved by 
\begin{equation}
  \Psi_1^{(1)} = ab ,\qquad \Phi_1^{(1)} = h + \frac{Q_1}{4h_0}.
\end{equation}
This solution can be used to construct the other linearly independent solution to \eqref{eq:Alfven-pair1-radialeom} via the Wronskian trick. Applying this method to $\Psi_1$, we find that 
\begin{equation}
  \Psi_1^{(2)}(r) = ab \int_{s=0}^r ds \frac{s^3}{a^2 b\sqrt{c}}.
\end{equation}
It is convenient to use the relation \eqref{eq:Alfven-pair1-radialeom} to solve for $\Phi_1^{(2)}$ in terms of $\Psi_1^{(2)}$ in order to avoid the unnecessary appearance of a logarithmic divergence. We find 
\begin{equation}
  \Phi_1^{(2)}(r) = \frac{1}{4h_0} + \int_{s=0}^r ds \frac{h_0\sqrt{c}}{sa} \Psi_1^{(2)}(s),
\end{equation} 
where the value of $\Phi_1^{(2)}(r=0)$ is fixed by the first derivative of $\Psi_1^{(2)}$ at $r\to 0$ through the first relation in Eq. \eqref{eq:Alfven-pair1-radialeom}. 
In summary, the solutions for $h_{ti}$ and $\delta B_{zi}$ in the outer region are
\begin{subequations}
\begin{align}
  h_{ti} &= -\frac{\delta \hat J_{zi}(x^\mu)}{h_0} + a(r)b(r)\, \hat C_1^{(1)}(x^\mu) + \psi_1^{(2)}(r) \hat C_1^{(2)}(x^\mu),\\
  \delta B_{zi} &= -\frac{\hat \pi_{ti}(x^\mu)}{4h_0} + \left( \frac{Q_1}{4h_0} + h \right) \hat C_1^{(1)} + \left[ \phi_1^{(2)}(r) + \left(\frac{h_0 \sqrt{c(r_h)}}{r_h a'(r_h)} \psi_1^{(2)}(r_h)  \right)\ln a  \right] \hat C_1^{(2)} ,
  \end{align}
\end{subequations}
where 
\begin{subequations}
\begin{align}
  \psi_1^{(2)} &= \Psi_1^{(2)},\\
  \phi_1^{(2)} &= \frac{1}{4h_0} + \int_{s=0}^r ds \frac{h_0 \psi_1^{(2)}(s)\sqrt{c(s)}}{sa(s)} \left( 1- \frac{s a'(r_h)\psi_1^{(2)}(r_h)\sqrt{c(r_h)}}{r_ha'(s)\psi_1^{(2)}(s)\sqrt{c(s)}} \right).
\end{align}  
\end{subequations}
To better understand the meaning of variables $\hat C_1^{(1)}$ and $\hat C_1^{(2)}$, it is helpful to rewrite them in terms of the zero magnetic field case. First, we need to consider how the fluctuations $h_{ti}$ and $\delta B_{zi}$ behave close to the boundary and define 
\begin{subequations}
\begin{align}
  \hat h_{ti} &= - \frac{\delta \hat J_{zi}}{h_0} + \hat C_1^{(1)},\\
  \delta \hat B_{zi} &= -\frac{\hat \pi_{ti}}{4h_0} + \frac{Q_1}{4h_0}\hat C^{(1)}_1 + \frac{C^{(2)}_1}{4h_0} \, .
\end{align}
\label{eq:Alfven-replaceC1}
\end{subequations}
Upon substituting the above definitions into the outer region solutions and expanding them near the boundary, we find 
\begin{subequations}
\begin{align}
h_{ti} &= \hat h_{ti} + \frac{1}{4}r^4 \left(\hat \pi_{ti} + 4h_0 \delta \hat B_{zi}  \right) + \CO(r^5),\\
\delta B_{zi} &= \delta \hat B_{zi} + \left(\delta \hat J_{zi}  + h_0 \hat h_{ti}\right) \ln(r/r_h) + \CO(r),
\end{align}
\end{subequations}
where, in terms of the radial conserved quantity \eqref{eq:Alfven-conservedBG-2},  
\begin{equation}
  ab = 1 + \int_{s=0}^r ds  \frac{s^3b}{\sqrt{c}}(Q_1+4h_0h)  \approx 1 + \frac{1}{4}r^4Q_1 + \CO(r^5).
\end{equation}
Note also that the above near-boundary expansion reduces to the one in the zero background magnetic field case of Section \ref{sec:photon} as we take $h_0 = 0$. 

An analogous sequence of manipulations can be applied to the second pair of fluctuations, $ h_{zi}$ and $\delta B_{ti}$ from Eqs. \eqref{eq:Alfven-pair2-1}--\eqref{eq:Alfven-pair2-2}. The solutions are then 
\begin{subequations}
\begin{align}
h_{zi}(r,x^\mu) &= -\frac{ \delta \hat J_{ti}(x^\mu)}{h_0} + \hat C_2^{(1)}(x^\mu) \Psi_2^{(1)}(r) + \hat C_2^{(2)}(x^\mu) \Psi_2^{(2)}(r),\\
\delta B_{ti}(r,x^\mu) &= -\frac{\hat \pi_{zi}(x^\mu)}{4h_0} + \hat C_2^{(1)}(x^\mu)\Phi_2^{(1)}(r) + \hat C_2^{(2)}(x^\mu) \Phi_2^{(2)}(r).
\end{align}
\end{subequations}
The functions $\Psi_2$ and $\Phi_2$ satisfy 
\begin{equation}
  \frac{a}{r^3\sqrt{c}} \Psi'_2 - 4h_0 \Psi_2 = 0,\qquad \frac{r\sqrt{c}}{b}\Phi_2 - h_0 \Psi_2 = 0,
\end{equation}
the solutions to which are  
\begin{subequations}
 \begin{align}
  \Psi_2^{(1)}(r) &= c(r),\\
  \Psi_2^{(2)}(r) &= c(r) \int_{s=0}^r ds \frac{s^3}{a(s) c(s)^{3/2}} ,\\
  \Phi_2^{(1)}(r) &= h(r) + \frac{Q_2}{4h_0}, \\
  \Phi_2^{(2)}(r) &= \frac{1}{4h_0} + \int_{s=0}^{r} ds \frac{h_0 b(s)}{s\sqrt{c(s)}}\Psi_2^{(2)}\,.
  \end{align}
\end{subequations}
Note that $\Psi_2^{(1)}$, $ \Phi_1^{(2)}$ and  $\Phi_2^{(2)}$ are finite at the horizon while $\Psi_2^{(2)}$ is not. The latter solution can again be split into a finite and a divergent part as 
\begin{equation}
  \Psi_2^{(2)}(r)  = \psi_2^{(2)} + \frac{r_h^3c(r)}{a'(r_h)c(r_h)^{3/2}} \ln a, 
  \end{equation}
where
\begin{equation} 
\psi_2^{(2)} = c(r)\int_{s=0}^r ds \frac{s^3}{a(s)c(s)^{3/2}}\left (1-\frac{r_h^3 a'(s)c(s)^{3/2}}{s^3a'(r_h)c(r_h)^{3/2}}  \right),
\end{equation}
and 
\begin{equation}
  \Psi_2^{(1)} = \psi_2^{(1)},\qquad \Phi_2^{(1)}= \phi_2^{(1)},\qquad \Phi_2^{(2)} = \phi_2^{(2)}\, . 
\end{equation}
Similarly, as before, we can relate the functions $\hat C_2^{(1)}$ and $\hat C_2^{(2)}$ to their $h_0 = 0$ counterparts as  
\begin{subequations}
\begin{align}
  \hat h_{zi} &= -\frac{\delta \hat J_{ti}}{h_0} + \hat C_2^{(1)},\\
  \delta \hat B_{ti} &= -\frac{\hat \pi_{zi}}{4h_0} + \frac{Q_2}{4h_0}\hat C_2^{(1)} + \frac{C_2^{(2)}}{4h_0}\, . 
  \end{align} 
  \label{eq:Alfven-replaceC2}
\end{subequations}
Hence, this gives the following two near-boundary expansions: 
\begin{subequations}
  \begin{align}
h_{zi} &= \hat h_{zi} + \frac{1}{4}r^4 \left( \hat \Pi_{zi} + 4h_0 \delta \hat B_{ti} \right) + \CO(r^5),\\
\delta B_{ti} &= \delta \hat B_{ti} + \left(\delta \hat J_{ti} +h_0 \hat h_{zi}  \right) \ln(r/r_h) + \CO(r),
  \end{align}
\end{subequations}
which reduce to the near-boundary expansions of metric and two-form field fluctuations in the zero magnetic field case (cf. Sec. \ref{sec:photon}) and decouple from each other upon setting $h_0 =0$.

%---------------------
\paragraph{Inner region:} The solutions in the near-horizon region can be written in the following form:
\begin{equation}\label{eq:Alfven-nearHorizonExpansion}
 \begin{pmatrix}
h_{ti}(r,x^\mu)\\
h_{zi}(r,x^\mu)\\
\delta B_{ti}(r,x^\mu)\\
\delta B_{zi}(r,x^\mu)
 \end{pmatrix} 
=
\begin{pmatrix}
\CH_{ti}^{(1)}(r,x^\mu)\\
\CH_{zi}^{(1)}(r,x^\mu)\\
\CB_{ti}^{(1)}(r,x^\mu)\\
\CB_{zi}^{(1)}(r,x^\mu)
\end{pmatrix}
+
\begin{pmatrix}
\CH_{ti}^{(2)}(r,x^\mu)\\
\CH_{zi}^{(2)}(r,x^\mu)\\
\CB_{ti}^{(2)}(r,x^\mu)\\
\CB_{zi}^{(2)}(r,x^\mu)
\end{pmatrix} a(r)^{-i\omega/4\pi T}, 
\end{equation}
where $\CH_{\mu \nu}^{(n)}(r,x^\mu)$ and $\CB_{\mu \nu}^{(n)}(r,x^\mu)$ are regular functions of $r$. The regularity at the horizon implies a number of non-trivial constraints among these functions. To see this, we substitute the near-horizon solutions \eqref{eq:Alfven-nearHorizonExpansion} into the equations of motion \eqref{eq:Alfven-dyeom1}, expand the equations near the horizon and then demand that the coefficients of divergent terms (i.e. $a(r)^{-1}$, $a(r)^{-i\omega/4\pi T}$, $a(r)^{-1-i\omega/4\pi T}$ and $a(r)^{-2-i\omega/4\pi T}$) vanish. Among many constraints that emerge from this procedure, demanding that the coefficients of $a(r)^{-1}$ vanish implies that
\begin{subequations}
\begin{align}
  \d_t \CH^{(1)}_{zi} -\d_z \CH^{(1)}_{ti} &=0,\label{eq:Alfven-horizonReg1}\\
  \d_t \CB^{(1)}_{zi} -\d_z \CB^{(1)}_{ti} &=0.\label{eq:Alfven-horizonReg2}
\end{align}
\end{subequations}
Similarly, requiring that the coefficients of $a(r)^{-2-i\omega/4\pi T}$ in \eqref{eq:Alfven-dyeom1} vanish implies that 
\begin{equation}
  \CH_{ti}^{(2)} = \CB^{(2)}_{ti} = 0,\label{eq:Alfven-horizonReg3}
\end{equation}
where we have used a shorthand notation $\CH_{\mu \nu} = \CH_{\mu \nu}(r_h,x^\mu)$. These are the essential ingredients required to derive the approximate conservation law of interest to this section. 

%-------------------------------------

\paragraph{Intermediate region:} We proceed by matching the two sets of solutions in the intermediate region. The matching conditions imply that the coefficient of the terms that scale as $\ln a$ have to match. In particular,  
\begin{subequations}
  \begin{align}
\d_t \CH_{zi}^{(2)} &= \frac{4\pi T r_h^3}{a'(r_h)\sqrt{c(r_h)}} \hat C_2^{(2)}, \\
\d_t \CB_{zi}^{(2)} &= \frac{4\pi T h_0 \sqrt{c(r_h)}}{r_h a'(r_h)} \psi_1^{(2)}(r_h) \hat C_1^{(2)}.
  \end{align}
\end{subequations}
Similarly, we find the following relations for the matching of the finite part of the solutions:
\begin{subequations}
\begin{align}
\CH_{ti}^{(1)} &= -\frac{\delta \hat J_{zi}}{h_0} + \psi_1^{(2)}(r_h)\,\hat C_1^{(2)}  ,\\
\CH_{zi}^{(1)} + \CH_{zi}^{(2)} &=  -\frac{\delta \hat J_{ti}}{h_0}+c(r_h)\,\hat C_2^{(1)} + \psi_2^{(2)}(r_h) \hat C_2^{(2)} ,\\
\CB_{ti}^{(1)} &= -\frac{\hat \pi_{zi}}{4h_0} + \phi_2^{(2)}(r_h)\,\hat C_{2}^{(2)} ,\\
\CB_{zi}^{(1)} + \CB_{zi}^{(2)} &= -\frac{\hat \pi_{ti}}{4h_0} + \frac{Q_1}{4h_0}\hat C_1^{(1)} + \phi_1^{(2)}(r_h)\hat C_1^{(2)}  .
\end{align}
\end{subequations}

To understand the constraints imposed by the regularity of $\CH_{\mu \nu}$ and $\CB_{\mu \nu}$, and consequently of $\hat C_{1,2}^{(n)}$, we replace $\hat C_{1,2}^{(n)}$ by variables in Eqs. \eqref{eq:Alfven-replaceC1}--\eqref{eq:Alfven-replaceC2}. Henceforth, for simplicity, we will also turn off the sources for both $\hat \pi_{\mu \nu}$ and $\delta \hat J_{\mu \nu}$. The first set of relations that follows from regularity conditions implies that
\begin{equation}
\begin{aligned}
  &\left(\d_t + \frac{1}{\tau_1}\right) \left(\pi_{zi} + 4h_0 \ln(\CM r_h) \delta \hat J_{ti}\right) - v_1^2 \d_z \left( \hat \pi_{ti} + 4h_0 \ln(\CM r_h)\delta \hat J_{zi} \right)
   - \CD_1\d_z \delta \hat J_{zi} = 0  ,
\end{aligned}\label{eq:Alfven-almostconserved-mixed-1}
\end{equation}
where
\begin{equation}
v_1^2 = \frac{\psi_1^{(2)}(r_h)}{\psi_2^{(2)}(r_h)},\quad 
  \CD_1 = \left( \frac{c(r_h) - Q_1\psi_1^{(2)}(r_h)}{h_0 \psi_2^{(2)}(r_h)}  \right),\quad \tau_1^{-1} = \frac{4\pi T r_h^3}{a'(r_h)\psi_2^{(2)}(r_h)\sqrt{c(r_h)}}\, .
\end{equation}
We briefly defer the physical interpretation of this result.    The other relation is significantly more complicated:
\begin{equation}
\begin{aligned}
&\left(\d_t + \frac{1}{\tau_2}\right)\left(\left( \ln(\CM r_h) - \frac{Q_1}{4h_0^2} \right)\delta \hat J_{zi} + \frac{\hat \pi_{ti}}{4h_0}\right) -\CD_2 \d_t \delta \hat J_{zi}  
 -v_2^2 \d_z \left( \ln(\CM r_h)\delta \hat J_{ti} + \frac{\hat \pi_{zi}}{4h_0} \right) = 0 ,
\end{aligned}  \label{eq:Alfven-almostconserved-mixed-2}
\end{equation}
where
\begin{equation}
  v_2^2 = \frac{\phi_2^{(2)}(r_h)}{\phi_1^{(2)}(r_h)},\quad \CD_2 = -\frac{Q_1/4h_0^2}{\left(4h_0\phi_1^{(2)}(r_h)\right)},\quad \tau_2^{-1} = \frac{4\pi Th_0\psi_1^{(2)}(r_h)\sqrt{c(r_h)}}{r_h a'(r_h)\phi_1^{(2)}(r_h)}\, .
\end{equation}

\paragraph{Summary:}  We now provide the physical interpretation of the above holographic results.   For simplicity, we will focus on the regime $h_0 \ll T^2$, where we will see how quasihydrodynamics emerges. The analysis is similar for more generic systems.  In this regime, $|Q_1| \gg 4h_0^2 \ln (\mathcal{M}r_h)$, and we find that $b(r_h)\approx c(r_h)\approx 1$, $\phi^{(2)}_1 \approx\phi^{(2)}_2 \approx 1/4h_0 $, and that $\psi^{(2)}_2 = \mathcal{O}(h_0)$, while \begin{equation}
    \psi^{(2)}_1  = -\frac{1}{Q_1} + \frac{\alpha h_0^2}{Q_1^2} + \mathcal{O}(h_0^3),
\end{equation}
where $\alpha$ is a dimensionless  coefficient which can be computed.

It is instructive to change variables to those used in Section \ref{sec:MHD}. Namely, the `velocity field'
  \begin{align}
\delta u^y &=\frac{1}{\chi}\pi^{ty},  \qquad \chi = \frac{4h_0^2}{e_r^2} +sT,
  \end{align}
along with $\delta B^y = \delta \hat J^{ty}$ and $\delta E^x = \delta \hat J^{yz}$;  recall the definition of $e_r$ in (\ref{eq:renormalized-e}). Here, $\chi$ denotes the susceptibility of the transverse momentum and is proportional to $(\varepsilon + p)$ for the system described in Section \ref{sec:MHD}. Eq. (\ref{eq:sec4bcons}) then implies that \begin{subequations}\label{eq:4bfincons}
  \begin{align}
  \chi \d_t \delta u^y + \d_z \pi^{zy} &=0,\\ 
  \partial_t \delta B^y + \partial_z \delta E^x &= 0.
  \end{align}
\end{subequations}
The first equation is the conservation of momentum. The second equation is Faraday's law.   

Finally, we turn our attention to the remaining two equations of motion.  Firstly, in the limit described above, we find that (\ref{eq:Alfven-almostconserved-mixed-1}) becomes 
\begin{equation}
\pi^{zy} \approx -\frac{1}{4\pi T} \d_z \left( \chi \delta u^y  +  \frac{4h_0}{e_r^2}\delta E^x \right).
\label{eq:4bviscosity}
\end{equation}   
Note that we have not included the term with $\partial_t  \pi^{zy}$ as it is subleading relative to the constant term in $ \pi^{zy}$ when $\omega \ll T$ (the regime of validity of our analysis).   Hence, $\pi^{zy}$ is \emph{not} a quasihydrodynamic degree of freedom.   Nevertheless, it was convenient to carry it through the calculation as if it was quasihydrodynamic.  Indeed, (\ref{eq:4bviscosity}) reduces to a first-order constitutive relation within hydrodynamics, and as $h_0\rightarrow 0$, we recover the classic holographic result that the shear viscosity $\eta$ obeys $\eta = s/4\pi $.   At first order in $h_0$, we observe a correction to the stress tensor arising from the dynamical electric field, which would not have been included  in conventional MHD.   The ellipsis in (\ref{eq:4bviscosity}) denotes higher-order corrections in $h_0 $, which we have neglected.

The second equation of motion follows from (\ref{eq:Alfven-almostconserved-mixed-2}): \begin{equation}
    \partial_t \delta E^x + \partial_z \delta B^y = -\frac{1}{\tau} \left(\delta E^x + h_0 \delta u^y\right) + \cdots , \label{eq:4bohmic}
\end{equation}
where $1/\tau$ is given by (\ref{eq:sec4atau}). This is precisely Ampere's law in a plasma.   Since the Ohmic current must be evaluated in the rest frame of the fluid, we obtain a velocity-dependent term in (\ref{eq:4bohmic}), exactly analogous to (\ref{MHD3}).   Again, $\cdots$ denotes corrections in $h_0$ which we have neglected for simplicity.

Combining (\ref{eq:4bfincons}), (\ref{eq:4bviscosity}) and (\ref{eq:4bohmic}) we thus arrived at a holographically derived theory of quasihydrodynamic MHD with dynamical electric fields in the transverse Alfv\'en channel, which is exactly analogous to our discussion in Section \ref{sec:MHD}.   In holography, we may carry out the computation to higher orders in $h_0$ in principle, although we will not systematically discuss the higher-order contributions here.    We expect that at sufficiently low $T$, the electric field will remain a quasihydrodynamic mode at higher orders in $h_0$, but have not found a simple analytic proof of this. 

As we have shown in this section, a holographic analysis of the bulk theory \eqref{eq:MHDintro-bulkaction} with a dynamical metric and a two-form gauge field of \cite{Grozdanov:2017kyl,Hofman:2017vwr} opens the door to a systematic derivation of not only MHD but also its various extensions that pertain to the physics of plasmas.

%%%%%%%%%%%%%%%%%%%%%%%%%%%%%%%%%%%%%%%%%
%%%%%%%%%%%%%%%%%%%%%%%%%%%%%%%%%%%%%%%%%

\section{Quasihydrodynamics from holography II: M\"uller-Israel-Stewart theory and higher-derivative gravity}\label{sec:MIS-Hol}

In this section, we show how the linearized conformal M\"uller-Israel-Stewart (MIS) equations from Section \ref{sec:MIS} can be derived from a systematic expansion in an example of a holographic higher-derivative theory.   In terms of the stress-energy tensor $T^{\mu\nu}$ and gapped modes $\Pi^{\mu\nu}$, the linearized fluctuations of the MIS equations \eqref{ISlin} can be written as
\begin{subequations} \label{eq:GB-deried}
\begin{align}
&\text{scalar:}  &   \d_t  \delta T^{xy}  &= - \frac{1}{\tau}  \delta T^{xy} , \label{eq:GB-deriedMIS1}\\
&\text{shear:}   &  \d_t \delta T^{xz} +\frac{\fD}{\tau} \d_z  \delta T^{tz}  &= -\frac{1}{\tau}  \delta T^{xz} ,\label{eq:GB-deriedMIS2}\\
&\text{sound:}   &  \d_t \Pi^{zz} + \frac{4}{3} \frac{\fD}{\tau} \d_z  \delta T^{tz} &= -\frac{1}{\tau} \Pi^{zz},\label{eq:GB-deriedMIS3}
\end{align}   
\end{subequations}
where the fluctuations are functions of $t$ and $z$ and the diffusion constant $\fD$ is defined in Eq. \eqref{MISDiffCoeff}. $\delta T^{\mu\nu}$ denotes the first-order perturbation of $T^{\mu\nu}$ and in the sound channel, $\Pi^{zz}$ is defined as $\Pi^{zz} \equiv \ \delta T^{zz} - \delta p$, where $\delta p$ is the perturbation of the pressure. As explained in Section \ref{sec:MIS}, the three equations in \eqref{eq:GB-deried} can be written in a covariant form of a single approximate conservation law. In particular, they are the $xy$, $xz$ and $zz$ components of Eq. \eqref{eq:MIS-3indexJ}. 

To show how the MIS equations arise from a dual gravitational description, we focus on the Einstein-Gauss-Bonnet theory  \cite{Brigante:2007nu,Grozdanov:2014kva,Grozdanov:2015asa,Grozdanov:2016fkt}, which is a useful holographic toy model for analyzing thermal field theories at intermediate coupling strengths. The crucial feature of the theory, and of other higher-derivative theories, is the emergence of quasinormal modes with a purely relaxing, imaginary dispersion relation $\omega(k)$ \cite{Grozdanov:2016vgg,Grozdanov:2016fkt,Casalderrey-Solana:2017zyh}. As was demonstrated in \cite{Grozdanov:2016vgg,Grozdanov:2016fkt}, these modes reproduce many expected spectral properties of thermal theories in transition from strong, towards weak coupling. They exhibit the emergence of quasiparticle-like excitations in the spectrum, formation of branch cuts \cite{WitczakKrempa:2012gn,Grozdanov:2016vgg,Grozdanov:2016fkt} and the quasiparticle transport peak in the spectral function at $k=0$ \cite{Solana:2018pbk}. Moreover, the existence of purely relaxing quasinormal modes leads to the breakdown of hydrodynamics, formally defined at the critical momentum $k_c$ at which the collision occurs, given some $\lgb$. In the shear channel, this is a results of the collision of diffusive and gapped poles. The breakdown of hydrodynamics also leads to a longer hydrodynamization time \cite{Grozdanov:2016zjj,DiNunno:2017obv}, relevant in heavy-ion collisions \cite{CasalderreySolana:2011us,Heller:2016gbp,Romatschke:2017ejr}, and a longer isotropization time \cite{Andrade:2016rln,Atashi:2016fai}.
 
The action of the Einstein-Gauss-Bonnet theory in five dimensions is
\begin{equation}\label{GBAction}
    S = \frac{1}{2\kappa_5} \int d^5x \sqrt{-g} \left[ R + \frac{12}{L^2} + \frac{\lgb L^2}{2}\left( R_{abcd}R^{abcd}-4R_{ab}R^{ab}+ R^2  \right)  \right].
\end{equation}
We set the Anti-de Sitter (AdS) radius to $L=1$. An especially useful feature of this theory lies in the fact that its equations of motion involve only first and second derivatives.  Thus, in principle, the coupling $\lgb \in (-\infty,1/4]$ can be treated non-perturbatively. The background solution for this theory can be written in the form
\begin{equation}\label{GB-Metric}
    ds^2 =\frac{L^2}{r^2}\left[-a(r)b(r) dt^2 + \frac{b(r)}{a(r)}dr^2 + dx^2 + dy^2 + dz^2  \right],
\end{equation}
where 
\begin{align}\label{eq:GB-background}
a(r) = N_{{\scriptscriptstyle GB}} f(r),&& b(r) = N_{{\scriptscriptstyle GB}},&&  f(r) = \frac{1-\sqrt{1-4\lgb  \left(1-r^4/r_h^4\right)}}{2\lgb }.
\end{align}
The symbol $r_h$ denotes the radial position of the horizon. The AdS boundary is at $r=0$. We set $N_{{\scriptscriptstyle GB}}^2 f(0) =1$ to ensure that the boundary speed of light equals to one. Moreover, the Hawking temperature is given by $T = N_{{\scriptscriptstyle GB}} / (\pi r_h )$. For further details regarding this theory, see \cite{Grozdanov:2016fkt}.

Increasing the {\em negative} value of the higher-derivative coupling constant $\lgb$, i.e. increasing $-\lgb$, can be though of as tuning the dual coupling constant away from infinity. In the regime of large $-\lgb$, the relaxational quasinormal mode comes into the regime of $|\omega| / T \ll 1$ and has a non-trivial interplay with the hydrodynamic modes \cite{Grozdanov:2016vgg,Grozdanov:2016fkt}. Its gap can be defined as
\begin{align}
\omega_\mathfrak{g} \equiv \omega(k=0) = - \frac{i}{\tau(\lgb)} .
\end{align}
This is the mode that plays the role of $\Pi^{\mu\nu}$ in the MIS theory. As we show below,  its associated relaxation time $\tau$ can be found analytically when $\tau(\lgb)T \gg 1$ . Moreover, we will also find the speed of propagation of modes at large $k$, $v$, which was discussed in Sections \ref{sec:DS} and \ref{sec:MIS}, cf. Eqs. \eqref{DS-speed} and \eqref{MIS-speed-Shear}. In particular, $v^2 = \fD / \tau$. The results derived from the action \eqref{GBAction} are 
\begin{subequations}\begin{align}\label{eq:GB-c2andtau}
    \tau &= \frac{\psi_2(r_h)}{r_h^3N_{{\scriptscriptstyle GB}} },  \\
    v^2 &= \frac{\psi_1(r_h)\psi_2'(0)}{\psi_2(r_h)\psi_1'(0)} \label{eq:GB-c2andD} ,
\end{align}\end{subequations}
where $\psi_1$ and $\psi_2$ are function of the background metric:
\begin{subequations}
\begin{align}
    \psi_1(r) &= r_h^4 \int_{s=0}^{r/r_h} ds\, \frac{s^3}{1-2\lgb f(r_hs)},\label{eq:GB-defpsi1}\\
    \psi_2(r) &= \frac{r_h^4}{4}\ln f(0) + r_h^4 \int_{s=0}^{r/r_h} ds\, \frac{s^3}{f(r_hs)}\left( 1-2\lgb f(r_hs) -\frac{f'(r_hs)}{s^3 f'(r_h )} \right).\label{eq:GB-defpsi2}
\end{align}    
\end{subequations}
Performing the integral $\psi_2$ explicitly, we obtain the relaxation time 
\begin{equation}
    \tau = \frac{1}{8\pi T}\left( \ggb (\ggb +2) -3 + 2 \ln\left( \frac{2}{\ggb+1 } \right) \right),
\end{equation}
where $\ggb  = \sqrt{1-4\lgb }$. The expression agrees precisely with the (zero-momentum) frequency of the gap of the non-hydrodynamic modes in all three channels of Einstein-Gauss-Bonnet gravity obtained from the analytic quasinormal mode calculations of Refs. \cite{Grozdanov:2016vgg,Grozdanov:2016fkt}:
\begin{align}
\wfr_{\mathfrak{g}} = \frac{\omega_\mathfrak{g}}{2\pi T} = - \frac{4 i }{\ggb (\ggb +2) -3 + 2 \ln\left( \frac{2}{\ggb+1 } \right)} .
\end{align}
In particular, see Eq. (2.44) of Ref. \cite{Grozdanov:2016fkt}. Importantly, we note that unlike in MIS theory, cf. Sec. \ref{sec:MIS}, the relaxation time $\tau$ that is derived from the gravity bulk, which encodes all higher-order corrections to hydrodynamics, does not equal the second-order transport coefficient $\tau_\Pi$ \cite{Grozdanov:2015asa,Grozdanov:2014kva,Grozdanov:2016fkt}:
\begin{align}
\tau_\Pi = \frac{1}{2\pi T} \left( \frac{1}{4} \left(1+\ggb\right) \left( 5+\ggb - \frac{2}{\ggb}\right) - \frac{1}{2} \ln \left[\frac{2 \left(\ggb+1\right)}{\ggb }\right]  \right).
\end{align}
However, in the large $-\lgb$ limit, as $\lgb \to - \infty$,
\begin{align}
\lim_{\lgb \to - \infty} \tau = \lim_{\lgb \to - \infty} \tau_\Pi = - \frac{\lgb}{2\pi T}.
\end{align}
Finally, the speed $v$ is given by
\begin{align}
v^2 = - \frac{8 \lgb }{\ggb (\ggb +2) -3 + 2 \ln\left( \frac{2}{\ggb+1 } \right)} . 
\end{align}
Note that the reason that $v$ is superluminal stems from the fact that it is computed from taking the limit of $k\to\infty$---the limit in which the Einstein-Gauss-Bonnet theory suffers from various known UV problems and instabilities \cite{Buchel:2009tt,Camanho:2014apa,Grozdanov:2016fkt,Andrade:2016yzc}. The theory should always be thought of as one with a UV cut-off on $\omega$ and $k$ for any non-zero $\lgb$. For this reason, acausal UV behavior is irrelevant for the $\omega \ll T$ and $k\ll T$ regime of interest to this work in which the theory of quasihydrodynamics is applicable.  

The remaining of this section is devoted to demonstrating how the equations \eqref{eq:GB-deried} emerge from the matching conditions in the bulk.

%--------------
\subsection{Scalar channel}
We begin by deriving the scalar channel equation \eqref{eq:GB-deriedMIS1}. To do this, we study the decoupled $\delta g_{xy}$ fluctuation of the black brane metric \eqref{GB-Metric}.
\begin{equation}
\delta(ds^2) = \delta g_{xy} (r,t,z) dx dy  \equiv \frac{2}{r^2} h_{xy}(r,t,z) dx dy . 
\end{equation}
The bulk equation of motion for this mode, in the Fourier basis $h_{xy}(r,t,z) \sim h_{xy}(r)e^{-i\omega t + i k z}$, can be written as
\begin{equation}\label{eq:GBscalar-eom}
  \d_r \left( \frac{f \d_r h_{xy}}{r^3(1-2\lgb f)} \right)= - \frac{\omega^2-k^2f\CA}{r^3N_{{\scriptscriptstyle GB}} ^2 f (1-2\lgb f)}h_{xy} ,
\end{equation}
where the coefficient $\CA$ is 
\begin{equation}\label{eq:GB-defCA}
  \CA = N_{{\scriptscriptstyle GB}} ^2 \left( 1- \frac{2\lgb rf'}{1-2\lgb f} \right).
  \end{equation}
The matching procedure then proceeds in the manner outlined in Section \ref{sec:Outline-Hol}.
\paragraph{Outer region:}
In the outer region, $e^{-4\pi T/\omega}/T \ll r_h -r$, where we neglect the right-hand side of \eqref{eq:GBscalar-eom}, we find that the solution for $h_{xy}$ is 
\begin{equation}
h_{xy}(r,x^\mu) = \hat h_{xy}(x^\mu) + \hat \pi_{xy}(x^\mu) \Psi_2(r) ,
\end{equation}
where $\Psi_2(r)$ can be written as 
\begin{equation}\label{eq:GBscalar-defPsi2}
  \Psi_2(r) = r_h^4 \int^{r/r_h}_{s=0}ds\,  \frac{s^3}{f(r_h s)} \Big(1-2\lgb  f(r_h s)\Big) ,
\end{equation}
which diverges at the horizon. We split $\Psi_2(r)$ into a finite part, $\psi_2$, and the part that is logarithmically divergent at the horizon:
\begin{equation}
\begin{aligned}
  \Psi_2(r) &=   \frac{r_h^3}{f'(r_h)}\int^{r/r_h}_{s=0}ds \frac{f'(r_h s)}{f(r_h s)} +r_h^4 \int^{r/r_h}_{s=0} ds \frac{s^3}{f(r_h s)}\left( 1- 2\lgb  f(r_h s) - \frac{ f'(r_h s)}{s^3 f'(r_h )} \right)\\
  &= -\frac{r_h^4}{4} \ln \left( f(r) \right) +\psi_2(r),
  \end{aligned} \label{eq:GB-splitPsi2topsi2}
\end{equation}
where $\psi_2(r)$ was stated in Eq. \eqref{eq:GB-defpsi2}. In terms of boundary operators, $\hat \pi_{xy} = \delta T_{xy}$.

\paragraph{Inner region:} 
The inner, near-horizon region is defined by $r_h-r \ll 1/ T$. There, we make the following ansatz for $h_{xy}$:
\begin{equation}
    h_{xy}(r,x^\mu) = \CH^{(1)}_{xy}(r,x^\mu) + \CH^{(2)}_{xy}(r,x^\mu) f(r)^{-i\omega/4\pi T},
\end{equation}
where $\CH^{(i)}_{\mu\nu}(r,x^\mu)$ are regular function at the horizon. We then substitute the above ansatz into the equation of motion \eqref{eq:GBscalar-eom} and demand that the coefficients of all divergent pieces vanish in the near-horizon expansion. For the coefficient of $f(r)^{-1}$, we find that 
\begin{equation}
    \CH^{(1)}_{xy} = 0,
\end{equation}
where we have used the shorthand notation $\CH^{(i)}_{xy} \equiv \CH^{(i)}_{xy} (r_h,x^\mu)$. 

\paragraph{Intermediate region:}
The intermediate region, defined as $e^{-4\pi T/\omega}/T \lesssim r_h-r \lesssim 1/T $, is where we match the two solution from outer and inner regions. Firstly, we expand the logarithmically divergent term of the inner region solution,
\begin{equation}
    h_{xy}(r\approx r_h) = \CH^{(2)}_{xy} - \frac{i\omega}{4\pi T} \CH^{(2)}_{xy} \ln f =\CH^{(2)}_{xy} + \frac{\ln f}{4\pi T}\d_t \CH^{(2)}_{xy} .
\end{equation}
Secondly, we demand that the two branches of solutions match:
\begin{align}
\frac{1}{4\pi T}\d_t \CH_{xy}^{(2)} = -\frac{r_h^4}{4} \hat \pi_{xy},&& \CH^{(2)}_{xy} = \hat h_{xy} + \hat \pi_{xy} \psi_2(r_h).
\end{align}
Thirdly, using the above relation for $\d_t \CH^{(2)}$ and the expression for the Hawking temperature $T$, we find that 
\begin{equation}
\psi_2(r_h)\d_t \hat\pi_{xy} = -\d_t \hat h_{xy}-N_{{\scriptscriptstyle GB}} r_h^3 \hat \pi_{xy} .   
\end{equation}
Finally, we can turn off the source $\hat h $ to show that the above matching condition takes the form of the first approximately conserved current in \eqref{eq:GB-deriedMIS1}, and thus of the linearized MIS theory, with the relaxation time $\tau$ stated in Eq. \eqref{eq:GB-c2andtau}. 

\subsection{Shear channel}
In this shear channel, we turn on the metric fluctuations
\begin{equation}
    \delta (ds^2) = \sum_{i=x,y}\frac{2}{r^2}\left( h_{ti}(r,t,z) dt dx^i + h_{zi} (r,t,z)dz dx^i \right) ,
\end{equation}
where we have used the radial gauge $h_{r\mu}=0$. The relevant set of coupled dynamical equations of motions is
\begin{subequations}\begin{align}
    \d_r \left(\frac{1-2\lgb f}{r^3}\d_r h_{ti}  \right) - \frac{k\CA}{N_{{\scriptscriptstyle GB}} ^2r^3f}(1-2\lgb f)\left(\omega h_{iz} + k h_{ti}  \right) & = 0,\\
   \d_r \left( \frac{f}{r^3(1-2\lgb f)}  \d_r h_{iz}\right) + \frac{\omega(1-2\lgb f)}{N_{{\scriptscriptstyle GB}} ^2r^3f}\left(\omega h_{iz} + k h_{ti}   \right) & = 0 ,
\end{align}\label{eq:GB-shear-eom2ndorder}
\end{subequations}
and a first-order constraint equation
\begin{equation}\label{eq:GB-shear-eom1storder}
  \omega \d_r h_{ti} + k  f \CA \, \d_r h_{iz} = 0 ,
\end{equation}
where the coefficient $\CA$ was defined in \eqref{eq:GB-defCA}. As before, it is the first-order constraint \eqref{eq:GB-shear-eom1storder} that implies the conservation of the transverse momentum $T^{ti}$. This is not the case for the approximate conservation of $\delta T^{iz}$, which instead arises from the matching condition in the bulk. Note that the outer, the inner and the intermediate matching regions are defined in the same way as in the scalar channel.
 
%-------
\paragraph{Outer region:}
As in the scalar channel, first consider the outer region, where we write the metric fluctuations as
\begin{subequations}
\begin{align}
    h_{ti}(r,x^\mu) &= \hat h_{ti}(x^\mu) + \hat \pi_{ti}(x^\mu) \Psi_1(r), \\
  h_{iz}(r,x^\mu) &= \hat h_{zi}(x^\mu) + \hat \pi_{zi}(x^\mu) \Psi_2(r).
\end{align}  
\end{subequations}
The functions $\Psi_{1,2}$ are then split into a finite and a singular part at the horizon,
\begin{equation}
    \Psi_{1,2}(r) =  \psi_{1,2}(r) + \text{singular part} .
\end{equation}
The procedure for finding $\Psi_2$ is identical to the one the scalar channel, cf. Eq. \eqref{eq:GB-splitPsi2topsi2}. On the other hand, $\Psi_1$ contains no singular part and thus we can express it as 
\begin{equation}\label{eq:GB-defpsi1}
    \Psi_1(r) = \psi_1(r) = r_h^4 \int^{r/r_h}_{s=0} ds \, \frac{s^3}{1-2\lgb f(r_h s)} \, .
\end{equation}
Due to the constraint \eqref{eq:GB-shear-eom1storder}, $\hat\pi_{ti}$ and $\hat\pi_{iz}$ are not independent,
\begin{equation}\label{eq:GB-shear-derivedConservedP}
    \frac{\psi_1'(0)}{\psi_2'(0)} \d_t \hat\pi_{ti} - \d_z \hat\pi_{zi} = 0 ,
\end{equation}
which upon performing holographic renormalization yields the conserved transverse momentum $\d_\mu \delta T^{\mu}_{\;\; i} = 0$. 

\paragraph{Inner region:} 
We proceed by writing $h_{ti}$ and $h_{zi}$ at the horizon in terms of regular and infalling parts:
\begin{equation}
    \begin{pmatrix}
    \delta h_{ti}(r,x^\mu) \\ \delta h_{zi}(r,x^\mu)
    \end{pmatrix}
    = 
    \begin{pmatrix}
    \delta \CH^{(1)}_{ti}(r,x^\mu) \\ \delta \CH^{(1)}_{zi}(r,x^\mu)
    \end{pmatrix}
    + 
    \begin{pmatrix}
    \delta \CH^{(2)}_{ti}(r,x^\mu) \\ \delta \CH^{(2)}_{zi}(r,x^\mu)\, 
    \end{pmatrix} f(r)^{-i\omega/4\pi T} .
\end{equation}
The equation of motion near the horizon then implies
\begin{subequations}
\begin{align}\label{eq:GB-shear-regularity}
    \CH^{(2)}_{ti} &= 0, \\ 
    \d_t \CH_{zi}^{(1)} - \d_z \CH^{(1)}_{ti} & = 0. 
 \end{align}  
\end{subequations}

\paragraph{Intermediate region:}
Using the same matching procedure as in the scalar channel, the logarithmically divergent terms imply that
\begin{equation}
  \frac{1}{4\pi T}\d_t\CH^{(2)}_{zi} = - \frac{r_h^4}{4} \hat\pi_{zi}.
\end{equation}
Similarly, matching the finite pieces gives 
\begin{align}
  \CH_{ti}^{(1)} = \hat h_{ti}+ \psi_1(r_h)\hat\pi_{ti}(x^\mu),&& \CH^{(1)}_{zi} + \CH^{(2)}_{zi} = \hat h_{zi} + \psi_2(r_h) \hat\pi_{zi}(x^\mu) .
\end{align}
Turning off the source $\hat h$ and substituting the above matching conditions into Eq. \eqref{eq:GB-shear-regularity}, which is a consequence of horizon regularity, we find the following relation: 
\begin{equation}
   \d_t \hat\pi_{zi} - \frac{\psi_1(r_h)}{\psi_2(r_h)} \d_z \hat\pi_{ti} = -\left(\frac{N_{{\scriptscriptstyle GB}} r_h^3}{\psi_2(r_h)}\right) \hat\pi_{zi}.
\end{equation}
Finally, combining this result with the conservation of transverse momentum and using the holographic dictionary for the stress-energy tensor, we recover the conserved and the approximately conserved quasihydrodynamic MIS equations for $ \delta T^{ti}$ and $\delta T^{zi} $: 
\begin{subequations}\label{eq:Gb-shear-fullseteoms}\begin{align}
   \d_t \delta T^{ti} + \d_z \delta T^{zi} &=0,\\
   \d_t \delta T^{zi} + v^2 \d_z  \delta T^{ti} &= -\frac{1}{\tau} \delta T^{zi},
\end{align}\end{subequations}
where one can show after holographic renormalization that
\begin{align}
    \hat\pi_{zi} \propto \delta T^i_{\;\;z} ,&& \hat \pi_{ti} \propto \frac{\psi_2'(0)}{\psi_1'(0)}  \delta T^i_{\;\; t} .
\end{align}
Both expressions have the same proportionality constant. 

%------------------
\subsection{Sound channel}

Lastly, in the sound channel, we turn on the following perturbations the metric, again in the radial gauge, 
\begin{equation}
 \delta(ds^2) = \frac{1}{r^2}\left( h_{tt} dt^2 + 2h_{tz}dt\,dz +h_{xx}dx^2 + h_{yy}dy^2 + h_{zz}dz^2 \right)   .
\end{equation}
There are four second-order equations of motion for $h_{tt}$, $h_{tz}$, $h_{ii} \equiv h_{xx}+h_{yy}$ and $h_{zz}$. The equation of motion for $h_{tt}$ does not play a direct role in the computation (the relevant part of it is a derivative of a constraint). The remaining equations are 
\begin{subequations}\label{eq:GB-sound-2ndordereom}
\begin{align}
 \frac{d}{dr}\left( \frac{(1-2\lgb f}{r^3}h'_{tz} \right) + \frac{\omega k}{r^3f} \Big( 1-\lgb \left( 2f - rf' \right) \Big) h_{ii} & = 0 ,\label{eq:GB-sound-2ndordereom-1}\\
\frac{d}{dr}\left( \frac{\sqrt{f}(1-2\lgb  f)}{r^3} (h_{ii}'+h_{zz}') \right)- \frac{k^2}{r^3\sqrt{f}}\Big( 1-\lgb \left( 2f - rf' \right) \Big) h_{ii} &=0 ,\label{eq:GB-sound-2ndordereom-2}\\
 \frac{d}{dr} \left( \frac{f }{r^3}(1-2\lgb f)^{-1} (h_{ii}'-2h'_{zz})\right) - \frac{1}{N_{{\scriptscriptstyle GB}} ^2r^3f} \Bigg( 2N_{{\scriptscriptstyle GB}} ^2 k^2 f h_{tt} + 4\omega k h_{tz}& \nn 
+ 2\omega^2 h_{zz} + \left(\omega^2 - \CA k^2 f\right)h_{ii}\Bigg) &= 0  . \label{eq:GB-sound-2ndordereom-3}
 \end{align} 
 \end{subequations}
 In addition, we also have three first-order constraint equations
 \begin{subequations} \label{eq:GB-sound-1stodereom}
\begin{align}
 \omega \left( h_{ii}'+ h'_{zz} \right) + k h'_{tz} - \frac{f'}{f}\left(k h_{tz} +\frac{1}{2} \omega (h_{ii} + h_{zz}) \right) &=0, \label{eq:GB-sound-1stodereom1}\\
\omega h'_{tz} +  N_{{\scriptscriptstyle GB}} ^2k f\, h'_{tt} + \frac{1}{2} N_{{\scriptscriptstyle GB}} ^2k f' h_{tt} - N_{{\scriptscriptstyle GB}} ^2f k \CA h' &=0 , \label{eq:GB-sound-1stodereom2}\\
 \frac{r}{f}\left( \omega^2 h_{zz} +2\omega k h_{tz} + N^2f k^2 h_{tt} + (\omega^2-k^2 f\CA)h_{ii} \right) &\nn
+3 N_{{\scriptscriptstyle GB}} ^2f h_{tt}' - \left( 3f\CA - \frac{N_{{\scriptscriptstyle GB}} ^2f'}{2(1-2\lgb f)} \right)(h'_{zz} + h_{ii}' ) &= 0 . \label{eq:GB-sound-1stodereom3}
\end{align}  
\end{subequations}
As we shall see, the constraint equations in \eqref{eq:GB-sound-1stodereom} will becomes the conservation of energy, longitudinal momentum and the tracelessness condition of the stress-energy tensor, respectively. 

\paragraph{Outer region:}
The outer region solutions for $h_{tz}$, $h_{ii}-2h_{zz}$ and $h+h_{zz}$ can easily be obtained from Eqs. \eqref{eq:GB-sound-2ndordereom}. We find
\begin{subequations}\begin{align}
  h_{tz} &= \hat h_{tz}(x^\mu) + \hat \pi_{tz}(x^\mu) \Psi_1(r),\\
  h_{ii} - 2 h_{zz} &= \hat h_{ii}(x^\mu)- 2h_{zz}(x^\mu) + \hat \pi_2(x^\mu) \Psi_2(r),\\
  h_{ii} + h_{zz} &= \hat h + \hat h_{zz} + \hat \pi_3(x^\mu) \Psi_3(r).
\end{align}  
\end{subequations}
The functions $\Psi_1$ and $\Psi_3$ are finite everywhere in the bulk while $\Psi_2$ diverges at the horizon. The functions $\Psi_{1,2}$ are analogous to the ones that appear in the scalar and the shear channel computations. The new $\Psi_3$ is
\begin{equation}
  \Psi_3(r)=\psi_3(r) = r_h^4\int^{r/r_h}_{s=0} ds\, \frac{s^3}{\sqrt{f(r_h s)} (1-2 \lgb f(r_hs))}.
\end{equation}
We also write the solutions for $h$ and $h_{zz}$, which will prove convenient in the matching procedure:
\begin{subequations}
\begin{align}
  h_{ii} &= \hat h_{ii} +\frac{1}{3} \left(\hat\pi_2(x^\mu)\Psi_2(r) + 2\hat\pi_3(x^\mu) \Psi_3(r)  \right),\\
  h_{zz} &= \hat h_{zz} + \frac{1}{3} \left( \hat \pi_3 (x^\mu) \Psi_3(r) - \hat \pi_2(x^\mu) \Psi_2(r) \right).
\end{align}  
\end{subequations}

To better understand the role of the functions $\hat\pi_{\mu\nu}$, let us consider the first-order constraint equations in the outer region, which are 
\begin{subequations}
 \begin{align}
  \d_t (h_{ii}'+ h'_{zz}) - \d_z h'_{tz} &= 0,\\
  \d_t h'_{tz} - \d_z\left(h_{tt}'' -h_{ii}'\right) &=0 , \\
  h'_{tt} - (h'_{tz} + h_{ii}') &= 0 .
\end{align} 
\end{subequations}
After using the holographic dictionary, which is schematically $h_{\mu\nu}' \sim \delta T^{\mu}_{\;\;\nu}$, then, as claimed above, these equations become the conservation of energy $\d_\mu \delta T^{\mu}_{\;\; t}= 0$, momentum $\d_\mu  \delta T^{\mu}_{\;\;z} =0$, and the conformal tracelessness condition $ \delta T^{\mu}_{\;\;\mu} = 0$. In terms of the functions $\hat \pi_{\mu\nu}(x^\mu)$, these relations lead to
\begin{subequations}
  \begin{align}
  \psi_3'(0)\d_t \hat\pi_3 - \psi_1'(0)\d_z \hat\pi_{tz} &= 0 , \\ 
  \psi_1'(0) \d_t \hat\pi_{tz} -\frac{1}{3} \d_z \left( \psi_3'(0)\hat\pi_3 - \psi_2'(0)\hat\pi_2 \right) &= 0 .
\end{align}
\end{subequations}

\paragraph{Inner region:}
We again write down the ansatz 
\begin{equation}
  \begin{pmatrix}
h_{zz} \\
h_{ii}\\
h_{tx} 
  \end{pmatrix} = 
  \begin{pmatrix}
\CH^{(1)}_{zz}(r,x^\mu)\\
\CH^{(1)}_{ii}(r,x^\mu)\\
\CH^{(1)}_{tz}(r,x^\mu)\\
  \end{pmatrix}
  + \begin{pmatrix}
\CH^{(2)}_{zz}(r,x^\mu)\\
\CH^{(2)}_{ii}(r,x^\mu)\\
\CH^{(2)}_{tz}(r,x^\mu)\\
  \end{pmatrix} f(r)^{-i\omega/4\pi T}
\end{equation}
and substitute it into the equations of motion, which we expand in the near-horizon region. The functions $\CH^{(n)}$ are regular at the horizon. As a result, one finds a number of relations between the above near-horizon solutions. Among then, the ones relevant to the present derivation are  
\begin{align}\label{eq:soundGBregularity}
  \CH^{(1)}_{ii} = 0,&& \d_t\left(\d_t \CH^{(1)}_{zz} - 2\d_z \CH^{(1)}_{tz}\right) = 0, && \CH_{tz}^{(2)} = 0.
\end{align}
The second equation arose from the non-radial derivative part of \eqref{eq:GB-sound-2ndordereom-3}. It  implies that 
\begin{equation}
  \d_t \CH^{(1)}_{zz} - 2\d_z \CH^{(1)}_{tz} = \CF(t) ,
\end{equation}
where the function $\CF(t)$ is independent of spatial coordinates.  Assuming regularity at spatial infinity, we conclude that $\mathcal{F}(t)=0$ \cite{Chen:2017dsy}.  This will prove essential in recovering the approximate conservation law of the MIS theory. 

\paragraph{Intermediate region:}
Again, we expand the near-horizon solutions as 
\begin{equation}
  \begin{pmatrix}
h_{zz} \\
h_{ii}\\
h_{tx} 
  \end{pmatrix} = 
  \begin{pmatrix}
\CH^{(1)}_{zz} + \CH^{(2)}_{zz}\\
\CH^{(1)}_{ii}+\CH^{(2)}_{ii}\\
\CH^{(1)}_{tz} + \CH^{(2)}_{tz}\\
  \end{pmatrix}
  + \frac{1}{4\pi T}\d_t \begin{pmatrix}
\CH^{(2)}_{zz}\\
\CH^{(2)}\\
\CH^{(2)}_{tz}\\
  \end{pmatrix} \ln f , 
\end{equation}
where we have used the shorthand notation $\CH^{(n)}(r=r_h,x^\mu) = \CH^{(n)}$. The matching condition for the logarithmically divergent pieces implies that 
\begin{align}
  \d_t\CH^{(2)} = -\left( \frac{\pi T r_h^4}{3}\right) \hat\pi_2(x^\mu),&& \d_t \CH_{zz}^{(2)} = \left( \frac{\pi T r_h^4}{3} \right)\hat \pi_2(x^\mu) . \label{eq:soundGBmathcinglog}
\end{align}
For the finite terms, we find
\begin{subequations}
  \begin{align}
  \CH^{(1)}_{tz}  &= \hat h_{tz} + \hat \pi_{tz}\psi_1(r_h),\\
  \CH_{zz}^{(1)}+ \CH_{zz}^{(2)} &= \hat h_{zz} + \frac{1}{3} \left(\hat\pi_3 \Psi_3(r_h)-\hat \pi_2 \psi_2(r_h)  \right),\label{eq:soundGBmatchingzz}\\
   \CH^{(2)}_{ii} &= \hat h_{ii} + \frac{1}{3} \left( \hat\pi_2 \psi_2(r_h)  + 2 \hat \pi_3 \Psi_3(r_h)   \right),\label{eq:soundGBmatchingaa}
  \end{align}
\end{subequations}
where we used the fact that $\CH^{(2)}_{tz} = \CH^{(1)}=0$. Relations \eqref{eq:soundGBmatchingzz} and \eqref{eq:soundGBmatchingaa} can be combined into
\begin{equation}
\begin{aligned}
  \d_t \CH^{(1)}_{zz} &= \frac{1}{2} \d_t\CH^{(2)} - \d_t \CH^{(2)}_{zz}-\frac{1}{2} \psi_2(r_h) \d_t \hat \pi_2 \\
  &= -\frac{1}{2} \left(\pi T r_h^4  \right) \hat \pi_{tz} - \frac{1}{2} \psi_2(r_h) \d_t \hat \pi_2 , 
\end{aligned}
\end{equation}
where in the second line, we used the matching conditions for the divergent terms from Eq. \eqref{eq:soundGBmathcinglog}. The regularity condition \eqref{eq:soundGBregularity} then implies
\begin{equation}
  \d_t \CH^{(1)}_{zz} - 2 \d_z \CH^{(1)}_{tz} = -\frac{1}{2} \left( \pi T r_h^4\right) \hat \pi_2 - \frac{1}{2} \psi_2(r_h) \d_t \hat \pi_2 -2 \psi_1(r_h) \d_z \hat \pi_{tz} = 0 \label{eq:MISsoundFromGravity}.
\end{equation}
In other words, we find the approximate conservation law:
\begin{equation}
  \d_t \hat\pi_2 + 4\frac{\psi_1(r_h)}{\psi_2(r_h)} \d_z \hat \pi_{tz} = - \frac{ N_{{\scriptscriptstyle GB}} r_h^3}{\psi_2(r_h)}  \hat \pi_2. \label{eq:soundGBalmostconservedMIS}
\end{equation}

One can convert $\hat \pi_2$ and $\hat \pi_{tz}$ into the expectation values of the stress-energy tensor in the following way: first, we recall that the dissipative part of the stress-energy tensor can be written as $\hat \pi^{zz} =  \delta T^{zz}  - \delta p $ and that 
\begin{equation}
  \delta p = \frac{1}{3}\sum_{i=x,y,z}\delta T^{ii}   \propto \frac{1}{3} \hat \pi_3.
\end{equation}
It follows that the dissipative part of the stress-energy tensor in MIS theory is
\begin{equation}
  \Pi^{zz} = \delta T^{zz}  - \delta p \propto -\frac{1}{3} \hat \pi_2  .
\end{equation}
Similarly, one can write $\sigma^{zz}$ in terms of the $tz-$component of the stress-energy tensor as 
\begin{equation}
  \sigma^{zz} = \frac{4}{3} \d_z v^z = \frac{4}{3} \frac{1}{\varepsilon+p} \d_z  \delta T^{tz}  \propto \frac{4}{3} \frac{1}{\varepsilon+ p} \hat \pi_{tz},
\end{equation}
where $ \delta T^{tz}  \propto \hat \pi_{tz}$. Eq. \eqref{eq:soundGBalmostconservedMIS} expressing approximate conservation of $\hat \pi_2$ can finally be written as the linearized MIS equation in the sound channel (cf. Eq. \eqref{eq:GB-deriedMIS3}):
\begin{equation}
  \d_t \Pi^{zz} - \frac{4}{3} \frac{\fD}{\tau} \d_z \la \delta T^{tz}\ra = - \frac{1}{\tau} \Pi^{zz}.
\end{equation}
The relaxation time $\tau$  and the diffusion constant $\fD$ are again given by Eqs. \eqref{eq:GB-c2andtau} and \eqref{eq:GB-c2andD}.

%%%%%%%%%%%%%%%%%%%%%%%%%%%%%%%%%%%%%%%%%
%%%%%%%%%%%%%%%%%%%%%%%%%%%%%%%%%%%%%%%%%

\section{Conclusion}

In this paper, we have developed a general framework for discussing linearized hydrodynamic theories with additional approximately conserved currents, which we called  quasihydrodynamics. As shown, this framework can be used to understand a large number of  phenomenological theories discussed in previous literature, including our main two examples:  magnetohydrodynamics coupled to dynamical electric fields (which thus includes dynamical photons), and the relativistic M\"{u}ller-Israel-Stewart theory. Since a generic feature of such theories is the presence of not only massless hydrodynamic modes, but also of ``massive" long-lived modes, a systematic constructions of effective quasihydrodynamic theories is a formidable task, as the formal gradient expansion is not directly applicable.   An even more difficult task is a derivation of such effective theories from their underlying quantum microscopic description---for some recent progress in the hydrodynamic setting (without approximately conserved quantities) see \cite{bankslucas}.

With a view towards a long-term goal of building  systematic quasihydrodynamic theories, we studied the emergence of quasihydrodynamic theories in strongly coupled theories with holographic duals. As we have shown, holography can be used as a tool to systematically derive a low-energy description of systems with long-lived excitations. In particular, we have developed an explicit holographic algorithm for analytically computing the linearized quasihydrodynamic equations in a generic system, extending and simplifying earlier developments of \cite{Lucas:2015vna,Chen:2017dsy}. We first showed in detail how to carry out this procedure to unambiguosly demonstrate the existence of dynamical photons in a holographic dual of magnetohydrodynamics \cite{Grozdanov:2017kyl}. This explicitly confirms the claims made previously in \cite{Grozdanov:2017kyl,Hofman:2017vwr} that the holographic dual of a theory with a one-form symmetry encodes dynamical electromagnetism on the boundary. Moreover, it provides a systematic method for deriving extensions of MHD, which we anticipate will find a wealth of applications in plasma physics. In our second example, we uncovered the equations of MIS theory from a holographic higher-derivative Einstein-Gauss-Bonnet gravity theory, which was previously shown to be able to exhibit long-lived massive excitations within the low-energy (hydrodynamic) regime \cite{Grozdanov:2016vgg,Grozdanov:2016fkt}. Systematically derived extensions of the MIS theory may find future applications in the physics of heavy-ion collisions and other types of fluid dynamics at intermediate strength of the coupling constant. For example, with a view towards the description of a conjectured critical point of quantum chromodynamics, it would be interesting to understand the recently proposed {\em Hydro+} of \cite{Stephanov:2017ghc} from the point of view of our present work.

Beyond the ability to derive dynamical equations of motion and effective field content of quasihydrodynamic theories, we expect that our methods will be useful also for analytically recovering hydrodynamic dispersion relations in the holographic duals of ordinary fluids, superfluids, solids and more. The abstract nature of the method allows one to carry out the entire calculation in terms of gravitational background fields until the very end, be they analytically or numerically known. One is at most required to perform a set of simple final numerical integrals to obtain those dispersion relations.   

Finally, we anticipate that these methods will assist future research into the systematic development of dissipative nonlinear effective field theories for systems with weakly broken symmetries, following the series of recent advances in Refs. \cite{Grozdanov:2013dba,Haehl:2015foa,Crossley:2015evo,Torrieri:2016dko,Haehl:2015uoc,Glorioso:2016gsa,Jensen:2017kzi,Glorioso:2018wxw}.

\section*{Acknowledgements}
We thank Matteo Baggioli for discussions. S. G. would like to thank Andrei Starinets for his unwavering skepticism of the MIS theory. S. G. was supported by the U. S. Department of Energy under grant Contract Number DE-SC0011090. A. L. was supported by the Gordon and Betty Moore Foundation's EPiQS Initiative through Grant GBMF4302.  The work of N. P. was supported by Icelandic Research Fund grant 163422-052.  N. P. would like to thank the Stanford Institute for Theoretical Physics, NORDITA and CTP, Durham University for hospitality. N. P. would also like to acknowledge the support from COST Action MP1405 (QSPACE).  All authors thank NORDITA for hospitality during the program ``Bounding Transport and Chaos". 

\newpage
\bibliography{biblio}

\end{document}